\documentclass[aps,pra,reprint,amsmath,amssymb,superscriptaddress,onecolumn,longbibliography, notitlepage]{revtex4-2}
\usepackage{mathtools}
\usepackage{siunitx}
\usepackage[hidelinks]{hyperref}
\usepackage{caption}
\usepackage{float}
\usepackage{subcaption}
\usepackage{bbm}
\newcommand{\I}{\mathbbm{1}}

\begin{document}

\title{Experimental error mitigation using linear rescaling for variational quantum eigensolving with up to 20 qubits}
\author{Eliott~Rosenberg}
\email{enr27@cornell.edu}
\affiliation{Department of Physics, Cornell University, Ithaca, NY 14853, USA}
\author{Paul~Ginsparg}
\affiliation{Department of Physics, Cornell University, Ithaca, NY 14853, USA} 
\author{Peter~L.~McMahon}
\email{pmcmahon@cornell.edu}
\affiliation{School of Applied and Engineering Physics, Cornell University, Ithaca, NY 14853, USA}

\begin{abstract}
Quantum computers have the potential to help solve a range of physics and chemistry problems, but noise in quantum hardware currently limits our ability to obtain accurate results from the execution of quantum-simulation algorithms. Various methods have been proposed to mitigate the impact of noise on variational algorithms, including several that model the noise as damping expectation values of observables. In this work, we benchmark various methods, including a new method proposed here. We compare their performance in estimating the ground-state energies of several instances of the 1D mixed-field Ising model using the variational-quantum-eigensolver algorithm with up to 20 qubits on two of IBM's quantum computers. We find that several error-mitigation techniques allow us to recover energies to within 10\% of the true values for circuits containing up to about 25 ansatz layers, where each layer consists of CNOT gates between all neighboring qubits and Y-rotations on all qubits.

\end{abstract}

\maketitle

\section{Introduction}
\label{sec:intro}

Quantum computation promises to revolutionize physics and chemistry. Our ability to make quantitative predictions about many-body and strongly coupled quantum mechanical systems, such as atoms with many electrons, or systems of quarks and gluons, has, with classical computers, been severely hampered by the exponential memory requirements of storing and operating on many-body wavefunctions. Quantum computers overcome this difficulty by representing the wavefunction in the state of the qubits themselves. Noise, however, has prevented current quantum computers from surpassing the capabilities of classical computers in almost all applications, including quantum simulation.

In the long term, fault-tolerant quantum computation may be achieved using error-correction techniques that employ many physical qubits to encode each logical qubit. In the near term, there is much effort aimed at less costly methods to mitigate or compensate for the effects of noise in current quantum hardware. Some of these error-mitigation techniques involve artificially increasing the noise and extrapolating the output to zero noise \cite{Li_2017,temme2017error, endo2018practical,kandala2019error, Otten2019, He_2020}. Others involve representing the ideal circuit in terms of noisy ones (the so-called ``quasi-probability method'') \cite{temme2017error, endo2018practical}. Still others involve using replica copies of the system to effectively take powers of the density matrix and suppress errors \cite{koczor2021exponential, huggins2021virtual}, while others attempt to learn the effect of the noise by training on classically simulable circuits \cite{He_2020,czarnik2021error, montanaro2021error,shaw2021classicalquantum}.

Several proposed error-mitigation methods have successfully modeled the noise as damping the expectation values of observables towards their expectation values in maximally mixed states,
i.e., $\langle \hat O\rangle_{\rm noisy} \approx C \langle \hat O\rangle_{\rm exact} + (1-C) \mbox{Tr}[\hat O]/2^n$, where $0<C<1$, $\hat O$ is an observable, and $n$ is the number of qubits \cite{He_2020,czarnik2021error, montanaro2021error, arute2020observation, mi2021information,urbanek2021mitigating, vovrosh2021efficient, ville2021leveraging,shaw2021classicalquantum}. In this work, we consider traceless observables, so the last term is not present (although its inclusion would not present any significant difficulties). Knowing the damping factor, $C$, the exact expectation value can be recovered from the noisy one. Various methods for estimating the damping factor are proposed in the works cited above. Here we propose another method and compare the performance of these methods to each other and to zero noise extrapolation when applied to the test problem of measuring the ground state energy of a 1D mixed field Ising spin chain. We find that several methods offer significant improvements over no mitigation, enabling circuits of up to about 25 ansatz layers to yield expectation values within 10\% of the correct expectation value. Which method works best is problem-dependant, but for our test problem, we find that estimating the damping factor from the perturbative regime of the Hamiltonian works particularly well.

\section{Test problem}
\label{sec:prob_statement}
Our goal is to benchmark error mitigation methods, in order to get a sense for when excessive circuit depth impedes obtaining meaningful results. To benchmark these methods, we consider the problem of finding ground states of the 20-qubit 1D mixed-field Ising model.

The mixed-field Ising model is defined by the following Hamiltonian:
\begin{equation}
    \label{eq:ising}
    H = -J\sum_{i} Z_i Z_{i+1} - h_x \sum_i X_i - h_z \sum_i Z_i,
\end{equation}
where we impose cyclic boundary conditions on the spins. If $h_z = 0$, this reduces to the transverse-field Ising model, which is exactly solvable using a Jordan-Wigner transformation when the number of qubits is large. In this case, a term arising from the cyclic boundary conditions can be neglected \cite{PFEUTY197079}. It is, of course, also solvable in the limits $h^2_x/(J^2 + h_z^2) \ll 1$ and $J^2/h_x^2 \ll 1$. For intermediate values of the field strengths, however, it is non-integrable and quantum chaotic \cite{Kim_Huse_2013,Kim_Ikeda_Huse_2014}. Approximate methods have been developed to study this model outside of its integrable regime \cite{Wurtz_Polkonikov_2020}. For the system size studied here (20 qubits), such approximate methods are not necessary since the Hamiltonian can be exactly diagonalized on a classical computer. For a discussion of the perturbative regimes of this Hamiltonian, see Appendix~\ref{sec:appendix_pert}. Fig.~\ref{fig:ising_pert} compares the exact ground state energies of this model to those computable in second-order perturbation theory from the analytically solvable limits. Throughout this work, we set $J=1$, which sets the overall scale of $H$.

We use an ansatz to parameterize approximate ground states of this Hamiltonian. In particular, we consider a variant of the Alternating Layered Ansatz (ALT) \cite{cerezo2020costfunctiondependent, nakaji2020expressibility}, shown in Fig.~\ref{fig:ansatz} and described in Appendix \ref{sec:basis_gates}. We consider both the full ALT ansatz and a constrained version of it, with even cyclic permutation symmetry imposed. The permutation symmetry reduces the number of parameters from $n(l+1)$ to $2(l+1)$, where $n$ is the number of qubits and $l$ is the number of ansatz layers. Fig.~\ref{fig:ansatz_performance} shows that this ansatz is able to capture the ground states of these Hamiltonians as the number of layers is increased. In fact, we do not need to make $l$ very large to capture the ground state, but we do so anyway to study error mitigation.

For future applications of the Variational Quantum Eigensolver (VQE), in which the ansatz circuits are wide and deep enough that they are not classically simulable, the optimization step will need to be performed using measurements from the quantum computer. If the effect of the noise is indeed to damp expectation values of observables, and the parameter dependence of the damping factor does not noticeably shift the location of the minimum, then it is not necessary to apply error mitigation during the optimization routine. In this case, correcting the damping is more relevant after the optimization, when we evaluate the optimized circuit. We demonstrate that our symmetrized ansatz circuits can be optimized for the mixed-field Ising model on 20 qubits using unmitigated measurements from quantum computers (Figs.~\ref{fig:VQE_sydney} and \ref{fig:VQE_toronto}). Having shown this, and because our primary goal is to benchmark error-mitigation techniques, we proceed to optimize the remaining circuits classically. In all Figures other than \ref{fig:VQE_sydney} and \ref{fig:VQE_toronto}, the results were obtained with circuits that were optimized classically.

\section{Error mitigation techniques}

We consider the following questions: Given optimized circuits that approximate ground states of the Hamiltonian (\ref{eq:ising}), how can we best estimate the damping factor that characterizes the suppression of the expectation value of the Hamiltonian when evaluated on a quantum computer? How deep can we make our ansatz using this method? The remainder of this section contrasts techniques for estimating the damping factor. We do not include the method of Ref.~\cite{vovrosh2021efficient}, which requires simultaneous measurements of all of the qubits, because readout-error mitigation on a large number of qubits is a significant challenge beyond the scope of this work. We also do not include the motion-reversal method of Ref.~\cite{shaw2021classicalquantum} because, for circuits of the depths considered here, repeating them and their inverses several times leads to very small fidelities, to which it is impossible to accurately fit an exponential using a reasonable number of measurements. Further, this method relies on the assumption that the circuit and its inverse lead to similar decays in fidelity, which may not be true. We also do not include the quasi-probability method because it requires gateset tomography. We benchmark the following methods. ``Extrapolation in circuit depth" is a new method proposed here, whereas the others have been proposed in previous works. We also tested two methods that did not perform as well, which are described in Appendix \ref{sec:poorly_performing_methods}.

\subsection{Zero noise extrapolation}
\label{sec:ZNE}
In our implementation of zero noise extrapolation (ZNE) \cite{Li_2017,temme2017error, endo2018practical,kandala2019error, Otten2019, He_2020}, we artificially increase the noise by replacing CNOT gates with odd numbers of CNOT gates. For example, if we want to increase the noise by a factor of 3, we replace each CNOT gate by 3 CNOT gates. We can scale the noise by factors other than odd integers by randomly picking CNOT gates to repeat an odd number of times. We then use an exponential fit to extrapolate the measured energy to zero noise. We use three points (one of which is the unmodified circuit) to do this extrapolation.

\subsection{Calibrating using the perturbative regime}
\label{sec:pert}
Recently, the Google team observed that, at least for the specific system they studied, the damping factor was independent of the coupling strength \cite{arute2020observation}. They proposed, therefore, that one can measure the damping factor in a perturbative regime of the Hamiltonian, and then use the same damping factor outside of the perturbative regime to correct the results from the quantum computer. 

This method is, of course, limited to Hamiltonians that have a perturbative regime, which doesn't include all Hamiltonians of scientific interest. For example, instances of the Sachdev-Kitaev-Ye (SYK) Hamiltonian do not have a perturbative regime, but simulations of the model may give insights into holography and quantum gravity (e.g., \cite{Babbush_2019,Martyn_2019,Luo2019,brown2021quantum,nezami2021quantum}). Nevertheless, many Hamiltonians of interest do have a perturbative regime, so this technique may still prove useful.

The mixed-field Ising model has several different perturbative regimes, discussed in Appendix~\ref{sec:appendix_pert}. For the purposes of this error-mitigation technique, we consider only perturbation in small $h_x$; we fix $h_z = 0.1$ and consider $0 < h_x \leq 0.5$ to be the perturbative regime and $h_x = 1.5$ to be outside of the perturbative regime. While we might obtain a better estimate of the damping factor by averaging measurements over multiple perturbative regimes, approaching the desired field strength from both sides, multiple perturbative regimes is not a generic feature of Hamiltonians so we do not employ this trick.

\subsection{Extrapolation in circuit depth}
\label{sec:scaling_with_l}
In contrast to Ref.~\cite{arute2020observation}, but in agreement with the observation that exponentials work well for ZNE \cite{endo2018practical}, we expect that the damping factor should decay exponentially in the number of ansatz layers and generally observe this to be the case. Further, circuits with small depth may be classically simulable because only qubits in the backwards light cone of the measured qubits need to be included. For 1D ans\"atze with nearest-neighbor 2-qubit gates, such as the ansatz considered here, the number of qubits needed for exact classical simulation is $\mbox{min}(m+2l, n),$ where $l$ is the number of ansatz layers, $m$ is the number of qubits on which the observable acts, assumed to be adjacent, and $n$ is the total number of qubits. If 36 qubits are classically simulable, then for the Hamiltonian considered here (Eq.~\ref{eq:ising}), with $m=2$, we can simulate up to 17 ansatz layers for arbitrarily large $n$. More generally, one needs to consider the size of the backwards light cone as a function of the number of ansatz layers to determine the maximum number of layers that can be simulated classically. This method will not work as well or at all for observables that act on many qubits in the same term, such as the SYK Hamiltonian in either the Jordan-Wigner or Bravyi-Kitaev \cite{Bravyi_2002} representation. To test this mitigation method for our system, we fit an exponential to the damping factor for up to 15 ansatz layers and then use the fit to predict the damping factor for up to 50 layers.

\subsection{Calibrating by omitting 1-qubit gates}
\label{sec:zero_theta}
Another recent paper \cite{urbanek2021mitigating} proposed estimating the damping factor by eliminating all of the 1-qubit gates in the circuit. The remaining 2-qubit gates, which are all controlled-NOTs, leave the state in $|00\ldots0\rangle$, and  the damping factor is estimated as the fidelity of measuring all zeros. For our system, eliminating the single-qubit gates is achieved by setting $\vec \theta = 0$. In addition to estimating the damping factor using the fidelity, as suggested by Ref.~\cite{urbanek2021mitigating}, we also compare the measured energy to the energy in the ideal state to estimate the damping factor.

This method can be regarded as a special case of a method in which the ansatz is evaluated at classically solvable points. For our ansatz, if we pick the Y-rotations to be $\theta_i = n_i \pi/2$, for integers $n_i$, then the circuit is within the Clifford group and hence classically simulable. Using Clifford circuits to estimate the damping factor has been proposed previously \cite{czarnik2021error}. For generic choices of the $n_i$, however, the expectation value of the Hamiltonian becomes vanishingly small --- and hence impossible to measure with a finite number of shots --- as the number of ansatz layers becomes large. The fidelity also cannot be measured without adding more gates to the circuit because the final state is not generically a computational basis state, so estimating the predicted damping factor from the fidelity is not possible. Therefore, these circuits cannot be used to measure the damping factor using the method described in the previous paragraph. We might instead consider picking all of the $n_i$ to be even, so that the state remains a computational basis state. In generic computational basis states, at large $n$, the Hamiltonian will still have a small expectation value, although not as small as in a generic state. Nevertheless, we could estimate the damping factor by measuring the fidelity of the final state in such a circuit. In the present work, we focus on the case $\vec \theta = 0$, which produces the ground state of the free ($h_x = 0$) Hamiltonian, a state close to the states of interest. We leave the study of other Clifford circuits to future work.

In Ref.~ \cite{urbanek2021mitigating}, this method is combined with ZNE. We also benchmark the combination of this method with ZNE. There are two ways of combining these methods. One could apply ZNE to the original circuit and to the calibration circuit (the one with the single-qubit gates removed) separately and then divide the former result by the predicted damping factor from the latter result (we call this method ``ZNE first"), or we could divide the measured energy from each noise-scaled target circuit by the estimated damping factor from the corresponding noise-scaled calibration circuit and then extrapolate the quotient to zero noise (we call this method ``ZNE last"). We benchmark both methods.

\section{Experimental techniques}
\label{sec:expt_techniques}
Independent of the error-mitigation technique under consideration, we always do the following to reduce noise:
\begin{enumerate}
    \item We omit gates outside of the backwards light cone of the measured observable. This technique has previously been called a \textit{light-cone filter} and has been found to be beneficial \cite{mi2021information}.
    \item In order to implement the previous technique to its fullest advantage, and also to reduce the number of qubits measured simultaneously, we measure each Hamiltonian term separately.
    \item The qubits in our ansatz are connected in a loop, as are the qubits on the quantum computer. For a loop of $n$ qubits, there are $2n$ ways to map the circuit qubits to the physical qubits that preserve this connectivity (cyclic permutations and a reflection). We average over four of these mappings. Such an average has been called \textit{qubit assignment averaging} \cite{arute2020observation}. Instead of picking these mappings randomly, we pick configurations for which the damping factor, as predicted by multiplying fidelities, is the largest (closest to one).
    This ensures that we avoid particularly noisy qubits or gates to the extent possible while also averaging out systematic biases.
	\item Additionally, we implement randomized compiling \cite{Wallman_2016, Cai_2019, Cai_2020, urbanek2021mitigating}, which helps convert coherent errors into depolarizing-channel errors. In this method, every CNOT gate is randomly replaced by one of 16 dressed CNOT gates, to which it is equivalent in the absence of noise. The 16 dressed gates are described succinctly in Figure 1 of \cite{urbanek2021mitigating}. We apply this method when benchmarking error mitigation methods (i.e. Figs.~\ref{fig:fit_and_pert}-\ref{fig:heatmap_readout}) but not when running the VQE experiments (Figs.~\ref{fig:VQE_sydney}-\ref{fig:VQE_toronto}).
	\item Finally, we apply readout error mitigation, assuming that readout errors are uncorrelated and that the error rates are as measured by IBM and reported in the backend properties. For details, see Appendices~\ref{sec:readout_errors_intro}-\ref{sec:readout_errors}.
\end{enumerate}

\section{Results and Discussion}
First, we demonstrate that the training phase of VQE can converge even when we don't apply error-mitigation techniques to correct the damping. Then we compare the performance of the various error-mitigation techniques at estimating the damping factor for tuned circuits.

\subsection{VQE using unmitigated quantum circuit evaluations}
\label{sec:VQE}

\begin{figure}[h]
    \centering
    \begin{subfigure}{0.4\textwidth}
        \includegraphics[width=\textwidth]{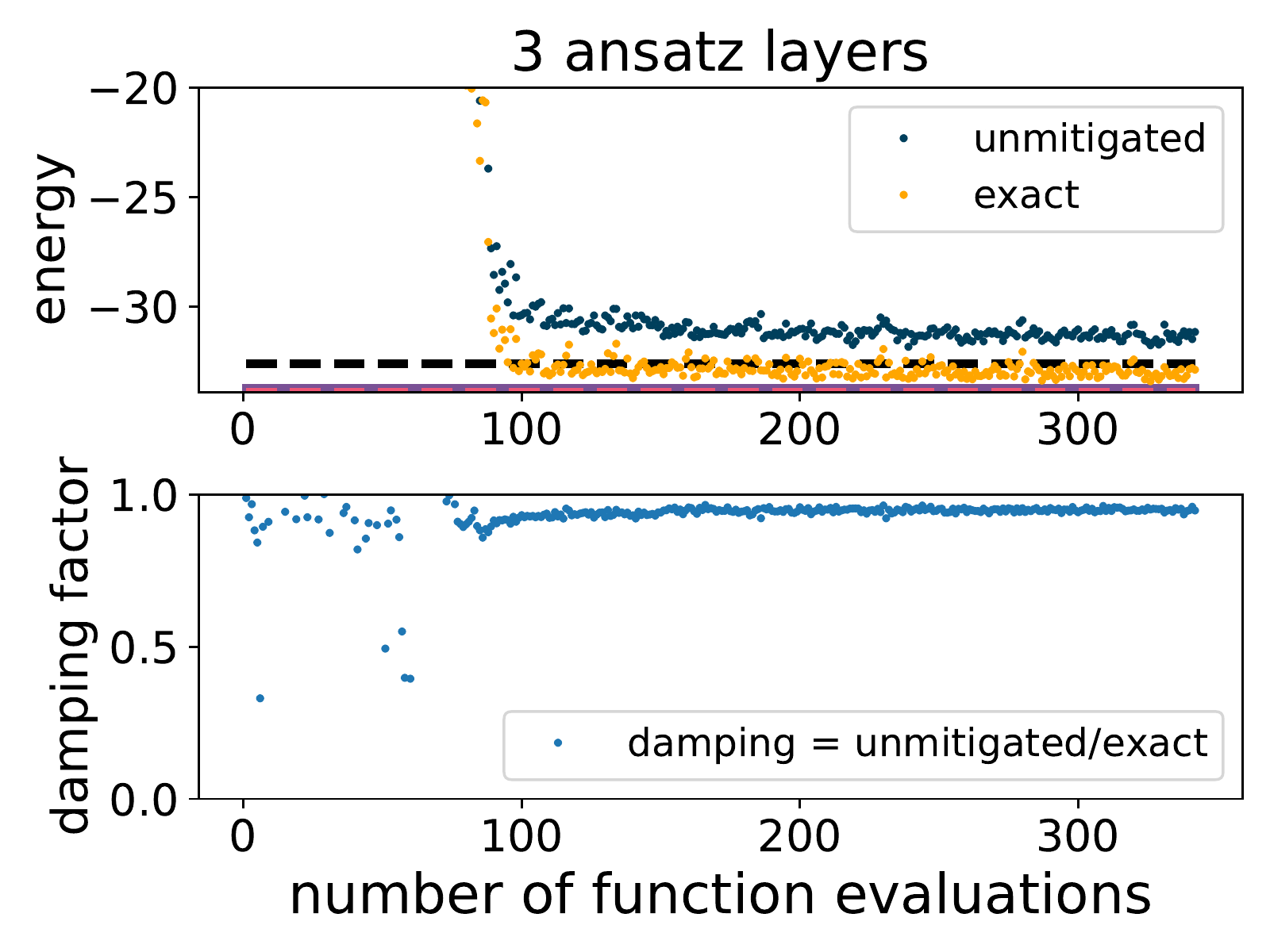}
        %\caption{February 5-6, 2021}
    \end{subfigure}
        \begin{subfigure}{0.4\textwidth}
        \includegraphics[width=\textwidth]{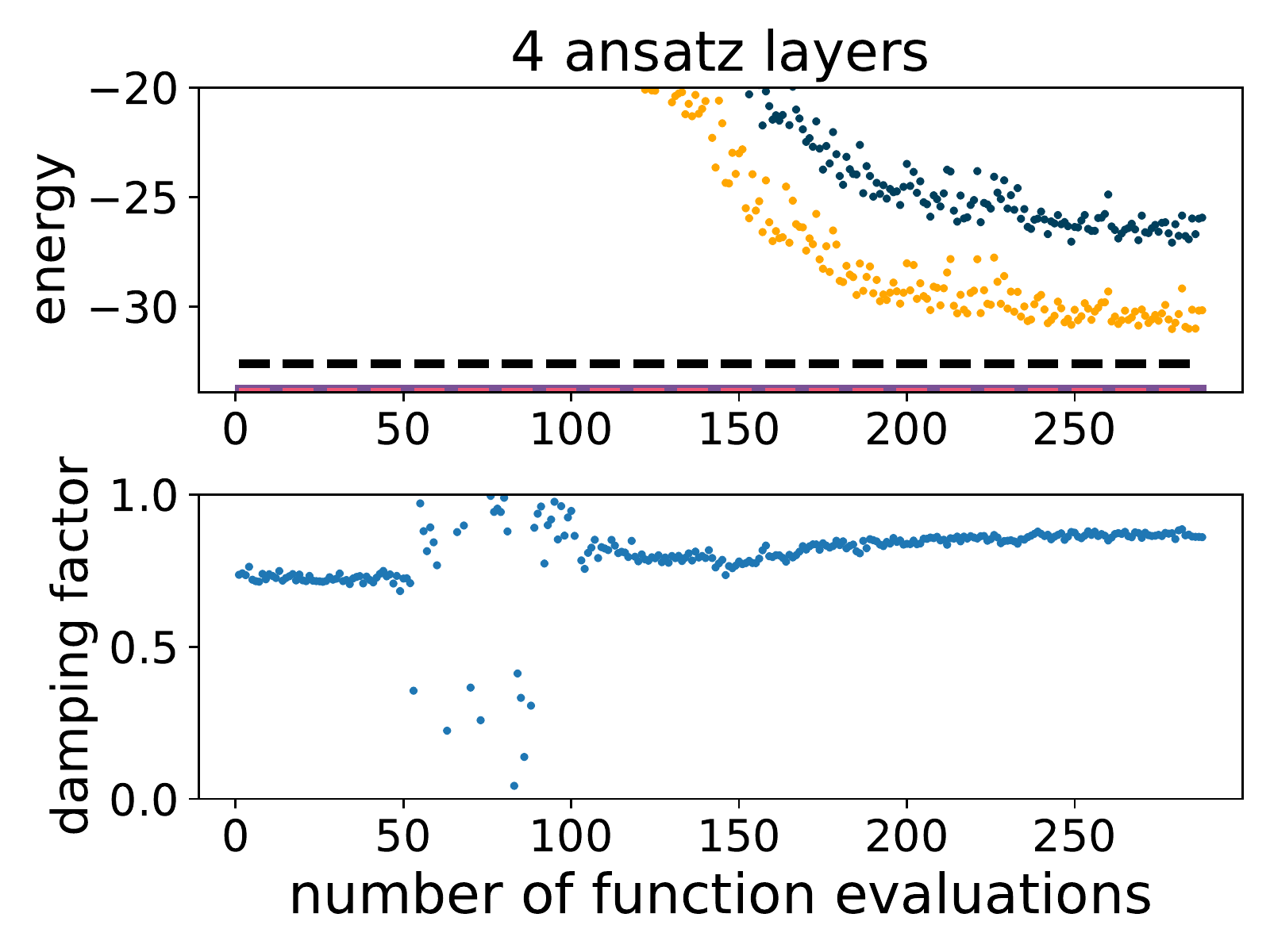}
        %\caption{February 2-3, 2021}
    \end{subfigure}
        \begin{subfigure}{0.4\textwidth}
        \includegraphics[width=\textwidth]{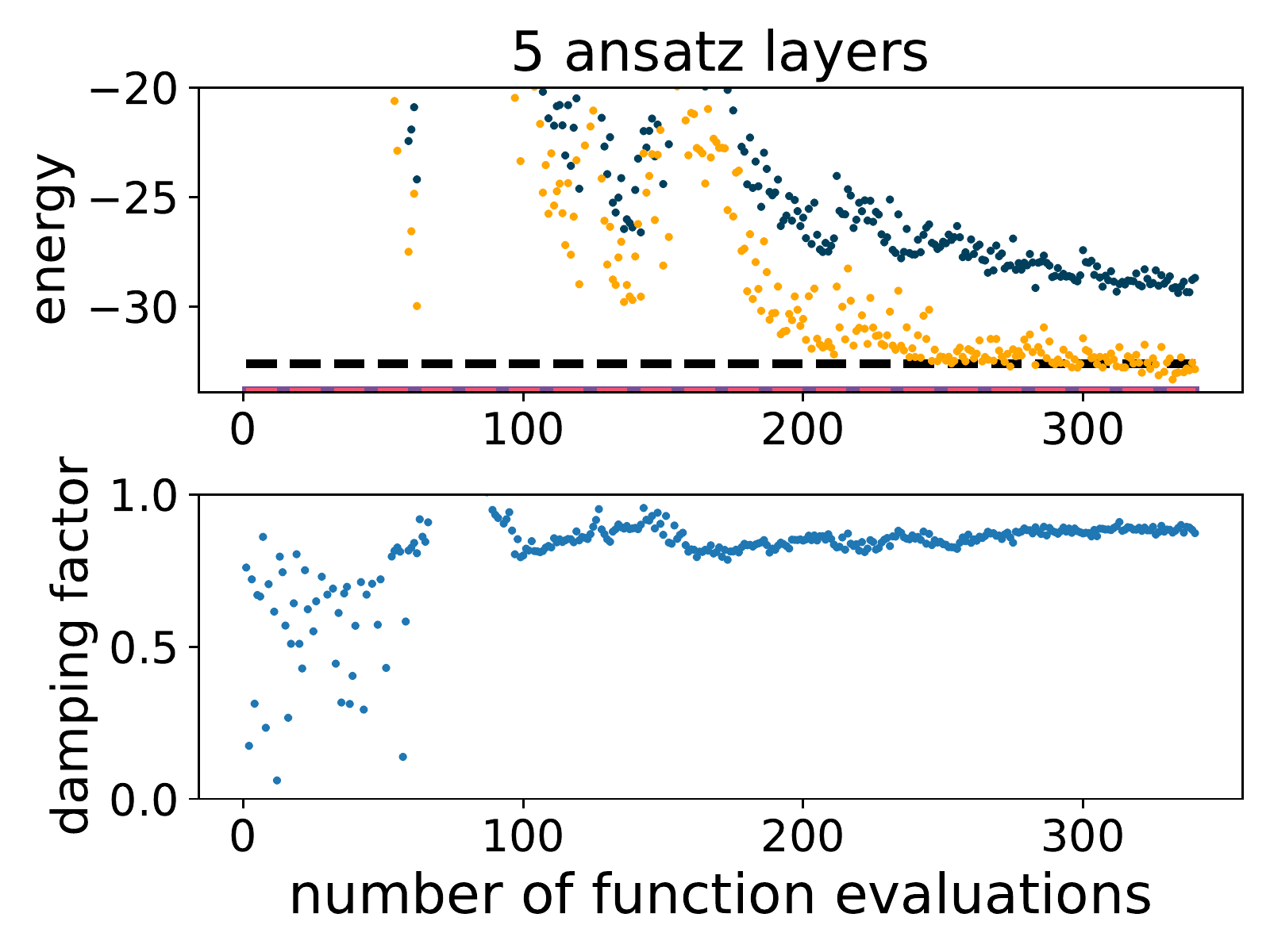}
        %\caption{February 3-4, 2021}
    \end{subfigure}
        \begin{subfigure}{0.4\textwidth}
        \includegraphics[width=\textwidth]{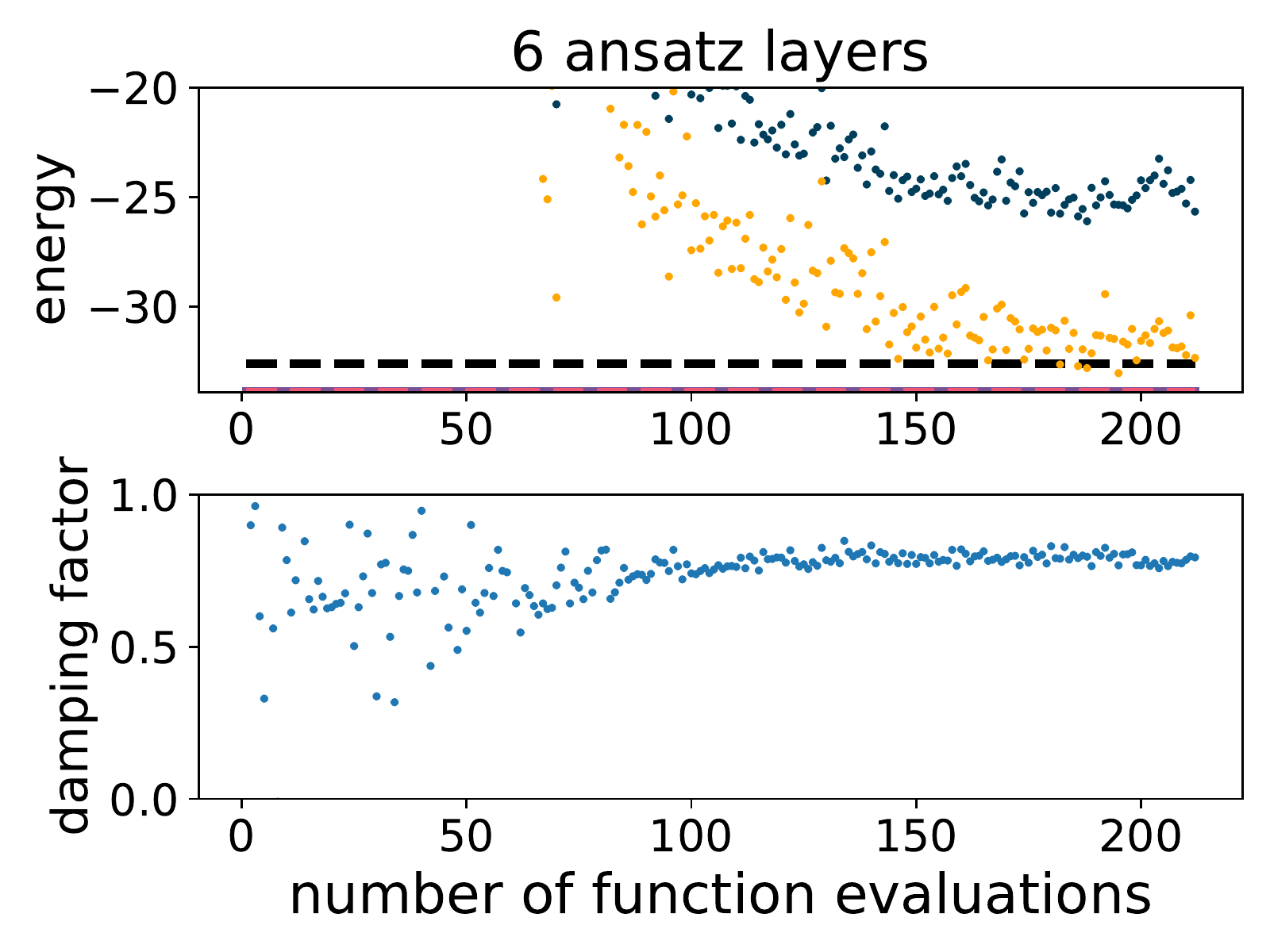}
        %\caption{February 6-8, 2021}
    \end{subfigure}
    \caption{VQE for the 20-qubit Ising Hamiltonian, with energy evaluations performed on \textit{ibmq\_sydney}, with $h_x = 1.5$ and $h_z = 0.1$. Unmitigated energies are fed into the SPSA algorithm. We impose cyclic permutation symmetry on the ansatz to reduce the number of parameters. Note that SPSA involves computing $E(\vec \theta_i \pm c_i\Delta_i)$, not $E(\vec \theta_i)$.
    In the upper panels, the dark-blue dots are the unmitigated energies measured on the quantum computer. These are damped compared to the dark-yellow dots, which show exact classical evaluations of the same circuits, from which we compute the observed damping factor (blue dots in lower panels). The purple horizontal line in each upper panel shows the energy of the classically optimized ansatz circuit, and the magenta dashed line shows the energy of the exact ground state. The horizontal black dashed line indicates the first excited state energy. In all trials except the 4-ansatz-layer trial, the optimizer succeeds in finding a state whose energy is below the first excited state energy (we would pick the point with the lowest energy, not necessarily the last point).}
    \label{fig:VQE_sydney}
    % created using plot_machine in optimize_library.py
\end{figure}

\begin{figure}[h]
    \centering
    \begin{subfigure}{0.4\textwidth}
        \includegraphics[width=\textwidth]{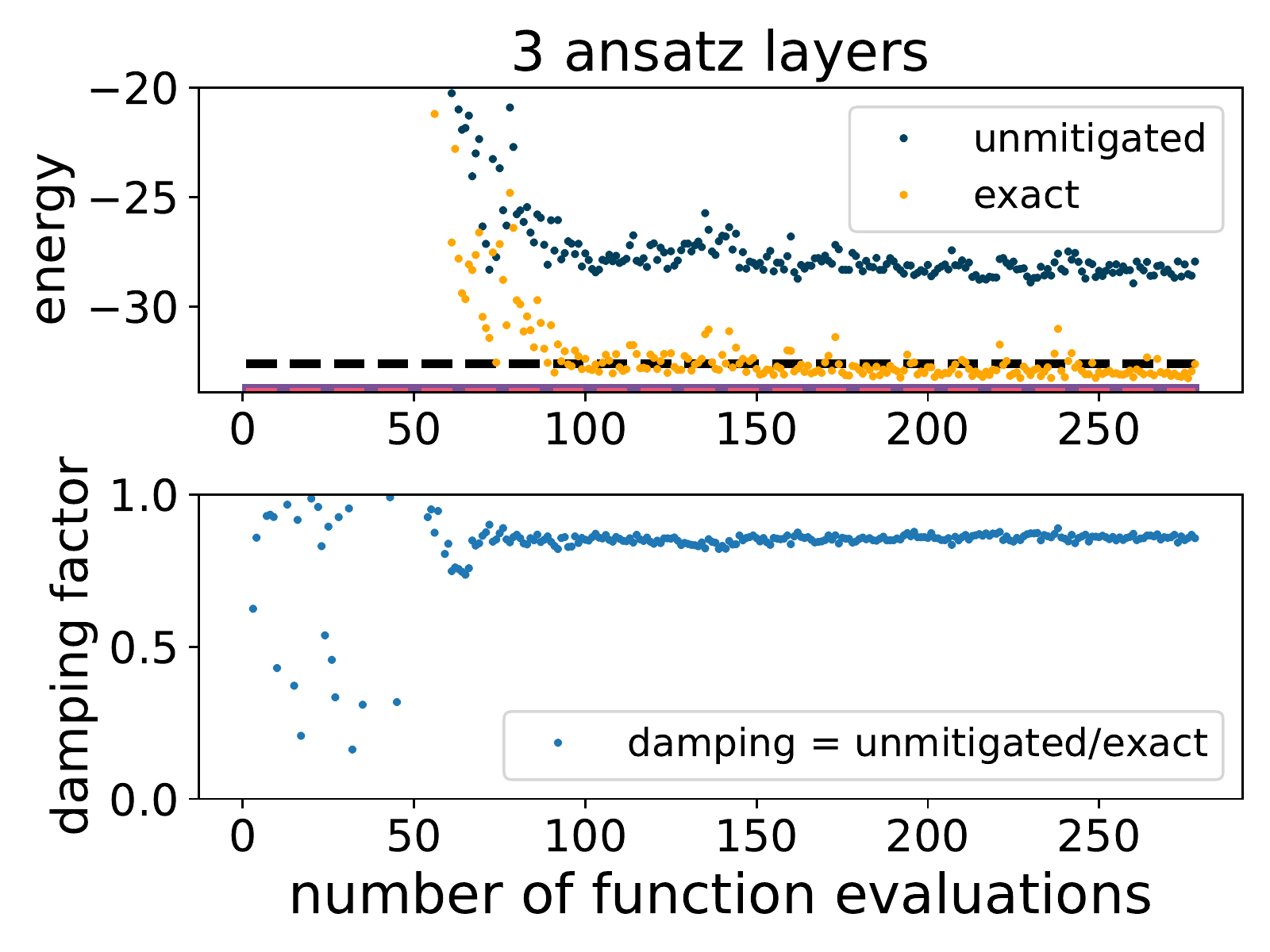}
        %\caption{February 7-8, 2021}
    \end{subfigure}
    \begin{subfigure}{0.4\textwidth}
        \includegraphics[width=\textwidth]{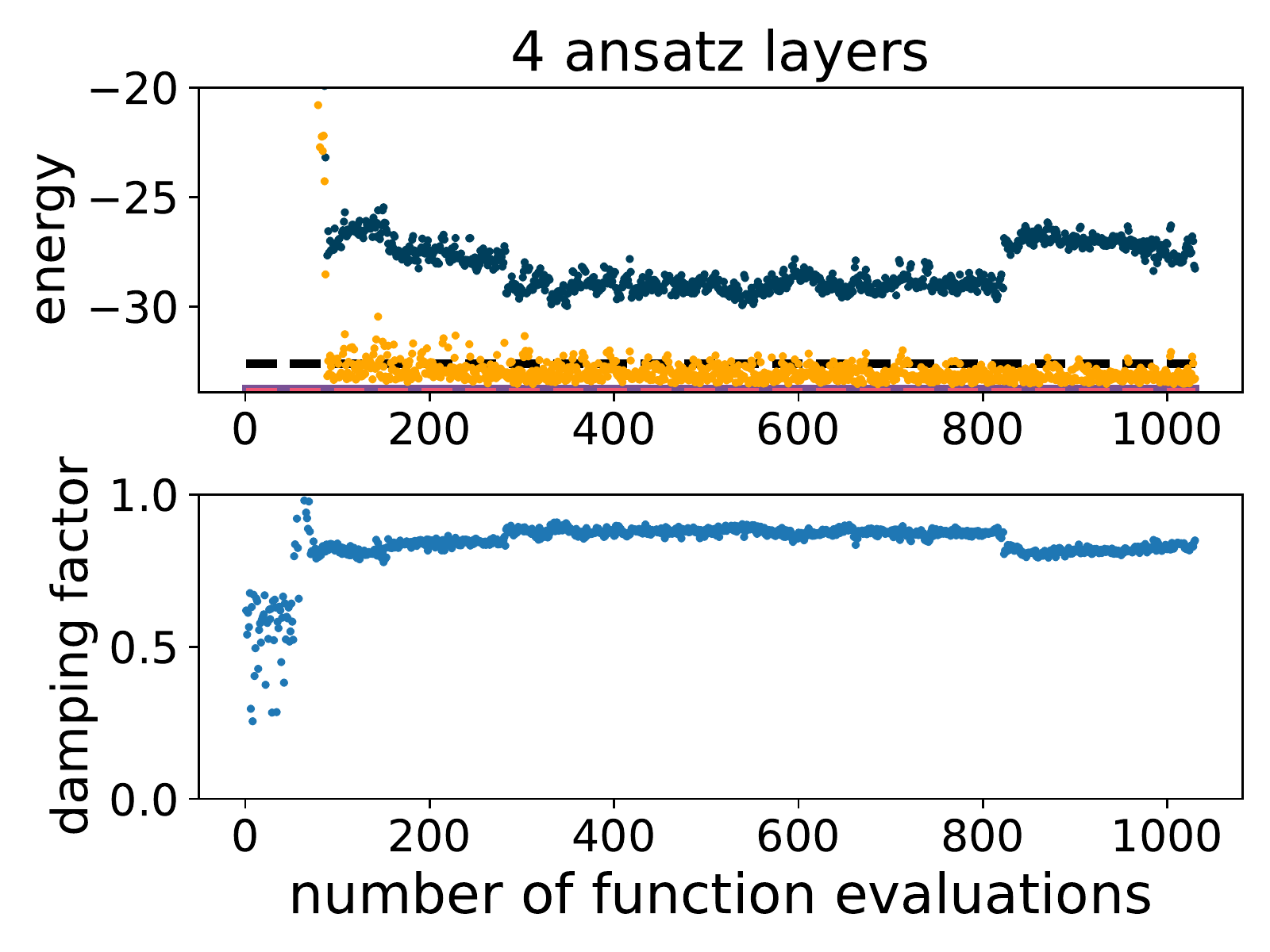}
        %\caption{February 3-6, 2021}
    \end{subfigure}
    \begin{subfigure}{0.4\textwidth}
        \includegraphics[width=\textwidth]{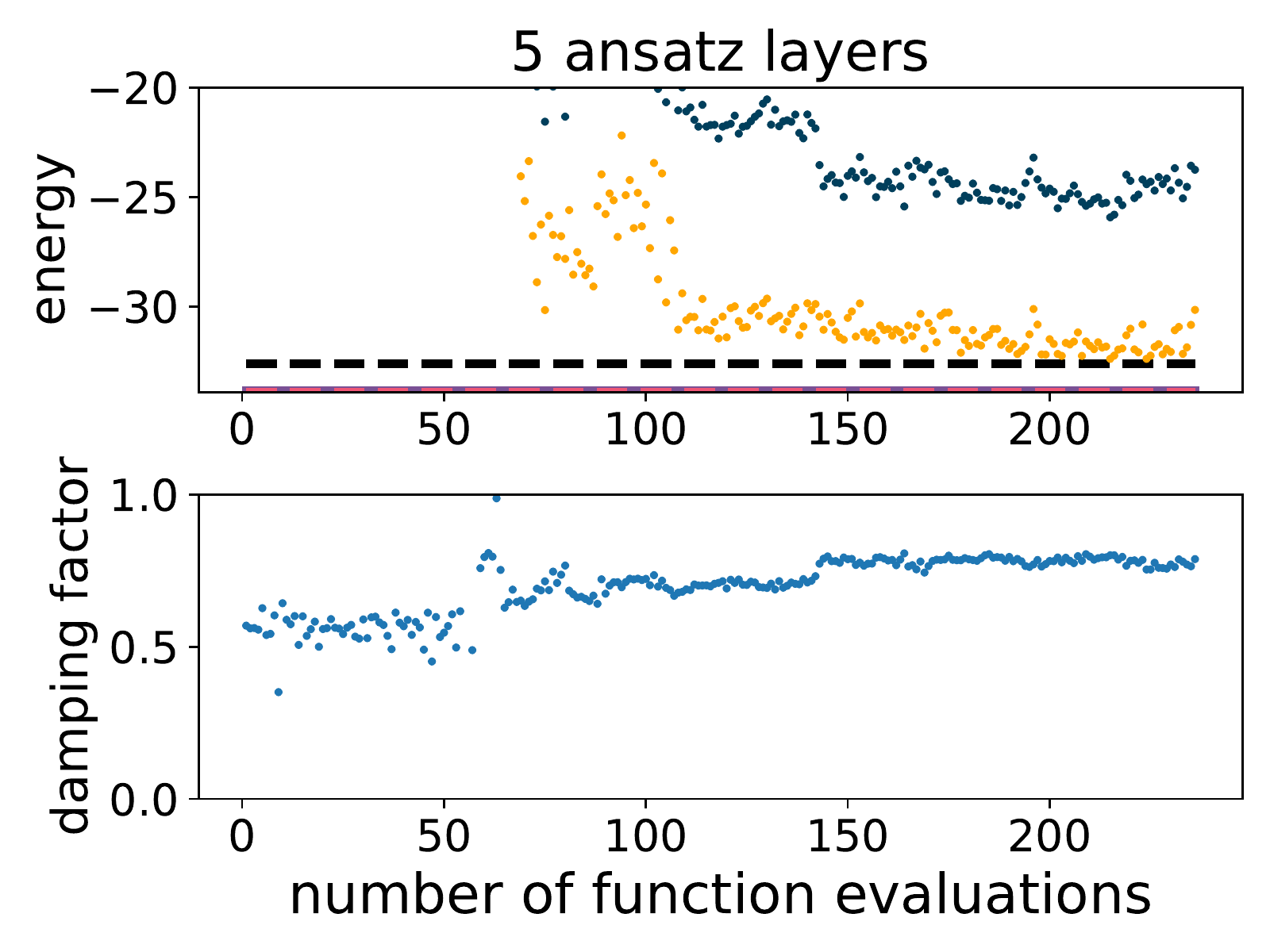}
        %\caption{February 6-7, 2021}
    \end{subfigure}
    \caption{Same as Fig.~\ref{fig:VQE_sydney} except on \textit{ibmq\_toronto}.
    The discontinuities, particularly apparent in the 4-layer experiment, coincide with daily recalibrations of the device. For the 3- and 4-layer trials the optimizer succeeds in obtaining states with energies below the first excited state energy.}
    \label{fig:VQE_toronto}
    % created using plot_machine in optimize_library.py
\end{figure}

Consider the training phase of VQE. In order to improve the convergence of our optimizer, we impose a cyclic permutation symmetry on the ansatz, so that each row of $Y$-rotations is described by two angles instead of $n$. This reduces the number of parameters from $n(l+1)$ to $2(l+1)$, where $n$ is the number of qubits and $l$ is the number of ansatz layers. We begin at a random point in parameter space and use the quantum computer to measure the energy. We feed the measured energy into the Simultaneous Perturbation Stochastic Approximation (SPSA) optimizer, running on a classical computer, which picks the updated parameters \cite{SPSA}. The first 50 function evaluations are used to calibrate an SPSA hyperparameter as in \cite{Kandala_2017}, following which the optimizer iterates through the SPSA algorithm. At each iteration, SPSA uses two evaluations of the energy, at $\vec \theta = \vec \theta_i \pm c_i\Delta_i$, and we submit both of these evaluations together in the same job for improved speed.

Figs.~\ref{fig:VQE_sydney} and \ref{fig:VQE_toronto} show the results of this VQE optimization for a 20-qubit mixed-field 1D Ising model on \textit{ibmq\_sydney} and \textit{ibmq\_toronto}, respectively, with 3--6 ansatz layers. We see that the optimizer is often able to converge to a state close to the true ground state, that is, below the energy of the first excited state. The lowest energy state achieved during our VQE runs has an energy of $-33.54$. For comparison, the true ground state energy is $-33.90$, and the first excited state energy is $-32.60$. We might worry that this convergence would be spoiled by a parameter-dependent damping factor. Indeed, in some of our runs, the damping factor can be seen to increase as the optimizer converges to the ground state, indicating its parameter dependence.

Although the parameter dependence of the damping factor does not prevent convergence in our system, we cannot guarantee that it never will. If the damping factor shifts the locations of the minima, techniques to minimize this effect will be necessary. These could include additional experimental techniques to symmetrize over measurement outcomes, such as exchanging the roles of ``1" and ``0", or applying an error mitigation technique during the optimization routine. Many of the error-mitigation techniques discussed here, however, assume either that the damping factor is parameter independent or that the circuit has been optimized. Of the methods that we benchmark, only ZNE could be used during the optimization phase.

One might naturally guess that some of the parameter dependence of the damping factor comes from asymmetric readout errors. We show in Appendix \ref{sec:readout_errors} that, indeed, the asymmetric readout errors lead to a predictable additive shift of expectation values, and hence a parameter-dependent damping factor (Eq.~\ref{eq:N1}). However, because the shift is constant, its effect is to shift the overall energy by a constant amount, leaving the location in parameter space of the minimum unchanged. For the mixed-field Ising Hamiltonian, this overall shift is approximately $-n(h_x + h_z)(e_1 - e_0) \approx -1.6$ for $n=20$, $h_x = 1.5$, $h_z = 0.1$, and $(e_1 - e_0) \approx 0.05.$ This is a larger relative shift when the energy is closer to zero. It should therefore have the effect of making the damping factor closer to unity at negative energies further from the minimum (closer to zero). We see instead the opposite effect, so this is not the source of the parameter-dependent damping factor.

\subsection{Comparison of mitigation techniques}

\begin{figure}[h]
    \centering
    \begin{subfigure}{0.4\textwidth}
        \includegraphics[width=\textwidth]{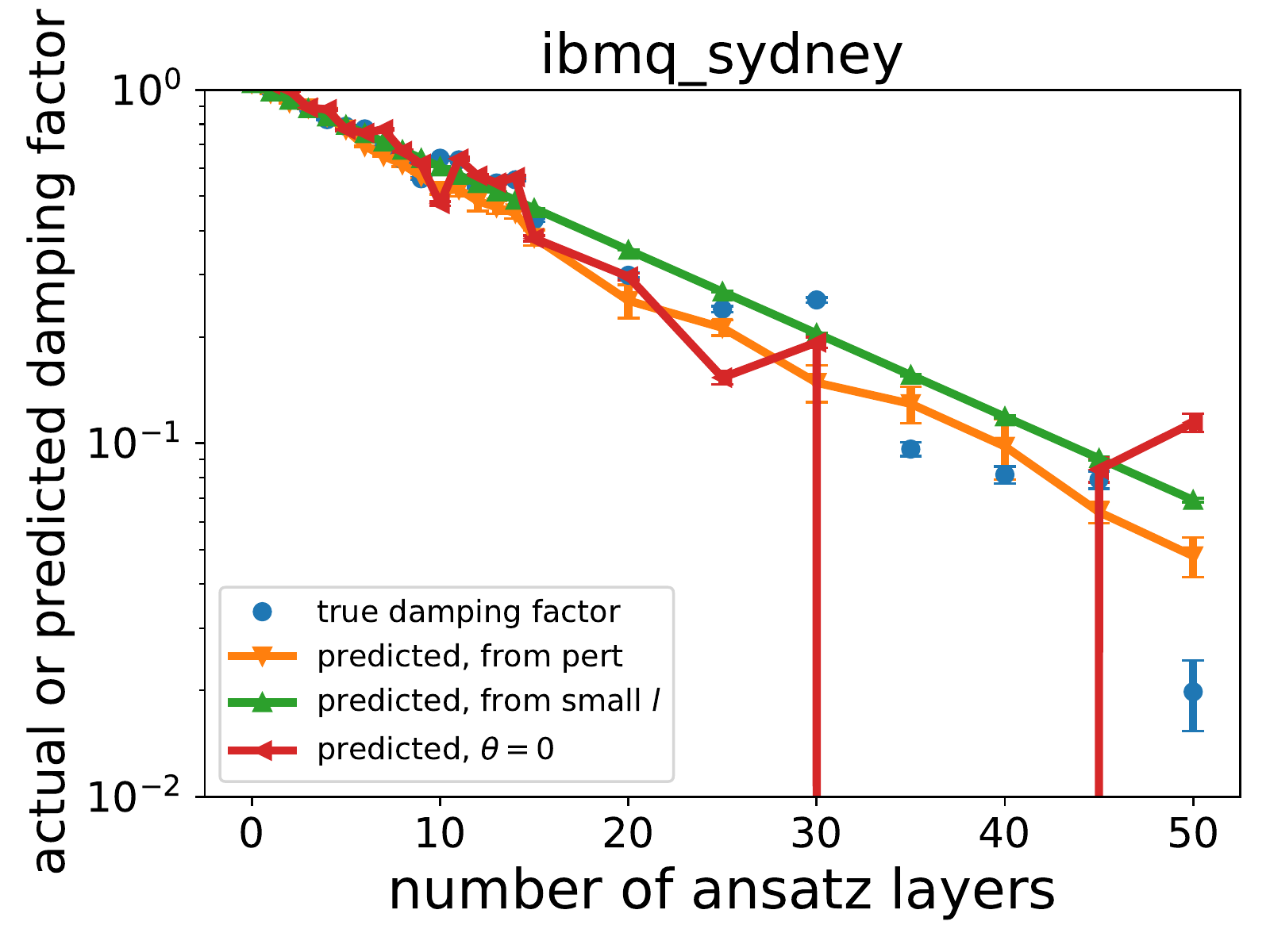}
    \end{subfigure}
    \begin{subfigure}{0.4\textwidth}
        \includegraphics[width=\textwidth]{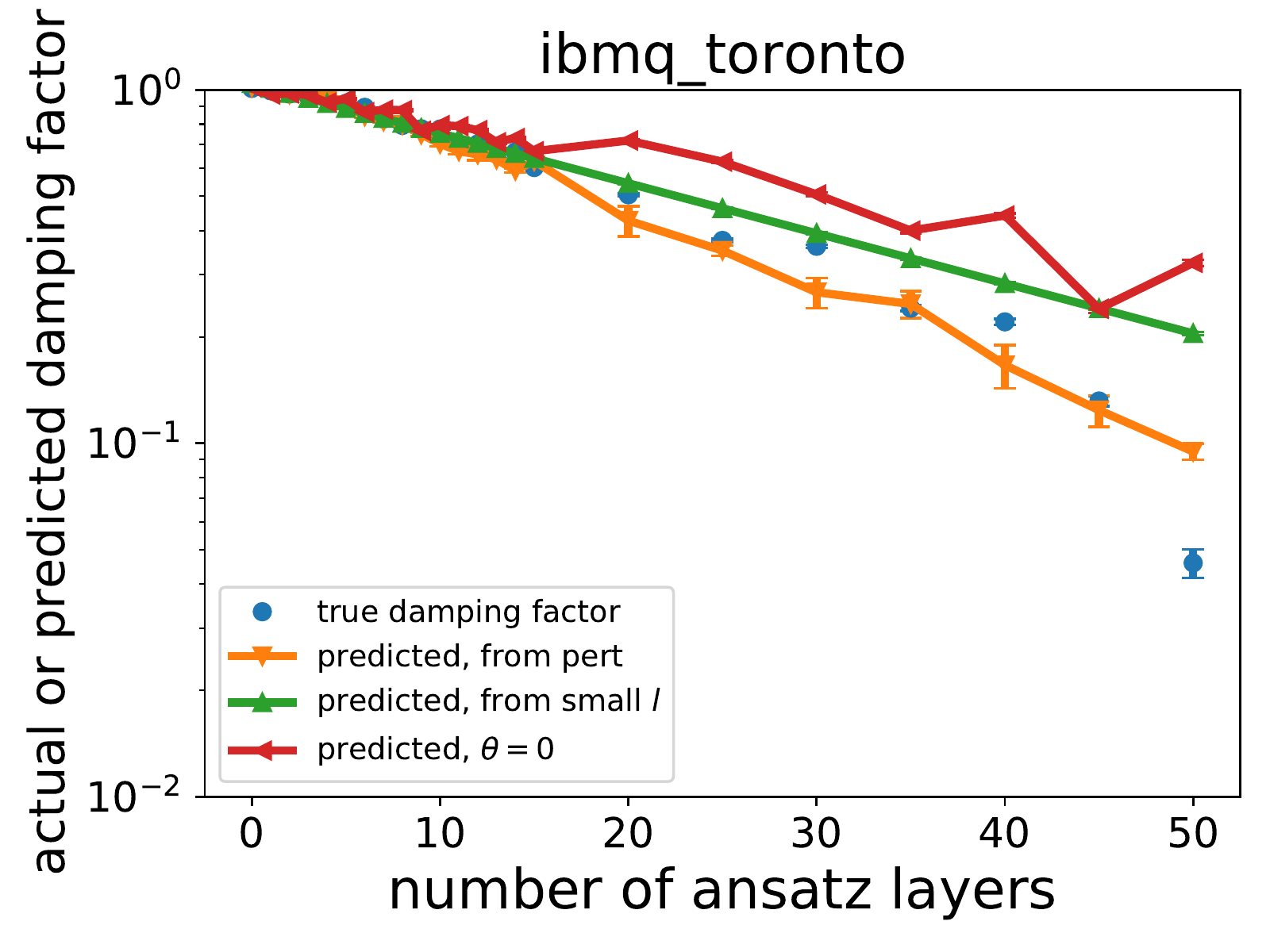}
    \end{subfigure}
    \caption{Measured and predicted damping factors for ground states of the 20-qubit mixed field Ising model with $h_x = 1.5$ and $h_z = 0.1$. The methods used to predict the damping factor are described in Secs.~\ref{sec:pert}, and \ref{sec:scaling_with_l}, and \ref{sec:zero_theta}, respectively. For the first method, we take $h_x = \{0.1, 0.2, 0.3, 0.4, 0.5\}$ to represent the perturbative regime and average the measured damping factors for those five circuits. ZNE is absent from this plot because it does not predict a damping factor.  When computing energies, we apply readout error mitigation using the reported readout error rates, as discussed in Appendix \ref{sec:readout_errors}. We see that the ``from pert" method correctly predicts the damping factor for circuits up to about 25 layers and that all of the methods fail by 50 layers. Error bars indicate statistical errors only (see Appendix \ref{sec:errorbars}).}
    \label{fig:fit_and_pert}
\end{figure}

\begin{figure}[h]
	\centering
	\begin{subfigure}{0.4\textwidth}
		\includegraphics[width=\textwidth]{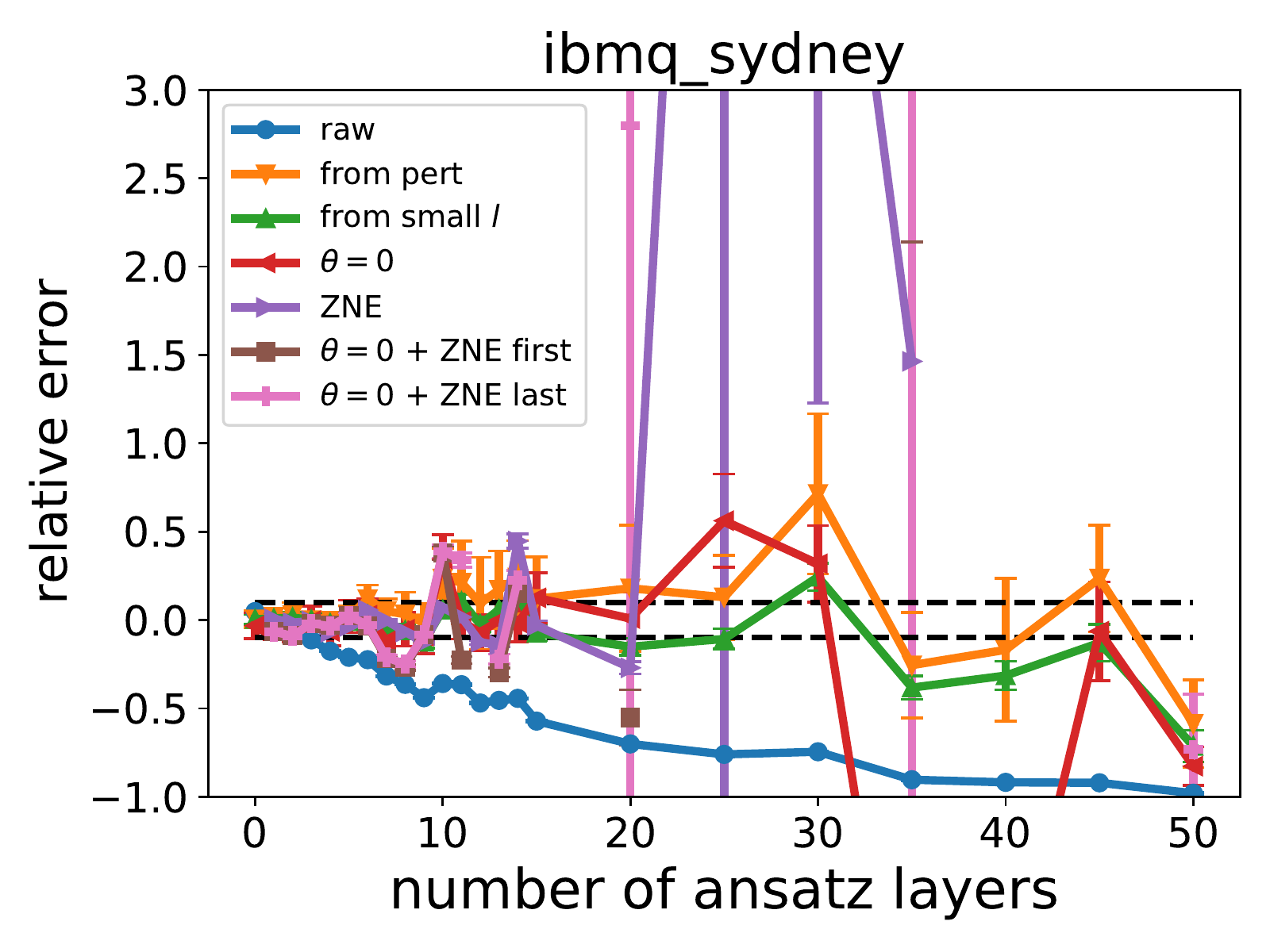}
	\end{subfigure}
	\begin{subfigure}{0.4\textwidth}
		\includegraphics[width=\textwidth]{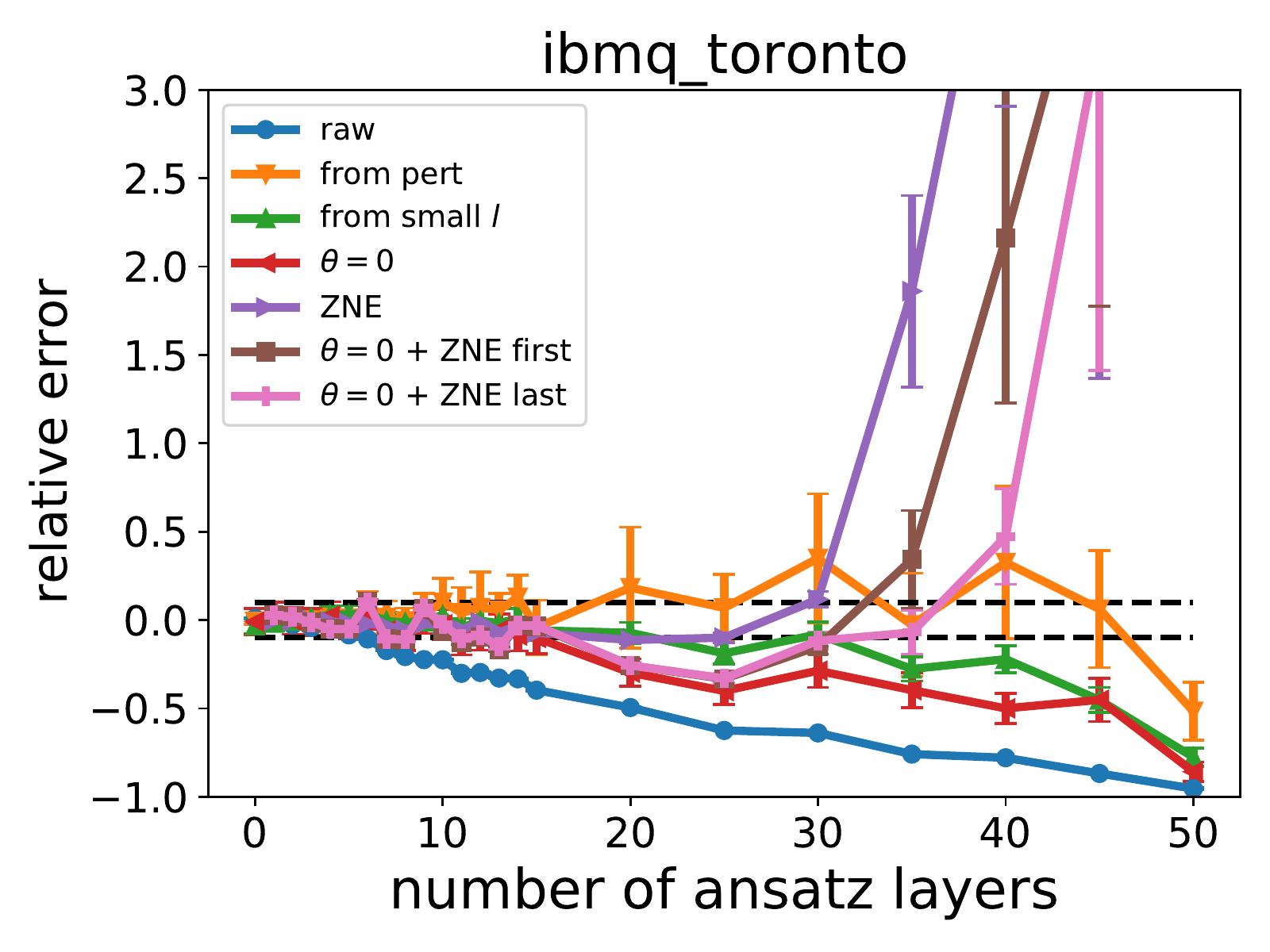}
	\end{subfigure}
	\caption{Relative error in the measured ground state energy of the 20-qubit mixed field Ising model with $h_x = 1.5$ and $h_z = 0.1$, mitigated using each of the methods studied in this paper.}
	\label{fig:rel_error}
\end{figure}

Having established that, in at least some cases, error mitigation is not needed to tune the ansatz parameters, we now compare the performance of the error-mitigation techniques, applied to optimized circuits. For the methods that work by estimating a damping factor, we plot the observed damping factor on the same axes as those predicted by the various methods (Fig.~\ref{fig:fit_and_pert}). Additionally, for all of the methods, we plot the relative error of the reconstructed energy as a function of the number of ansatz layers (Fig.~\ref{fig:rel_error}). Finally, we show to how many layers each method allows us to extend our ansatz circuits while successfully mitigating the energy to within 10\% of its true value (Figs.~\ref{fig:heatmap}-\ref{fig:heatmap_readout}).

\section{Discussion and Conclusion}
\begin{figure}[h]
    \centering
       \begin{subfigure}{0.4\textwidth}
        \includegraphics[width=\textwidth]{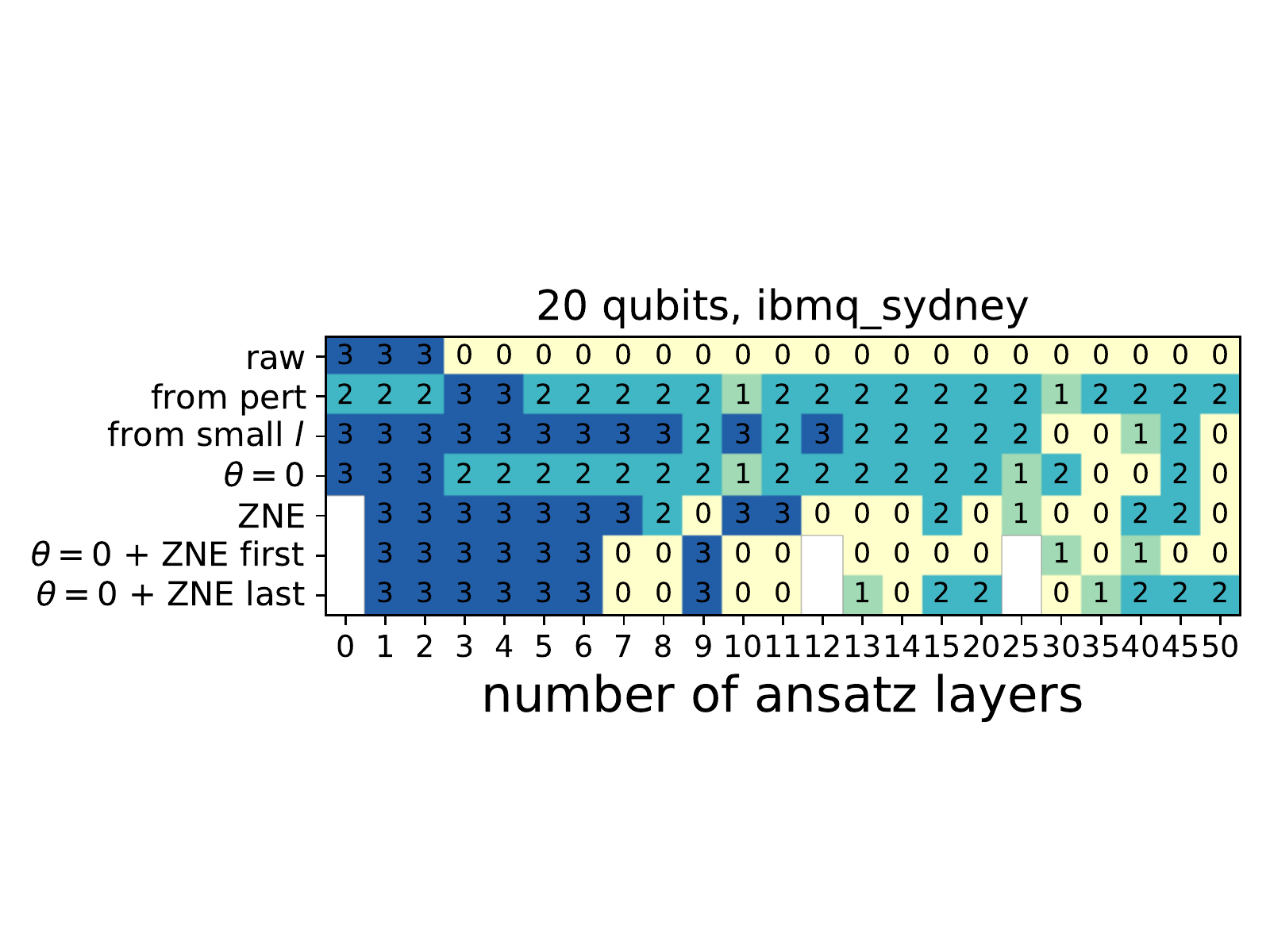}
        \end{subfigure}
        \begin{subfigure}{0.4\textwidth}
        \includegraphics[width=\textwidth]{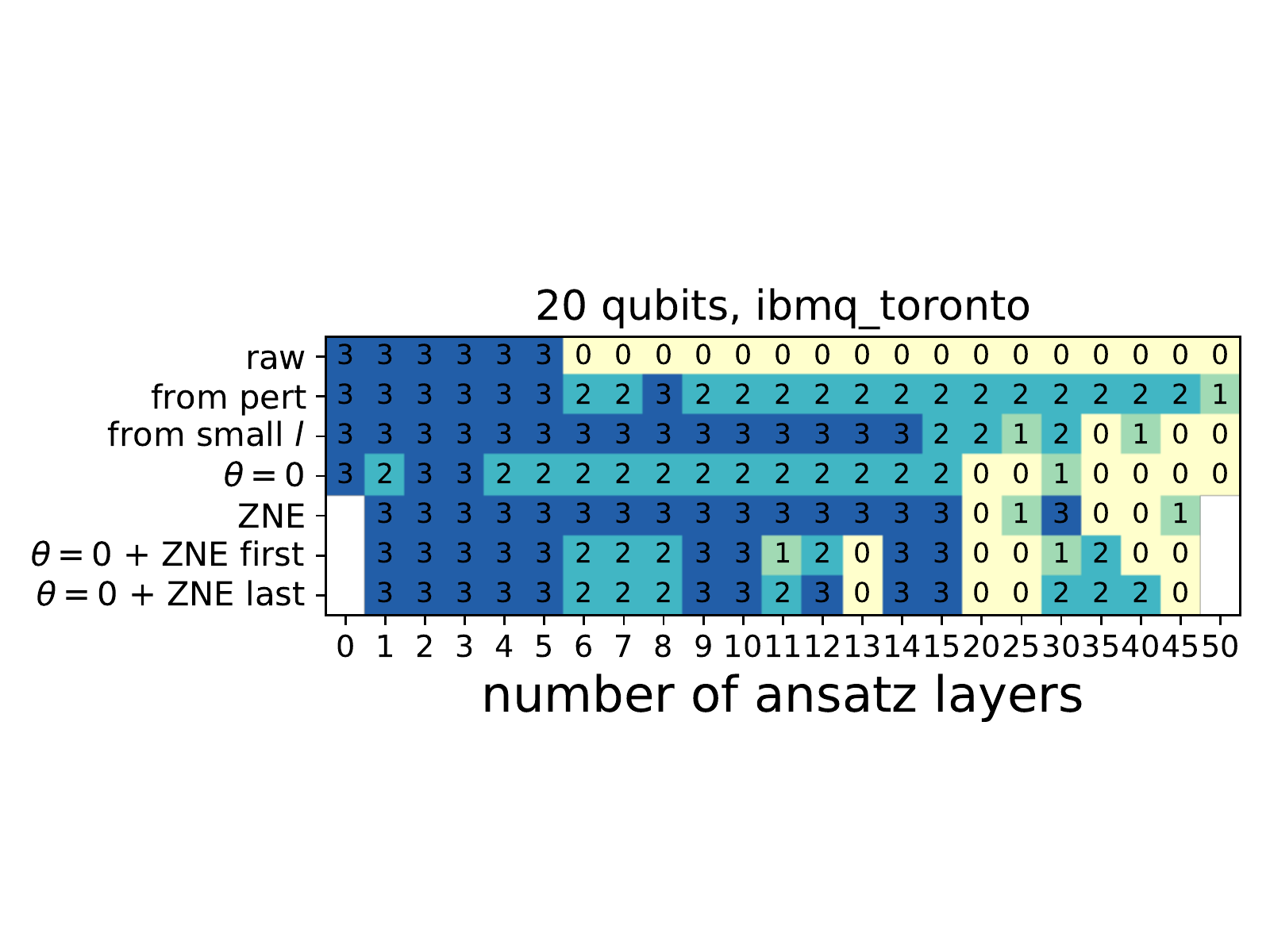}
        \end{subfigure}
    \caption{Effectiveness of error mitigation. ``3" means that the reconstructed energy is within 10\% of the correct value. ``2" means statistically indeterminate at 1 standard deviation. ``1" means statistically indeterminate at 2 standard deviations. ``0" means that the reconstructed energy deviated by more than 10\% from the true energy. In this figure, the methods involving calibration circuits do not include separate readout error mitigation.}
    \label{fig:heatmap}
\end{figure}

\begin{figure}[h]
    \centering
        \includegraphics[width=0.4\textwidth]{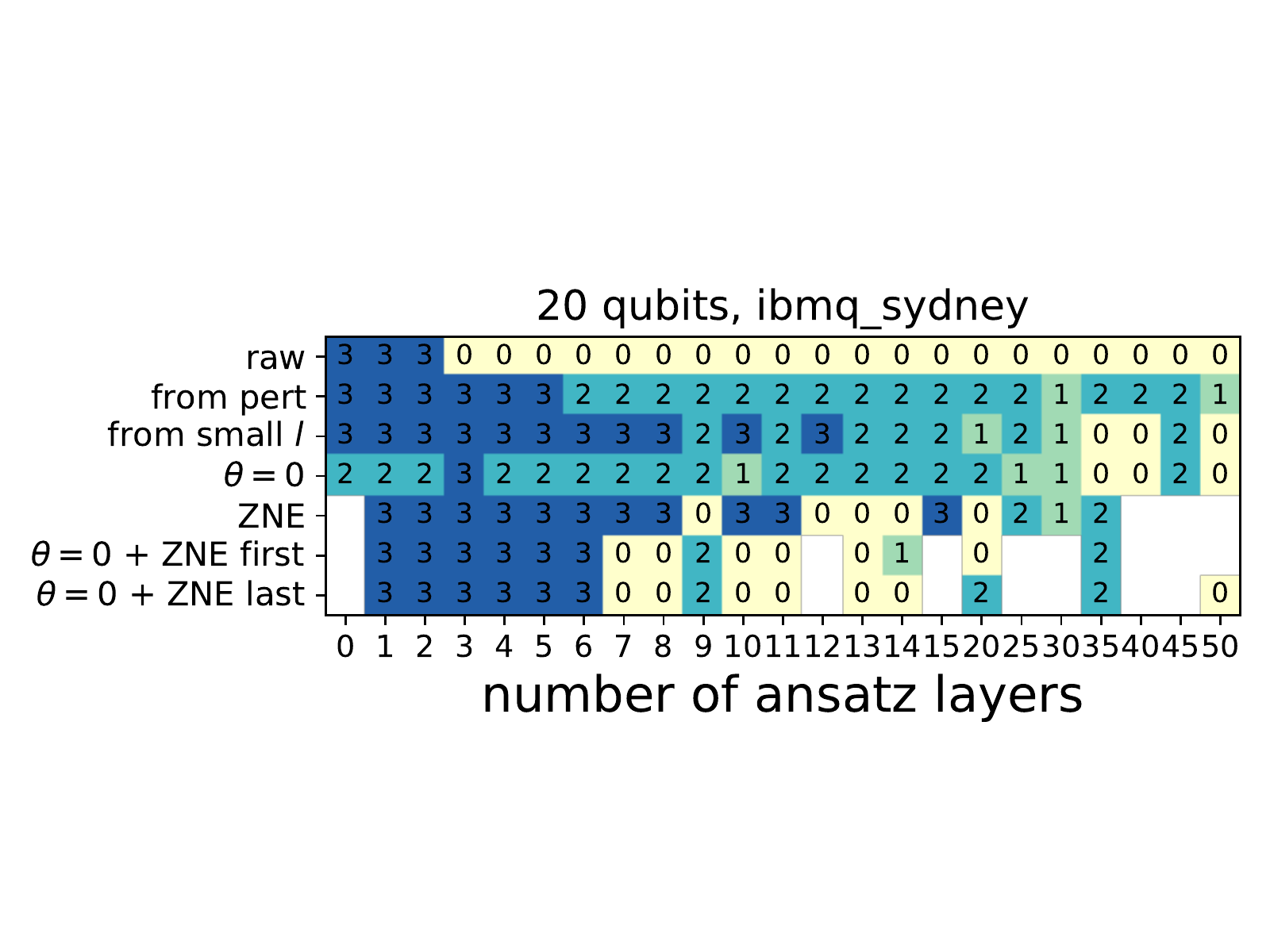}
        \includegraphics[width=0.4\textwidth]{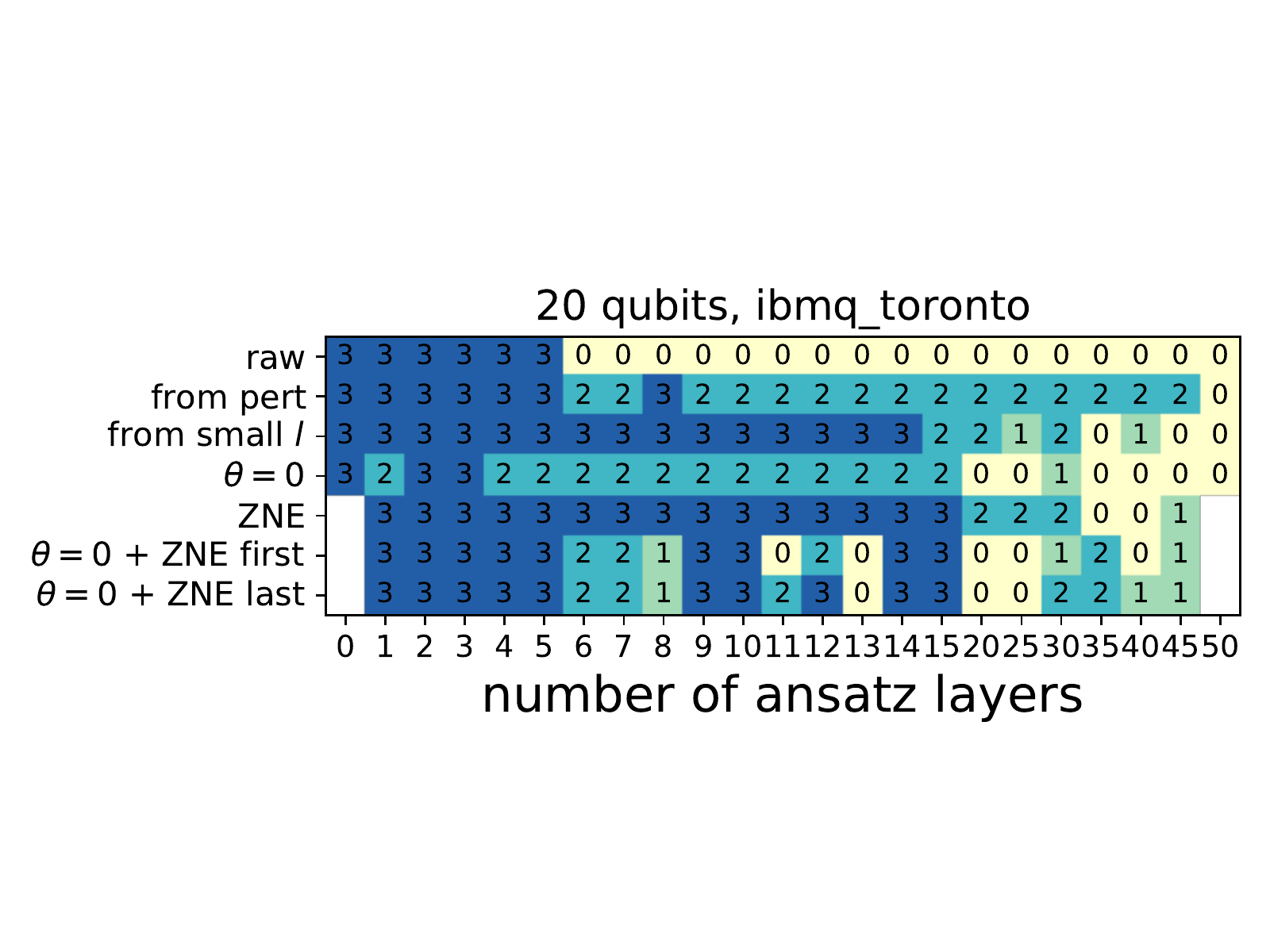}
    \caption{Same as Fig.~\ref{fig:heatmap}, but applying readout error mitigation (Appendix \ref{sec:readout_errors}) to all energy evaluations, including in calibration circuits.}
    \label{fig:heatmap_readout}
\end{figure}

\begin{figure}[h]
\centering
	\begin{subfigure}{0.3\textwidth}
	\includegraphics[width=\textwidth]{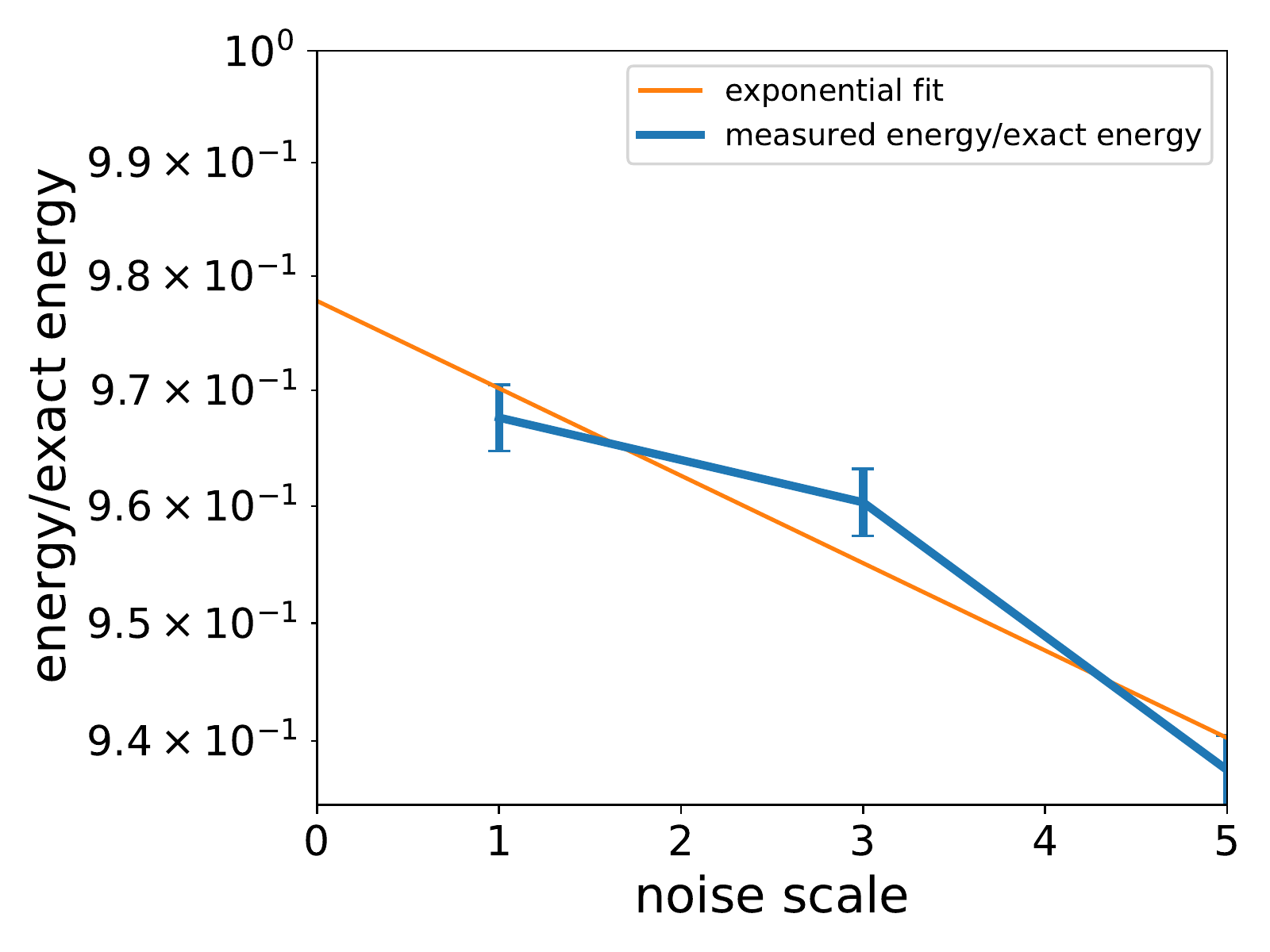}
	\caption{1 ansatz layer, without readout error mitigation}
	\end{subfigure}
	\begin{subfigure}{0.3\textwidth}
	\includegraphics[width=\textwidth]{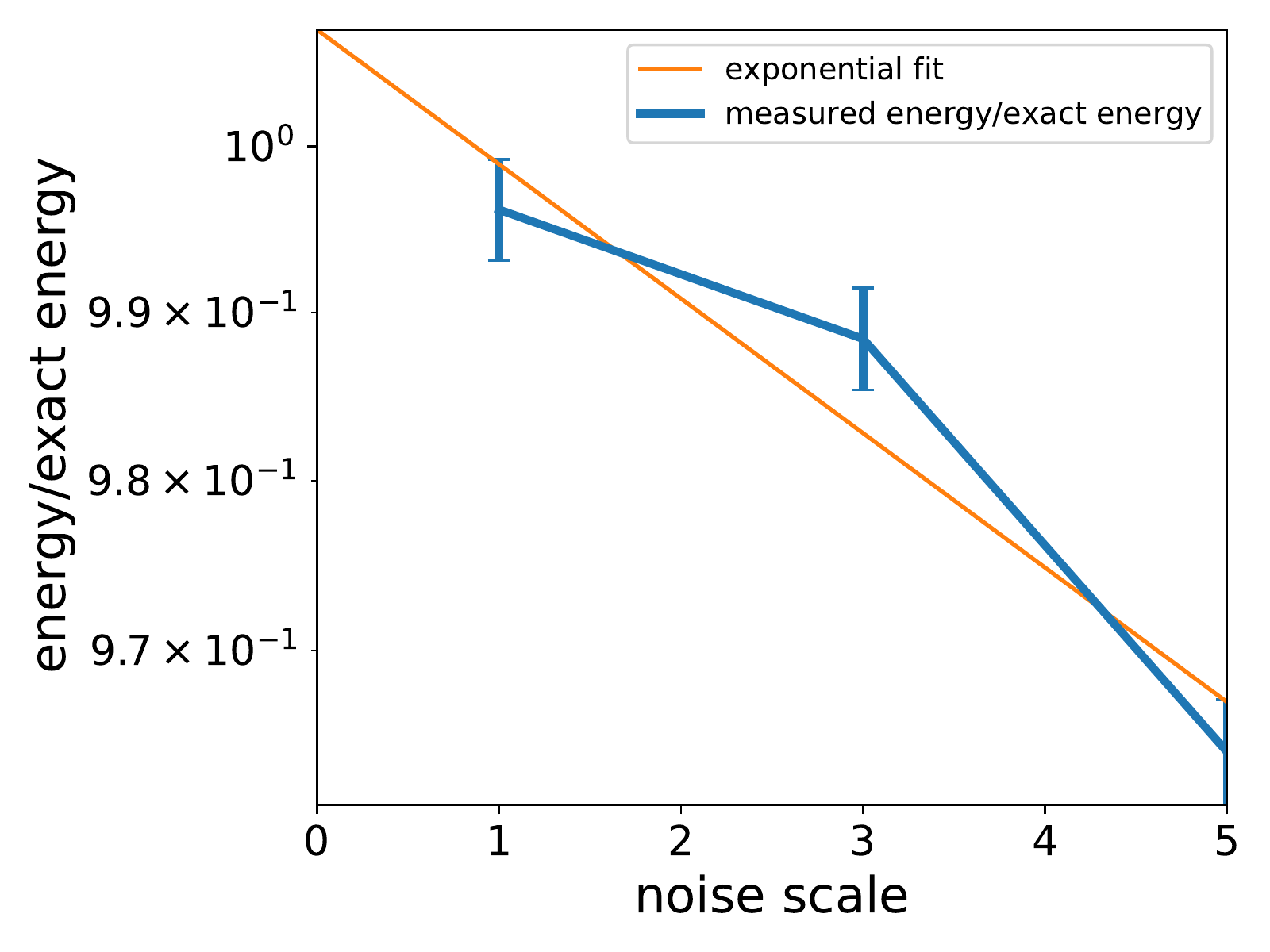}
	\caption{1 ansatz layer, with readout error mitigation}
	\end{subfigure}
	\begin{subfigure}{0.3\textwidth}
	\includegraphics[width=\textwidth]{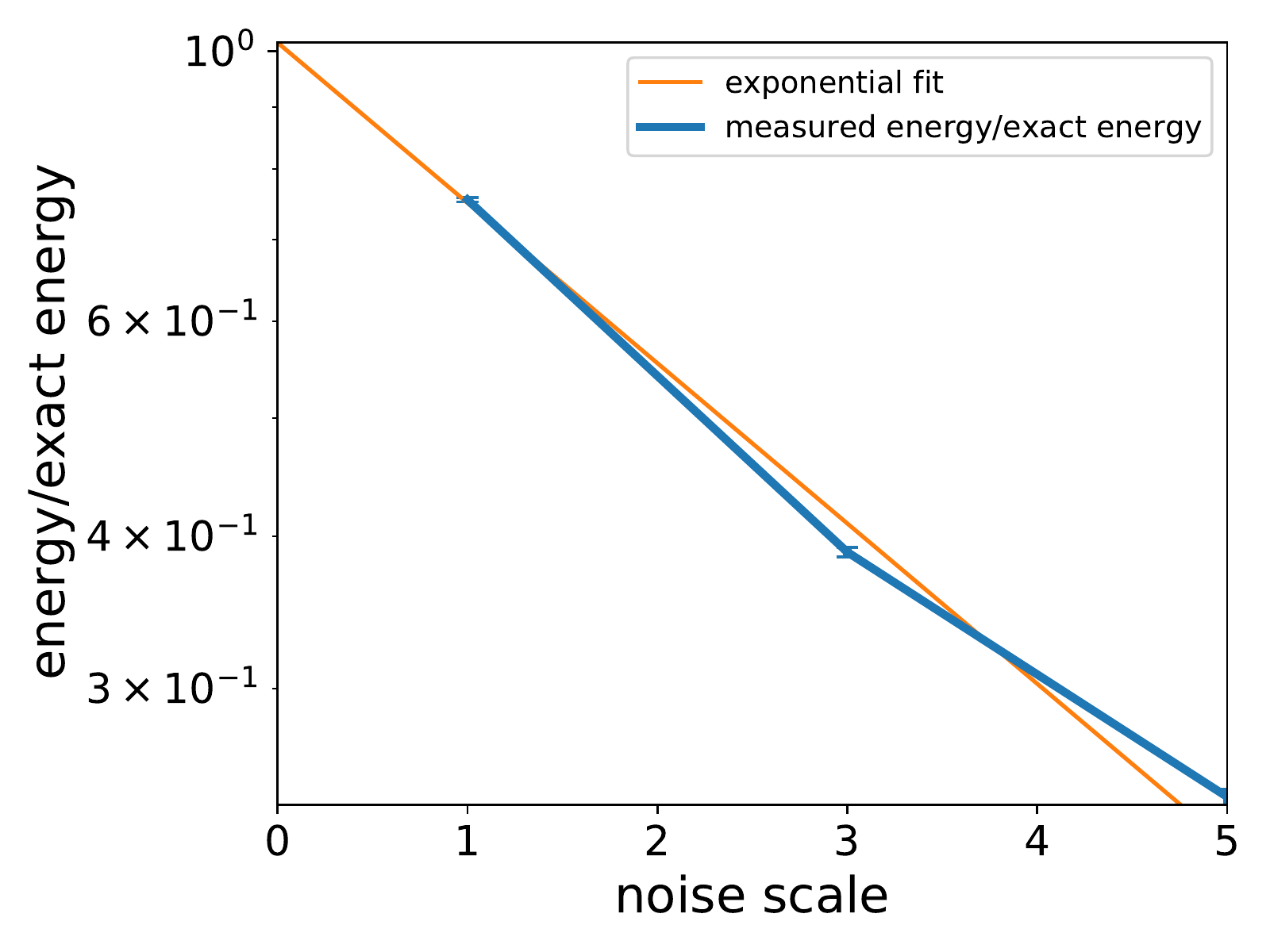}
	\caption{6 ansatz layers, without readout error mitigation}
	\end{subfigure}
	\begin{subfigure}{0.3\textwidth}
	\includegraphics[width=\textwidth]{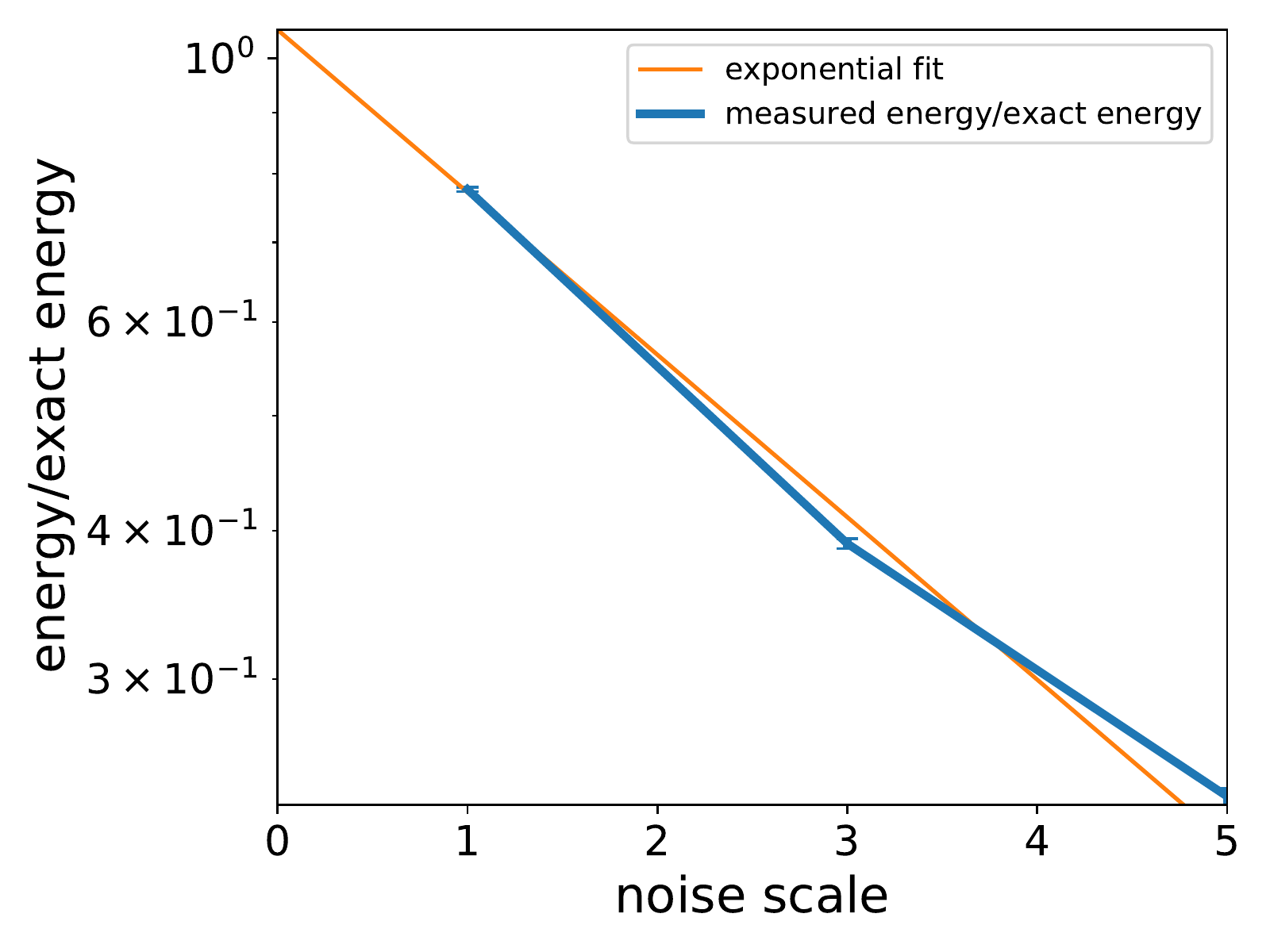}
	\caption{6 ansatz layers, with readout error mitigation}
	\end{subfigure}
	\begin{subfigure}{0.3\textwidth}
	\includegraphics[width=\textwidth]{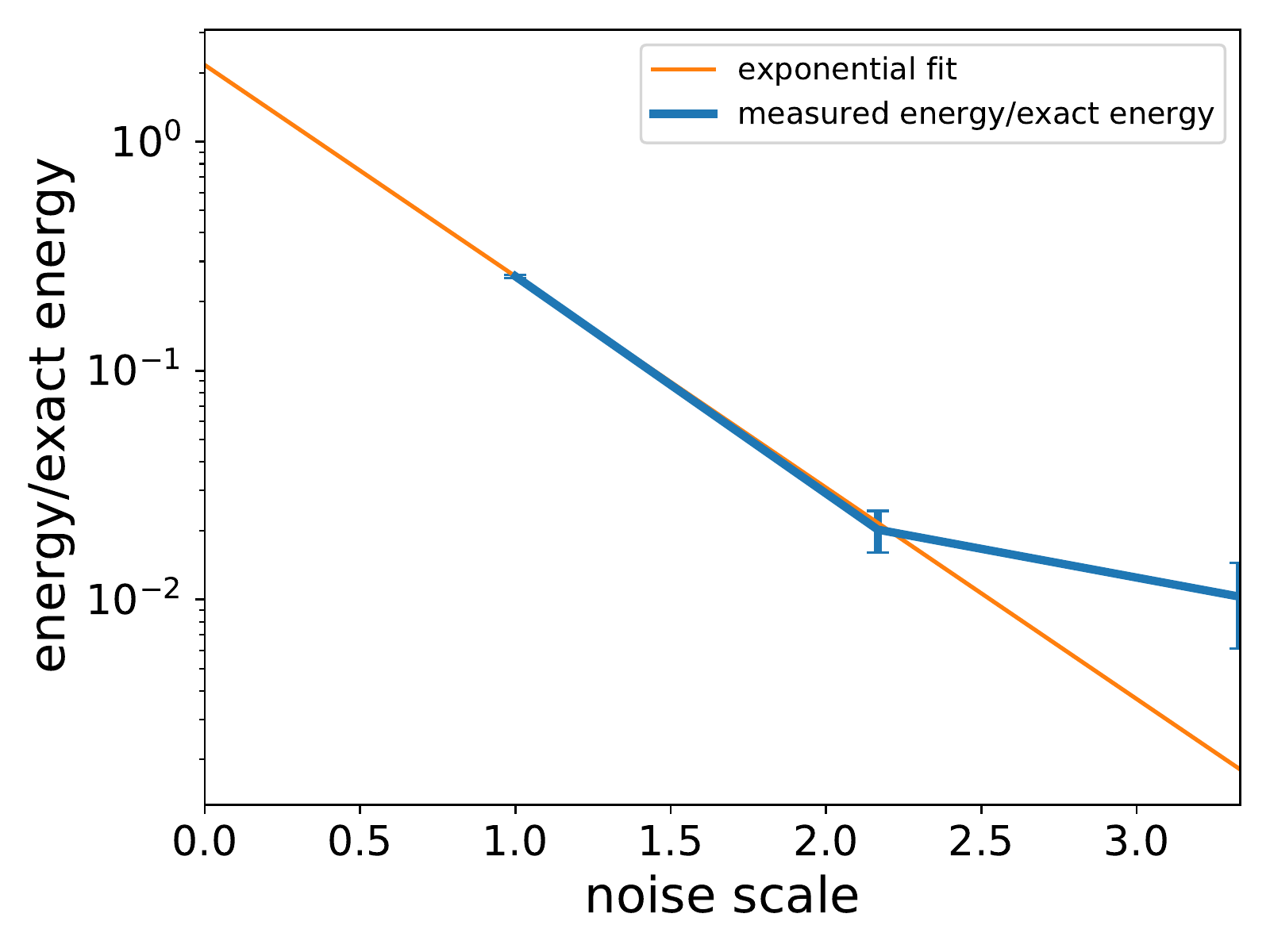}
	\caption{30 ansatz layers, without readout error mitigation}
	\end{subfigure}
	\begin{subfigure}{0.3\textwidth}
	\includegraphics[width=\textwidth]{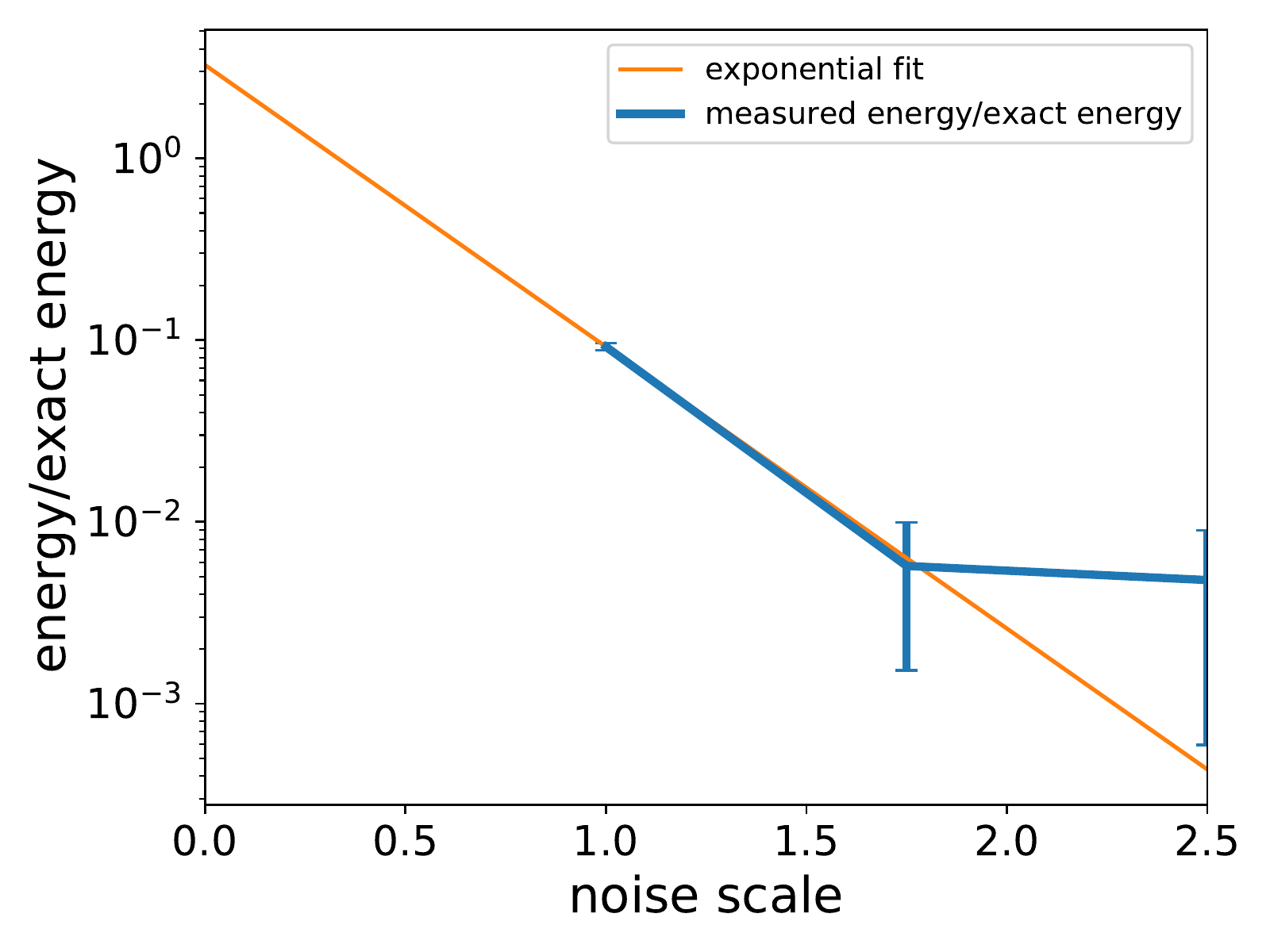}
	\caption{40 ansatz layers, without readout error mitigation}
	\end{subfigure}
	\caption{Some of the exponential fits used for zero noise extrapolation on \textit{ibmq\_sydney}. It is not clear that readout error mitigation, as we perform it, is beneficial for this method. In fact, for 30 and 40 ansatz layers, some of the readout-mitigated energies were negative, causing the exponential fits to fail. In future work, we intend to to explore better ways of performing readout error mitigation. It is also clear why this method fails at large circuit depth. The measured energy is very small even before we artifically increase the noise, and it is therefore difficult to extrapolate back to zero.}
	\label{fig:ZNE}
\end{figure}

We have used the variational-quantum-eigensolver algorithm to solve a quantum spin model with up to 20 qubits (spins), and have compared various error-mitigation techniques that can be applied to the VQE output to correct the measured ground-state energy. We see that, for studying ground states of the mixed-field Ising model on 20 qubits on \textit{ibmq\_toronto} and \textit{ibmq\_sydney}, all of these methods provide improvements over no mitigation. Zero noise extrapolation fails for large circuit depths because the measured energy, even with no added noise, is damped very close to zero, and it is difficult to extrapolate these measurements to zero noise. For the run on \textit{ibmq\_sydney}, ZNE first fails at 9 layers (Figs.\ref{fig:heatmap}-\ref{fig:heatmap_readout}). Some of the fits used in our implementation of ZNE are shown in Fig.~\ref{fig:ZNE} and illustrate the difficulty of extrapolating to zero noise when the initial circuit is deep.

\begin{figure}[h]
	\centering
	\includegraphics[width=0.3\textwidth]{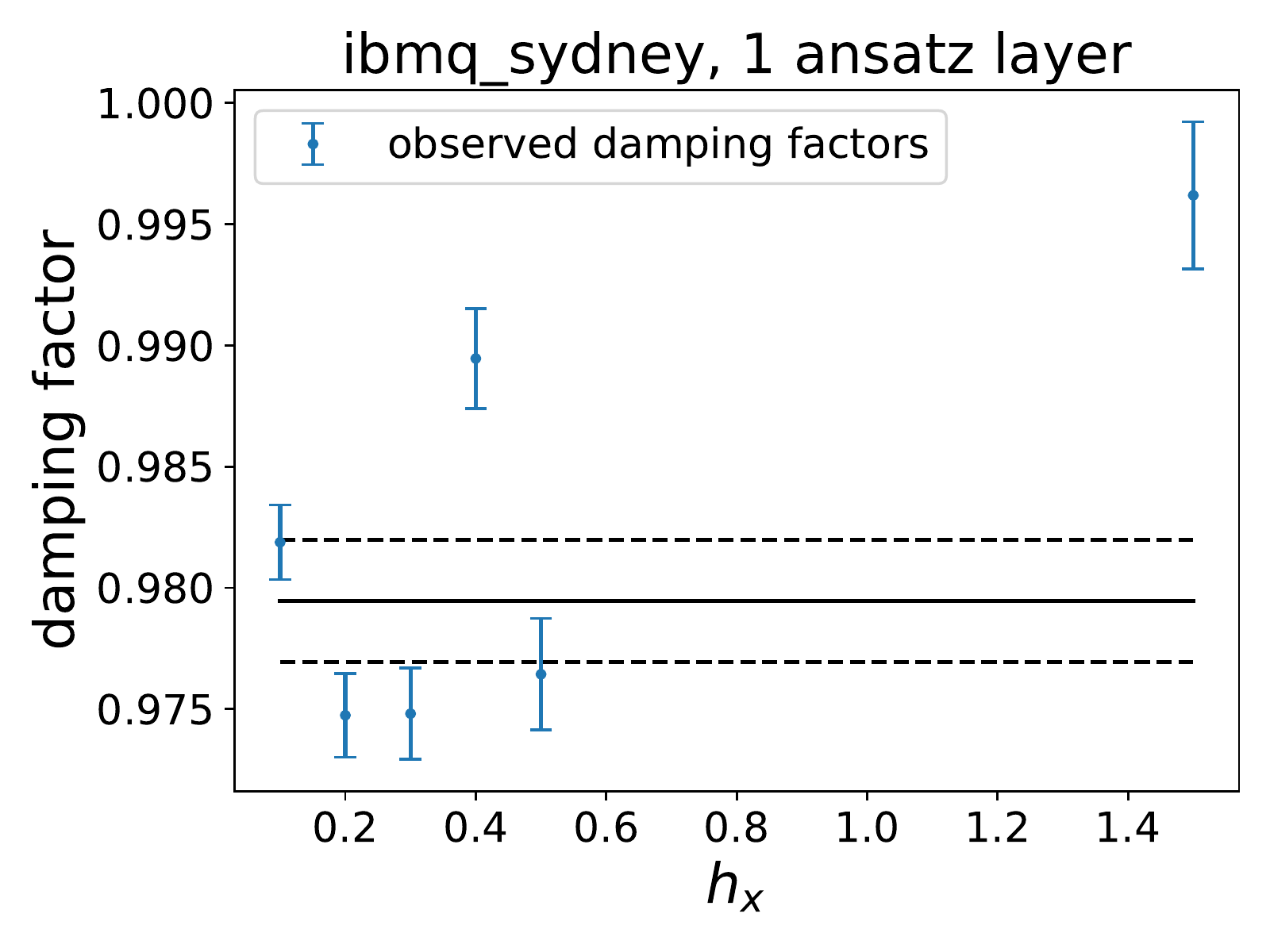}
	\includegraphics[width=0.3\textwidth]{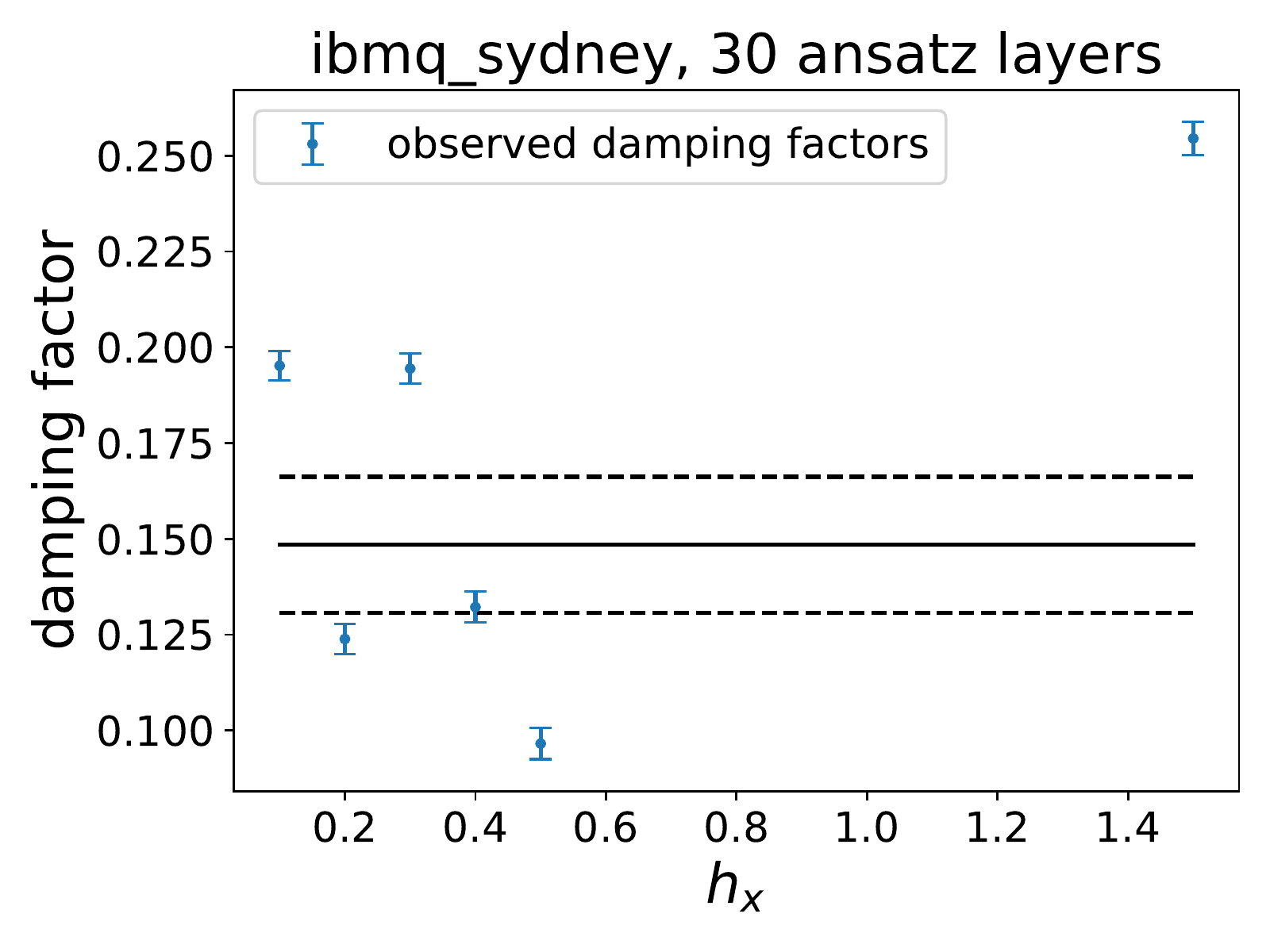}
	\includegraphics[width=0.3\textwidth]{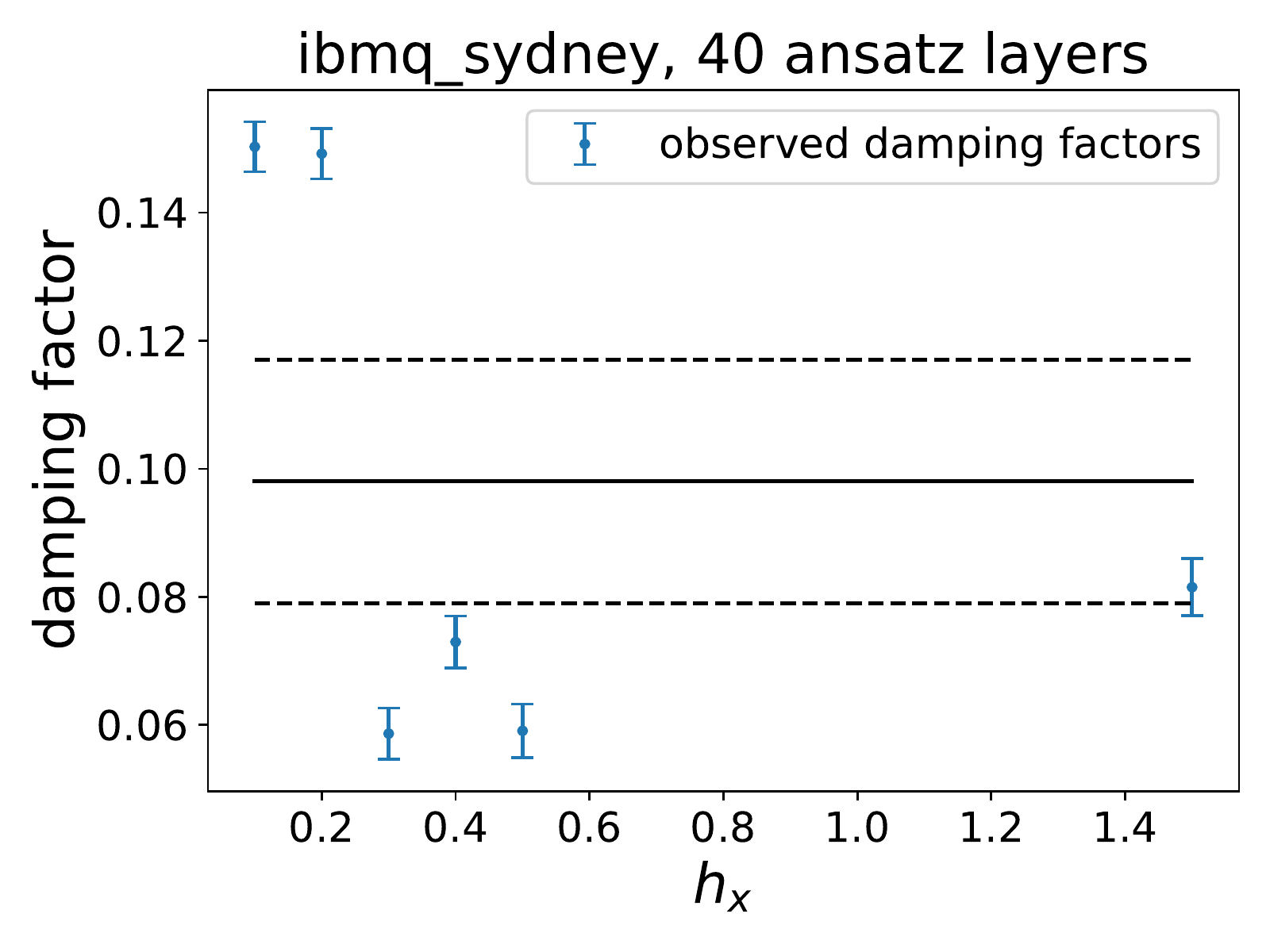}
	\caption{Some of the extrapolations used when estimating the damping factor from the perturbative regime. The solid horizontal line indicates the predicted damping factor (the mean of the damping factors for $h_x \leq 0.5$), and the dashed horizontal lines indicate the 1$\sigma$ uncertainty of the mean. Readout error mitigation has already been applied in these plots.}
	\label{fig:from_pert}
\end{figure}

Using the Hamiltonian's perturbative regime to estimate the damping factor worked relatively well, perhaps best out of the methods that we tested. In Fig.~\ref{fig:fit_and_pert}, we see that it correctly predicts the damping factor on both devices most of the time even though it, too, seems to fail when the circuit extends to 50 ansatz layers. The reader may have noticed that the error bars in Fig.~\ref{fig:rel_error} are larger for this method than for the others. This is because the variance of the damping factor over the five Hamiltonians in the perturbative regime contributes to the statistical uncertainty of the predicted damping factor (see Fig.~\ref{fig:from_pert}). This could be reduced by picking more points in the perturbative regime. However, the fact that the spread of the measured damping factors is larger than one would expect from their individual error bars indicates that this method is unlikely to predict the correct damping factor exactly. Further, this method has a significant overhead, requiring the optimization of the ansatz circuits for each of the perturbative regime Hamiltonians in addition to the Hamiltonian of interest. Its applicability is also limited because, unlike ZNE, it can only be applied to optimized circuits and only to systems that have a perturbative regime.

As seen in Fig.~\ref{fig:fit_and_pert}, the damping factor did decay approximately exponentially in the number of ansatz layers. Therefore, the exponential fit to the points with at most 15 layers worked reasonably well out to about 25 layers, which is better than ZNE performed on \textit{ibmq\_sydney} but not as well as the ``from pert" method worked. Extrapolating in the number of ansatz layers also only works for optimized circuits, but it does not require the Hamiltonian to have a perturbative regime. It does, however, require the Hamiltonian and the ansatz to be local in order for the shallow-depth ansatz circuits to be exactly classically simulable.

The method of removing the single-qubit gates, like ZNE, does not require the ansatz to be optimized and does not place any restrictions on the Hamiltonian or the ansatz. However, it does assume that the damping factor is independent of the ansatz parameters, which we have seen is not true. Consequently, this method, whether combined with ZNE or not, does not  perform as well as the other methods (c.f. Fig.~\ref{fig:fit_and_pert}).

In summary, we have seen that error-mitigation methods that work by estimating a damping factor can outperform zero noise extrapolation and extend feasible circuit depths to about 25 ansatz layers. However, these methods are also more specialized, best suited to specific types of Hamiltonians and ans\"atze and poorly suited to the optimization phase, during which one could still apply ZNE. The Hamiltonian studied in this work was particularly simple and did not require error mitigation during the optimization phase. However, that will not always be the case.

By carefully studying error mitigation techniques and the circuit depths that they enable on current devices, we are progressing toward the day when quantum comptuers will  be a tool for learning new things about physics and chemistry systems.

\section*{Data and code availability}

The code used to perform the numerical simulations, the experimental runs on IBM's hardware, and the data analysis (including error mitigation) is available at \url{https://github.com/mcmahon-lab/error_mitigation_vqe}. The data used to produce the figures may be made available upon request.

\section*{Acknowledgements}
We thank Thomas Hartman for helpful discussions and Mandar Sohoni for providing feedback on a draft of this paper. Additionally, we thank Eun-Ah Kim and three anonymous referees for useful feedback. We also thank Sarah Kaiser for finding a bug in an earlier (unpublished) version of the accompanying code. We gratefully acknowledge financial support from US DOE grant DE-SC0020397 and access to IBM quantum-computer time via the \textit{AFRL IBM Q Network Hub}, facilitated by Laura Wessing and Jon Maggiolino. The views expressed are those of the authors, and do not reflect the official policy or position of IBM or the IBM Quantum team. We used Qiskit \cite{Qiskit} to interact with IBM's devices. PLM acknowledges membership in the CIFAR Quantum Information Science Program as an Azrieli Global Scholar.

\newpage

\bibliography{references}

\begin{thebibliography}{37}%
\makeatletter
\providecommand \@ifxundefined [1]{%
 \@ifx{#1\undefined}
}%
\providecommand \@ifnum [1]{%
 \ifnum #1\expandafter \@firstoftwo
 \else \expandafter \@secondoftwo
 \fi
}%
\providecommand \@ifx [1]{%
 \ifx #1\expandafter \@firstoftwo
 \else \expandafter \@secondoftwo
 \fi
}%
\providecommand \natexlab [1]{#1}%
\providecommand \enquote  [1]{``#1''}%
\providecommand \bibnamefont  [1]{#1}%
\providecommand \bibfnamefont [1]{#1}%
\providecommand \citenamefont [1]{#1}%
\providecommand \href@noop [0]{\@secondoftwo}%
\providecommand \href [0]{\begingroup \@sanitize@url \@href}%
\providecommand \@href[1]{\@@startlink{#1}\@@href}%
\providecommand \@@href[1]{\endgroup#1\@@endlink}%
\providecommand \@sanitize@url [0]{\catcode `\\12\catcode `\$12\catcode
  `\&12\catcode `\#12\catcode `\^12\catcode `\_12\catcode `\%12\relax}%
\providecommand \@@startlink[1]{}%
\providecommand \@@endlink[0]{}%
\providecommand \url  [0]{\begingroup\@sanitize@url \@url }%
\providecommand \@url [1]{\endgroup\@href {#1}{\urlprefix }}%
\providecommand \urlprefix  [0]{URL }%
\providecommand \Eprint [0]{\href }%
\providecommand \doibase [0]{https://doi.org/}%
\providecommand \selectlanguage [0]{\@gobble}%
\providecommand \bibinfo  [0]{\@secondoftwo}%
\providecommand \bibfield  [0]{\@secondoftwo}%
\providecommand \translation [1]{[#1]}%
\providecommand \BibitemOpen [0]{}%
\providecommand \bibitemStop [0]{}%
\providecommand \bibitemNoStop [0]{.\EOS\space}%
\providecommand \EOS [0]{\spacefactor3000\relax}%
\providecommand \BibitemShut  [1]{\csname bibitem#1\endcsname}%
\let\auto@bib@innerbib\@empty
%</preamble>
\bibitem [{\citenamefont {Li}\ and\ \citenamefont {Benjamin}(2017)}]{Li_2017}%
  \BibitemOpen
  \bibfield  {author} {\bibinfo {author} {\bibfnamefont {Y.}~\bibnamefont
  {Li}}\ and\ \bibinfo {author} {\bibfnamefont {S.~C.}\ \bibnamefont
  {Benjamin}},\ }\bibfield  {title} {\bibinfo {title} {Efficient variational
  quantum simulator incorporating active error minimization},\ }\href
  {https://doi.org/10.1103/PhysRevX.7.021050} {\bibfield  {journal} {\bibinfo
  {journal} {Physical Review X}\ }\textbf {\bibinfo {volume} {7}},\ \bibinfo
  {pages} {021050} (\bibinfo {year} {2017})}\BibitemShut {NoStop}%
\bibitem [{\citenamefont {Temme}\ \emph {et~al.}(2017)\citenamefont {Temme},
  \citenamefont {Bravyi},\ and\ \citenamefont {Gambetta}}]{temme2017error}%
  \BibitemOpen
  \bibfield  {author} {\bibinfo {author} {\bibfnamefont {K.}~\bibnamefont
  {Temme}}, \bibinfo {author} {\bibfnamefont {S.}~\bibnamefont {Bravyi}},\ and\
  \bibinfo {author} {\bibfnamefont {J.~M.}\ \bibnamefont {Gambetta}},\
  }\bibfield  {title} {\bibinfo {title} {Error mitigation for short-depth
  quantum circuits},\ }\href {https://doi.org/10.1103/physrevlett.119.180509}
  {\bibfield  {journal} {\bibinfo  {journal} {Physical Review Letters}\
  }\textbf {\bibinfo {volume} {119}},\ \bibinfo {pages} {180509} (\bibinfo
  {year} {2017})}\BibitemShut {NoStop}%
\bibitem [{\citenamefont {Endo}\ \emph {et~al.}(2018)\citenamefont {Endo},
  \citenamefont {Benjamin},\ and\ \citenamefont {Li}}]{endo2018practical}%
  \BibitemOpen
  \bibfield  {author} {\bibinfo {author} {\bibfnamefont {S.}~\bibnamefont
  {Endo}}, \bibinfo {author} {\bibfnamefont {S.~C.}\ \bibnamefont {Benjamin}},\
  and\ \bibinfo {author} {\bibfnamefont {Y.}~\bibnamefont {Li}},\ }\bibfield
  {title} {\bibinfo {title} {Practical quantum error mitigation for near-future
  applications},\ }\href {https://doi.org/10.1103/physrevx.8.031027} {\bibfield
   {journal} {\bibinfo  {journal} {Physical Review X}\ }\textbf {\bibinfo
  {volume} {8}},\ \bibinfo {pages} {031027} (\bibinfo {year}
  {2018})}\BibitemShut {NoStop}%
\bibitem [{\citenamefont {Kandala}\ \emph {et~al.}(2019)\citenamefont
  {Kandala}, \citenamefont {Temme}, \citenamefont {C{\'o}rcoles}, \citenamefont
  {Mezzacapo}, \citenamefont {Chow},\ and\ \citenamefont
  {Gambetta}}]{kandala2019error}%
  \BibitemOpen
  \bibfield  {author} {\bibinfo {author} {\bibfnamefont {A.}~\bibnamefont
  {Kandala}}, \bibinfo {author} {\bibfnamefont {K.}~\bibnamefont {Temme}},
  \bibinfo {author} {\bibfnamefont {A.~D.}\ \bibnamefont {C{\'o}rcoles}},
  \bibinfo {author} {\bibfnamefont {A.}~\bibnamefont {Mezzacapo}}, \bibinfo
  {author} {\bibfnamefont {J.~M.}\ \bibnamefont {Chow}},\ and\ \bibinfo
  {author} {\bibfnamefont {J.~M.}\ \bibnamefont {Gambetta}},\ }\bibfield
  {title} {\bibinfo {title} {Error mitigation extends the computational reach
  of a noisy quantum processor},\ }\href
  {https://doi.org/10.1038/s41586-019-1040-7} {\bibfield  {journal} {\bibinfo
  {journal} {Nature}\ }\textbf {\bibinfo {volume} {567}},\ \bibinfo {pages}
  {491} (\bibinfo {year} {2019})}\BibitemShut {NoStop}%
\bibitem [{\citenamefont {Otten}\ and\ \citenamefont {Gray}(2019)}]{Otten2019}%
  \BibitemOpen
  \bibfield  {author} {\bibinfo {author} {\bibfnamefont {M.}~\bibnamefont
  {Otten}}\ and\ \bibinfo {author} {\bibfnamefont {S.~K.}\ \bibnamefont
  {Gray}},\ }\bibfield  {title} {\bibinfo {title} {Recovering noise-free
  quantum observables},\ }\href {https://doi.org/10.1103/PhysRevA.99.012338}
  {\bibfield  {journal} {\bibinfo  {journal} {Physical Review A}\ }\textbf
  {\bibinfo {volume} {99}},\ \bibinfo {pages} {012338} (\bibinfo {year}
  {2019})}\BibitemShut {NoStop}%
\bibitem [{\citenamefont {He}\ \emph {et~al.}(2020)\citenamefont {He},
  \citenamefont {Nachman}, \citenamefont {de~Jong},\ and\ \citenamefont
  {Bauer}}]{He_2020}%
  \BibitemOpen
  \bibfield  {author} {\bibinfo {author} {\bibfnamefont {A.}~\bibnamefont
  {He}}, \bibinfo {author} {\bibfnamefont {B.}~\bibnamefont {Nachman}},
  \bibinfo {author} {\bibfnamefont {W.~A.}\ \bibnamefont {de~Jong}},\ and\
  \bibinfo {author} {\bibfnamefont {C.~W.}\ \bibnamefont {Bauer}},\ }\bibfield
  {title} {\bibinfo {title} {Zero-noise extrapolation for quantum-gate error
  mitigation with identity insertions},\ }\href
  {https://doi.org/10.1103/PhysRevA.102.012426} {\bibfield  {journal} {\bibinfo
   {journal} {Physical Review A}\ }\textbf {\bibinfo {volume} {102}},\ \bibinfo
  {pages} {012426} (\bibinfo {year} {2020})}\BibitemShut {NoStop}%
\bibitem [{\citenamefont {Koczor}(2021)}]{koczor2021exponential}%
  \BibitemOpen
  \bibfield  {author} {\bibinfo {author} {\bibfnamefont {B.}~\bibnamefont
  {Koczor}},\ }\bibfield  {title} {\bibinfo {title} {Exponential error
  suppression for near-term quantum devices},\ }\bibfield  {journal} {\bibinfo
  {journal} {Physical Review X}\ }\textbf {\bibinfo {volume} {11}},\ \href
  {https://doi.org/10.1103/physrevx.11.031057} {10.1103/physrevx.11.031057}
  (\bibinfo {year} {2021})\BibitemShut {NoStop}%
\bibitem [{\citenamefont {Huggins}\ \emph {et~al.}(2021)\citenamefont
  {Huggins}, \citenamefont {McArdle}, \citenamefont {O'Brien}, \citenamefont
  {Lee}, \citenamefont {Rubin}, \citenamefont {Boixo}, \citenamefont {Whaley},
  \citenamefont {Babbush},\ and\ \citenamefont {McClean}}]{huggins2021virtual}%
  \BibitemOpen
  \bibfield  {author} {\bibinfo {author} {\bibfnamefont {W.~J.}\ \bibnamefont
  {Huggins}}, \bibinfo {author} {\bibfnamefont {S.}~\bibnamefont {McArdle}},
  \bibinfo {author} {\bibfnamefont {T.~E.}\ \bibnamefont {O'Brien}}, \bibinfo
  {author} {\bibfnamefont {J.}~\bibnamefont {Lee}}, \bibinfo {author}
  {\bibfnamefont {N.~C.}\ \bibnamefont {Rubin}}, \bibinfo {author}
  {\bibfnamefont {S.}~\bibnamefont {Boixo}}, \bibinfo {author} {\bibfnamefont
  {K.~B.}\ \bibnamefont {Whaley}}, \bibinfo {author} {\bibfnamefont
  {R.}~\bibnamefont {Babbush}},\ and\ \bibinfo {author} {\bibfnamefont {J.~R.}\
  \bibnamefont {McClean}},\ }\href@noop {} {\bibinfo {title} {Virtual
  distillation for quantum error mitigation}} (\bibinfo {year} {2021}),\
  \Eprint {https://arxiv.org/abs/2011.07064} {arXiv:2011.07064 [quant-ph]}
  \BibitemShut {NoStop}%
\bibitem [{\citenamefont {Czarnik}\ \emph {et~al.}(2021)\citenamefont
  {Czarnik}, \citenamefont {Arrasmith}, \citenamefont {Coles},\ and\
  \citenamefont {Cincio}}]{czarnik2021error}%
  \BibitemOpen
  \bibfield  {author} {\bibinfo {author} {\bibfnamefont {P.}~\bibnamefont
  {Czarnik}}, \bibinfo {author} {\bibfnamefont {A.}~\bibnamefont {Arrasmith}},
  \bibinfo {author} {\bibfnamefont {P.~J.}\ \bibnamefont {Coles}},\ and\
  \bibinfo {author} {\bibfnamefont {L.}~\bibnamefont {Cincio}},\ }\href@noop {}
  {\bibinfo {title} {Error mitigation with {C}lifford quantum-circuit data}}
  (\bibinfo {year} {2021}),\ \Eprint {https://arxiv.org/abs/2005.10189}
  {arXiv:2005.10189 [quant-ph]} \BibitemShut {NoStop}%
\bibitem [{\citenamefont {Montanaro}\ and\ \citenamefont
  {Stanisic}(2021)}]{montanaro2021error}%
  \BibitemOpen
  \bibfield  {author} {\bibinfo {author} {\bibfnamefont {A.}~\bibnamefont
  {Montanaro}}\ and\ \bibinfo {author} {\bibfnamefont {S.}~\bibnamefont
  {Stanisic}},\ }\href@noop {} {\bibinfo {title} {Error mitigation by training
  with fermionic linear optics}} (\bibinfo {year} {2021}),\ \Eprint
  {https://arxiv.org/abs/2102.02120} {arXiv:2102.02120 [quant-ph]} \BibitemShut
  {NoStop}%
\bibitem [{\citenamefont {Shaw}(2021)}]{shaw2021classicalquantum}%
  \BibitemOpen
  \bibfield  {author} {\bibinfo {author} {\bibfnamefont {A.}~\bibnamefont
  {Shaw}},\ }\href@noop {} {\bibinfo {title} {Classical-quantum noise
  mitigation for {NISQ} hardware}} (\bibinfo {year} {2021}),\ \Eprint
  {https://arxiv.org/abs/2105.08701} {arXiv:2105.08701 [quant-ph]} \BibitemShut
  {NoStop}%
\bibitem [{\citenamefont {{Google AI Quantum}}\ and\ \citenamefont
  {collaborators}(2020)}]{arute2020observation}%
  \BibitemOpen
  \bibfield  {author} {\bibinfo {author} {\bibnamefont {{Google AI Quantum}}}\
  and\ \bibinfo {author} {\bibnamefont {collaborators}},\ }\href@noop {}
  {\bibinfo {title} {Observation of separated dynamics of charge and spin in
  the {Fermi-Hubbard} model}} (\bibinfo {year} {2020}),\ \Eprint
  {https://arxiv.org/abs/2010.07965} {arXiv:2010.07965 [quant-ph]} \BibitemShut
  {NoStop}%
\bibitem [{\citenamefont {Mi}\ \emph {et~al.}(2021)\citenamefont {Mi},
  \citenamefont {Roushan}, \citenamefont {Quintana}, \citenamefont {Mandrà},
  \citenamefont {Marshall}, \citenamefont {Neill}, \citenamefont {Arute},
  \citenamefont {Arya}, \citenamefont {Atalaya}, \citenamefont {Babbush},\ and\
  \citenamefont {et~al.}}]{mi2021information}%
  \BibitemOpen
  \bibfield  {author} {\bibinfo {author} {\bibfnamefont {X.}~\bibnamefont
  {Mi}}, \bibinfo {author} {\bibfnamefont {P.}~\bibnamefont {Roushan}},
  \bibinfo {author} {\bibfnamefont {C.}~\bibnamefont {Quintana}}, \bibinfo
  {author} {\bibfnamefont {S.}~\bibnamefont {Mandrà}}, \bibinfo {author}
  {\bibfnamefont {J.}~\bibnamefont {Marshall}}, \bibinfo {author}
  {\bibfnamefont {C.}~\bibnamefont {Neill}}, \bibinfo {author} {\bibfnamefont
  {F.}~\bibnamefont {Arute}}, \bibinfo {author} {\bibfnamefont
  {K.}~\bibnamefont {Arya}}, \bibinfo {author} {\bibfnamefont {J.}~\bibnamefont
  {Atalaya}}, \bibinfo {author} {\bibfnamefont {R.}~\bibnamefont {Babbush}},\
  and\ \bibinfo {author} {\bibnamefont {et~al.}},\ }\bibfield  {title}
  {\bibinfo {title} {Information scrambling in quantum circuits},\ }\bibfield
  {journal} {\bibinfo  {journal} {Science}\ }\href
  {https://doi.org/10.1126/science.abg5029} {10.1126/science.abg5029} (\bibinfo
  {year} {2021})\BibitemShut {NoStop}%
\bibitem [{\citenamefont {Urbanek}\ \emph {et~al.}(2021)\citenamefont
  {Urbanek}, \citenamefont {Nachman}, \citenamefont {Pascuzzi}, \citenamefont
  {He}, \citenamefont {Bauer},\ and\ \citenamefont
  {de~Jong}}]{urbanek2021mitigating}%
  \BibitemOpen
  \bibfield  {author} {\bibinfo {author} {\bibfnamefont {M.}~\bibnamefont
  {Urbanek}}, \bibinfo {author} {\bibfnamefont {B.}~\bibnamefont {Nachman}},
  \bibinfo {author} {\bibfnamefont {V.~R.}\ \bibnamefont {Pascuzzi}}, \bibinfo
  {author} {\bibfnamefont {A.}~\bibnamefont {He}}, \bibinfo {author}
  {\bibfnamefont {C.~W.}\ \bibnamefont {Bauer}},\ and\ \bibinfo {author}
  {\bibfnamefont {W.~A.}\ \bibnamefont {de~Jong}},\ }\href@noop {} {\bibinfo
  {title} {Mitigating depolarizing noise on quantum computers with
  noise-estimation circuits}} (\bibinfo {year} {2021}),\ \Eprint
  {https://arxiv.org/abs/2103.08591} {arXiv:2103.08591 [quant-ph]} \BibitemShut
  {NoStop}%
\bibitem [{\citenamefont {Vovrosh}\ \emph {et~al.}(2021)\citenamefont
  {Vovrosh}, \citenamefont {Khosla}, \citenamefont {Greenaway}, \citenamefont
  {Self}, \citenamefont {Kim},\ and\ \citenamefont
  {Knolle}}]{vovrosh2021efficient}%
  \BibitemOpen
  \bibfield  {author} {\bibinfo {author} {\bibfnamefont {J.}~\bibnamefont
  {Vovrosh}}, \bibinfo {author} {\bibfnamefont {K.~E.}\ \bibnamefont {Khosla}},
  \bibinfo {author} {\bibfnamefont {S.}~\bibnamefont {Greenaway}}, \bibinfo
  {author} {\bibfnamefont {C.}~\bibnamefont {Self}}, \bibinfo {author}
  {\bibfnamefont {M.~S.}\ \bibnamefont {Kim}},\ and\ \bibinfo {author}
  {\bibfnamefont {J.}~\bibnamefont {Knolle}},\ }\bibfield  {title} {\bibinfo
  {title} {Simple mitigation of global depolarizing errors in quantum
  simulations},\ }\bibfield  {journal} {\bibinfo  {journal} {Physical Review
  E}\ }\textbf {\bibinfo {volume} {104}},\ \href
  {https://doi.org/10.1103/physreve.104.035309} {10.1103/physreve.104.035309}
  (\bibinfo {year} {2021})\BibitemShut {NoStop}%
\bibitem [{\citenamefont {Ville}\ \emph {et~al.}(2021)\citenamefont {Ville},
  \citenamefont {Morvan}, \citenamefont {Hashim}, \citenamefont {Naik},
  \citenamefont {Mitchell}, \citenamefont {Kreikebaum}, \citenamefont
  {O'Brien}, \citenamefont {Wallman}, \citenamefont {Hincks}, \citenamefont
  {Emerson}, \citenamefont {Smith}, \citenamefont {Younis}, \citenamefont
  {Iancu}, \citenamefont {Santiago},\ and\ \citenamefont
  {Siddiqi}}]{ville2021leveraging}%
  \BibitemOpen
  \bibfield  {author} {\bibinfo {author} {\bibfnamefont {J.-L.}\ \bibnamefont
  {Ville}}, \bibinfo {author} {\bibfnamefont {A.}~\bibnamefont {Morvan}},
  \bibinfo {author} {\bibfnamefont {A.}~\bibnamefont {Hashim}}, \bibinfo
  {author} {\bibfnamefont {R.~K.}\ \bibnamefont {Naik}}, \bibinfo {author}
  {\bibfnamefont {B.}~\bibnamefont {Mitchell}}, \bibinfo {author}
  {\bibfnamefont {J.-M.}\ \bibnamefont {Kreikebaum}}, \bibinfo {author}
  {\bibfnamefont {K.~P.}\ \bibnamefont {O'Brien}}, \bibinfo {author}
  {\bibfnamefont {J.~J.}\ \bibnamefont {Wallman}}, \bibinfo {author}
  {\bibfnamefont {I.}~\bibnamefont {Hincks}}, \bibinfo {author} {\bibfnamefont
  {J.}~\bibnamefont {Emerson}}, \bibinfo {author} {\bibfnamefont
  {E.}~\bibnamefont {Smith}}, \bibinfo {author} {\bibfnamefont
  {E.}~\bibnamefont {Younis}}, \bibinfo {author} {\bibfnamefont
  {C.}~\bibnamefont {Iancu}}, \bibinfo {author} {\bibfnamefont {D.~I.}\
  \bibnamefont {Santiago}},\ and\ \bibinfo {author} {\bibfnamefont
  {I.}~\bibnamefont {Siddiqi}},\ }\href@noop {} {\bibinfo {title} {Leveraging
  randomized compiling for the {QITE} algorithm}} (\bibinfo {year} {2021}),\
  \Eprint {https://arxiv.org/abs/2104.08785} {arXiv:2104.08785 [quant-ph]}
  \BibitemShut {NoStop}%
\bibitem [{\citenamefont {Pfeuty}(1970)}]{PFEUTY197079}%
  \BibitemOpen
  \bibfield  {author} {\bibinfo {author} {\bibfnamefont {P.}~\bibnamefont
  {Pfeuty}},\ }\bibfield  {title} {\bibinfo {title} {The one-dimensional
  {I}sing model with a transverse field},\ }\href
  {https://doi.org/10.1016/0003-4916(70)90270-8} {\bibfield  {journal}
  {\bibinfo  {journal} {Annals of Physics}\ }\textbf {\bibinfo {volume} {57}},\
  \bibinfo {pages} {79} (\bibinfo {year} {1970})}\BibitemShut {NoStop}%
\bibitem [{\citenamefont {Kim}\ and\ \citenamefont
  {Huse}(2013)}]{Kim_Huse_2013}%
  \BibitemOpen
  \bibfield  {author} {\bibinfo {author} {\bibfnamefont {H.}~\bibnamefont
  {Kim}}\ and\ \bibinfo {author} {\bibfnamefont {D.~A.}\ \bibnamefont {Huse}},\
  }\bibfield  {title} {\bibinfo {title} {Ballistic spreading of entanglement in
  a diffusive nonintegrable system},\ }\href
  {https://doi.org/10.1103/PhysRevLett.111.127205} {\bibfield  {journal}
  {\bibinfo  {journal} {Physical Review Letters}\ }\textbf {\bibinfo {volume}
  {111}},\ \bibinfo {pages} {127205} (\bibinfo {year} {2013})}\BibitemShut
  {NoStop}%
\bibitem [{\citenamefont {Kim}\ \emph {et~al.}(2014)\citenamefont {Kim},
  \citenamefont {Ikeda},\ and\ \citenamefont {Huse}}]{Kim_Ikeda_Huse_2014}%
  \BibitemOpen
  \bibfield  {author} {\bibinfo {author} {\bibfnamefont {H.}~\bibnamefont
  {Kim}}, \bibinfo {author} {\bibfnamefont {T.~N.}\ \bibnamefont {Ikeda}},\
  and\ \bibinfo {author} {\bibfnamefont {D.~A.}\ \bibnamefont {Huse}},\
  }\bibfield  {title} {\bibinfo {title} {Testing whether all eigenstates obey
  the eigenstate thermalization hypothesis},\ }\href
  {https://doi.org/10.1103/PhysRevE.90.052105} {\bibfield  {journal} {\bibinfo
  {journal} {Physical Review E}\ }\textbf {\bibinfo {volume} {90}},\ \bibinfo
  {pages} {052105} (\bibinfo {year} {2014})}\BibitemShut {NoStop}%
\bibitem [{\citenamefont {Wurtz}\ and\ \citenamefont
  {Polkovnikov}(2020)}]{Wurtz_Polkonikov_2020}%
  \BibitemOpen
  \bibfield  {author} {\bibinfo {author} {\bibfnamefont {J.}~\bibnamefont
  {Wurtz}}\ and\ \bibinfo {author} {\bibfnamefont {A.}~\bibnamefont
  {Polkovnikov}},\ }\bibfield  {title} {\bibinfo {title} {Emergent conservation
  laws and nonthermal states in the mixed-field {I}sing model},\ }\href
  {https://doi.org/10.1103/PhysRevB.101.195138} {\bibfield  {journal} {\bibinfo
   {journal} {Physical Review B}\ }\textbf {\bibinfo {volume} {101}},\ \bibinfo
  {pages} {195138} (\bibinfo {year} {2020})}\BibitemShut {NoStop}%
\bibitem [{\citenamefont {Cerezo}\ \emph {et~al.}(2021)\citenamefont {Cerezo},
  \citenamefont {Sone}, \citenamefont {Volkoff}, \citenamefont {Cincio},\ and\
  \citenamefont {Coles}}]{cerezo2020costfunctiondependent}%
  \BibitemOpen
  \bibfield  {author} {\bibinfo {author} {\bibfnamefont {M.}~\bibnamefont
  {Cerezo}}, \bibinfo {author} {\bibfnamefont {A.}~\bibnamefont {Sone}},
  \bibinfo {author} {\bibfnamefont {T.}~\bibnamefont {Volkoff}}, \bibinfo
  {author} {\bibfnamefont {L.}~\bibnamefont {Cincio}},\ and\ \bibinfo {author}
  {\bibfnamefont {P.~J.}\ \bibnamefont {Coles}},\ }\bibfield  {title} {\bibinfo
  {title} {Cost function dependent barren plateaus in shallow parametrized
  quantum circuits},\ }\bibfield  {journal} {\bibinfo  {journal} {Nature
  Communications}\ }\textbf {\bibinfo {volume} {12}},\ \href
  {https://doi.org/10.1038/s41467-021-21728-w} {10.1038/s41467-021-21728-w}
  (\bibinfo {year} {2021})\BibitemShut {NoStop}%
\bibitem [{\citenamefont {Nakaji}\ and\ \citenamefont
  {Yamamoto}(2021)}]{nakaji2020expressibility}%
  \BibitemOpen
  \bibfield  {author} {\bibinfo {author} {\bibfnamefont {K.}~\bibnamefont
  {Nakaji}}\ and\ \bibinfo {author} {\bibfnamefont {N.}~\bibnamefont
  {Yamamoto}},\ }\bibfield  {title} {\bibinfo {title} {Expressibility of the
  alternating layered ansatz for quantum computation},\ }\href
  {https://doi.org/10.22331/q-2021-04-19-434} {\bibfield  {journal} {\bibinfo
  {journal} {Quantum}\ }\textbf {\bibinfo {volume} {5}},\ \bibinfo {pages}
  {434} (\bibinfo {year} {2021})}\BibitemShut {NoStop}%
\bibitem [{\citenamefont {Babbush}\ \emph {et~al.}(2019)\citenamefont
  {Babbush}, \citenamefont {Berry},\ and\ \citenamefont
  {Neven}}]{Babbush_2019}%
  \BibitemOpen
  \bibfield  {author} {\bibinfo {author} {\bibfnamefont {R.}~\bibnamefont
  {Babbush}}, \bibinfo {author} {\bibfnamefont {D.~W.}\ \bibnamefont {Berry}},\
  and\ \bibinfo {author} {\bibfnamefont {H.}~\bibnamefont {Neven}},\ }\bibfield
   {title} {\bibinfo {title} {Quantum simulation of the {S}achdev-{Y}e-{K}itaev
  model by asymmetric qubitization},\ }\href
  {https://doi.org/10.1103/PhysRevA.99.040301} {\bibfield  {journal} {\bibinfo
  {journal} {Physical Review A}\ }\textbf {\bibinfo {volume} {99}},\ \bibinfo
  {pages} {040301} (\bibinfo {year} {2019})}\BibitemShut {NoStop}%
\bibitem [{\citenamefont {Martyn}\ and\ \citenamefont
  {Swingle}(2019)}]{Martyn_2019}%
  \BibitemOpen
  \bibfield  {author} {\bibinfo {author} {\bibfnamefont {J.}~\bibnamefont
  {Martyn}}\ and\ \bibinfo {author} {\bibfnamefont {B.}~\bibnamefont
  {Swingle}},\ }\bibfield  {title} {\bibinfo {title} {Product spectrum ansatz
  and the simplicity of thermal states},\ }\href
  {https://doi.org/10.1103/PhysRevA.100.032107} {\bibfield  {journal} {\bibinfo
   {journal} {Physical Review A}\ }\textbf {\bibinfo {volume} {100}},\ \bibinfo
  {pages} {032107} (\bibinfo {year} {2019})}\BibitemShut {NoStop}%
\bibitem [{\citenamefont {Luo}\ \emph {et~al.}(2019)\citenamefont {Luo},
  \citenamefont {You}, \citenamefont {Li}, \citenamefont {Jian}, \citenamefont
  {Lu}, \citenamefont {Xu}, \citenamefont {Zeng},\ and\ \citenamefont
  {Laflamme}}]{Luo2019}%
  \BibitemOpen
  \bibfield  {author} {\bibinfo {author} {\bibfnamefont {Z.}~\bibnamefont
  {Luo}}, \bibinfo {author} {\bibfnamefont {Y.-Z.}\ \bibnamefont {You}},
  \bibinfo {author} {\bibfnamefont {J.}~\bibnamefont {Li}}, \bibinfo {author}
  {\bibfnamefont {C.-M.}\ \bibnamefont {Jian}}, \bibinfo {author}
  {\bibfnamefont {D.}~\bibnamefont {Lu}}, \bibinfo {author} {\bibfnamefont
  {C.}~\bibnamefont {Xu}}, \bibinfo {author} {\bibfnamefont {B.}~\bibnamefont
  {Zeng}},\ and\ \bibinfo {author} {\bibfnamefont {R.}~\bibnamefont
  {Laflamme}},\ }\bibfield  {title} {\bibinfo {title} {Quantum simulation of
  the non-fermi-liquid state of {S}achdev-{Y}e-{K}itaev model},\ }\href
  {https://doi.org/10.1038/s41534-019-0166-7} {\bibfield  {journal} {\bibinfo
  {journal} {npj Quantum Information}\ }\textbf {\bibinfo {volume} {5}},\
  \bibinfo {pages} {53} (\bibinfo {year} {2019})}\BibitemShut {NoStop}%
\bibitem [{\citenamefont {Brown}\ \emph {et~al.}(2021)\citenamefont {Brown},
  \citenamefont {Gharibyan}, \citenamefont {Leichenauer}, \citenamefont {Lin},
  \citenamefont {Nezami}, \citenamefont {Salton}, \citenamefont {Susskind},
  \citenamefont {Swingle},\ and\ \citenamefont {Walter}}]{brown2021quantum}%
  \BibitemOpen
  \bibfield  {author} {\bibinfo {author} {\bibfnamefont {A.~R.}\ \bibnamefont
  {Brown}}, \bibinfo {author} {\bibfnamefont {H.}~\bibnamefont {Gharibyan}},
  \bibinfo {author} {\bibfnamefont {S.}~\bibnamefont {Leichenauer}}, \bibinfo
  {author} {\bibfnamefont {H.~W.}\ \bibnamefont {Lin}}, \bibinfo {author}
  {\bibfnamefont {S.}~\bibnamefont {Nezami}}, \bibinfo {author} {\bibfnamefont
  {G.}~\bibnamefont {Salton}}, \bibinfo {author} {\bibfnamefont
  {L.}~\bibnamefont {Susskind}}, \bibinfo {author} {\bibfnamefont
  {B.}~\bibnamefont {Swingle}},\ and\ \bibinfo {author} {\bibfnamefont
  {M.}~\bibnamefont {Walter}},\ }\href@noop {} {\bibinfo {title} {Quantum
  gravity in the lab: Teleportation by size and traversable wormholes}}
  (\bibinfo {year} {2021}),\ \Eprint {https://arxiv.org/abs/1911.06314}
  {arXiv:1911.06314 [quant-ph]} \BibitemShut {NoStop}%
\bibitem [{\citenamefont {Nezami}\ \emph {et~al.}(2021)\citenamefont {Nezami},
  \citenamefont {Lin}, \citenamefont {Brown}, \citenamefont {Gharibyan},
  \citenamefont {Leichenauer}, \citenamefont {Salton}, \citenamefont
  {Susskind}, \citenamefont {Swingle},\ and\ \citenamefont
  {Walter}}]{nezami2021quantum}%
  \BibitemOpen
  \bibfield  {author} {\bibinfo {author} {\bibfnamefont {S.}~\bibnamefont
  {Nezami}}, \bibinfo {author} {\bibfnamefont {H.~W.}\ \bibnamefont {Lin}},
  \bibinfo {author} {\bibfnamefont {A.~R.}\ \bibnamefont {Brown}}, \bibinfo
  {author} {\bibfnamefont {H.}~\bibnamefont {Gharibyan}}, \bibinfo {author}
  {\bibfnamefont {S.}~\bibnamefont {Leichenauer}}, \bibinfo {author}
  {\bibfnamefont {G.}~\bibnamefont {Salton}}, \bibinfo {author} {\bibfnamefont
  {L.}~\bibnamefont {Susskind}}, \bibinfo {author} {\bibfnamefont
  {B.}~\bibnamefont {Swingle}},\ and\ \bibinfo {author} {\bibfnamefont
  {M.}~\bibnamefont {Walter}},\ }\href@noop {} {\bibinfo {title} {Quantum
  gravity in the lab: Teleportation by size and traversable wormholes, {P}art
  {II}}} (\bibinfo {year} {2021}),\ \Eprint {https://arxiv.org/abs/2102.01064}
  {arXiv:2102.01064 [quant-ph]} \BibitemShut {NoStop}%
\bibitem [{\citenamefont {Bravyi}\ and\ \citenamefont
  {Kitaev}(2002)}]{Bravyi_2002}%
  \BibitemOpen
  \bibfield  {author} {\bibinfo {author} {\bibfnamefont {S.~B.}\ \bibnamefont
  {Bravyi}}\ and\ \bibinfo {author} {\bibfnamefont {A.~Y.}\ \bibnamefont
  {Kitaev}},\ }\bibfield  {title} {\bibinfo {title} {Fermionic quantum
  computation},\ }\href {https://doi.org/10.1006/aphy.2002.6254} {\bibfield
  {journal} {\bibinfo  {journal} {Annals of Physics}\ }\textbf {\bibinfo
  {volume} {298}},\ \bibinfo {pages} {210–226} (\bibinfo {year}
  {2002})}\BibitemShut {NoStop}%
\bibitem [{\citenamefont {Wallman}\ and\ \citenamefont
  {Emerson}(2016)}]{Wallman_2016}%
  \BibitemOpen
  \bibfield  {author} {\bibinfo {author} {\bibfnamefont {J.~J.}\ \bibnamefont
  {Wallman}}\ and\ \bibinfo {author} {\bibfnamefont {J.}~\bibnamefont
  {Emerson}},\ }\bibfield  {title} {\bibinfo {title} {Noise tailoring for
  scalable quantum computation via randomized compiling},\ }\bibfield
  {journal} {\bibinfo  {journal} {Physical Review A}\ }\textbf {\bibinfo
  {volume} {94}},\ \href {https://doi.org/10.1103/physreva.94.052325}
  {10.1103/physreva.94.052325} (\bibinfo {year} {2016})\BibitemShut {NoStop}%
\bibitem [{\citenamefont {Cai}\ and\ \citenamefont
  {Benjamin}(2019)}]{Cai_2019}%
  \BibitemOpen
  \bibfield  {author} {\bibinfo {author} {\bibfnamefont {Z.}~\bibnamefont
  {Cai}}\ and\ \bibinfo {author} {\bibfnamefont {S.~C.}\ \bibnamefont
  {Benjamin}},\ }\bibfield  {title} {\bibinfo {title} {Constructing smaller
  {P}auli twirling sets for arbitrary error channels},\ }\bibfield  {journal}
  {\bibinfo  {journal} {Scientific Reports}\ }\textbf {\bibinfo {volume} {9}},\
  \href {https://doi.org/10.1038/s41598-019-46722-7}
  {10.1038/s41598-019-46722-7} (\bibinfo {year} {2019})\BibitemShut {NoStop}%
\bibitem [{\citenamefont {Cai}\ \emph {et~al.}(2020)\citenamefont {Cai},
  \citenamefont {Xu},\ and\ \citenamefont {Benjamin}}]{Cai_2020}%
  \BibitemOpen
  \bibfield  {author} {\bibinfo {author} {\bibfnamefont {Z.}~\bibnamefont
  {Cai}}, \bibinfo {author} {\bibfnamefont {X.}~\bibnamefont {Xu}},\ and\
  \bibinfo {author} {\bibfnamefont {S.~C.}\ \bibnamefont {Benjamin}},\
  }\bibfield  {title} {\bibinfo {title} {Mitigating coherent noise using
  {P}auli conjugation},\ }\bibfield  {journal} {\bibinfo  {journal} {npj
  Quantum Information}\ }\textbf {\bibinfo {volume} {6}},\ \href
  {https://doi.org/10.1038/s41534-019-0233-0} {10.1038/s41534-019-0233-0}
  (\bibinfo {year} {2020})\BibitemShut {NoStop}%
\bibitem [{SPS()}]{SPSA}%
  \BibitemOpen
  \href@noop {} {}\bibinfo {note} {We use Qiskit's implementation of SPSA,
  which is documented at
  \url{https://qiskit.org/documentation/stubs/qiskit.aqua.components.optimizers.SPSA.html}}\BibitemShut
  {NoStop}%
\bibitem [{\citenamefont {Kandala}\ \emph {et~al.}(2017)\citenamefont
  {Kandala}, \citenamefont {Mezzacapo}, \citenamefont {Temme}, \citenamefont
  {Takita}, \citenamefont {Brink}, \citenamefont {Chow},\ and\ \citenamefont
  {Gambetta}}]{Kandala_2017}%
  \BibitemOpen
  \bibfield  {author} {\bibinfo {author} {\bibfnamefont {A.}~\bibnamefont
  {Kandala}}, \bibinfo {author} {\bibfnamefont {A.}~\bibnamefont {Mezzacapo}},
  \bibinfo {author} {\bibfnamefont {K.}~\bibnamefont {Temme}}, \bibinfo
  {author} {\bibfnamefont {M.}~\bibnamefont {Takita}}, \bibinfo {author}
  {\bibfnamefont {M.}~\bibnamefont {Brink}}, \bibinfo {author} {\bibfnamefont
  {J.~M.}\ \bibnamefont {Chow}},\ and\ \bibinfo {author} {\bibfnamefont
  {J.~M.}\ \bibnamefont {Gambetta}},\ }\bibfield  {title} {\bibinfo {title}
  {Hardware-efficient variational quantum eigensolver for small molecules and
  quantum magnets},\ }\href {https://doi.org/10.1038/nature23879} {\bibfield
  {journal} {\bibinfo  {journal} {Nature}\ }\textbf {\bibinfo {volume} {549}},\
  \bibinfo {pages} {242–246} (\bibinfo {year} {2017})}\BibitemShut {NoStop}%
\bibitem [{\citenamefont {Abraham}\ \emph {et~al.}(2019)\citenamefont {Abraham}
  \emph {et~al.}}]{Qiskit}%
  \BibitemOpen
  \bibfield  {author} {\bibinfo {author} {\bibfnamefont {H.}~\bibnamefont
  {Abraham}} \emph {et~al.},\ }\href {https://doi.org/10.5281/zenodo.2562110}
  {\bibinfo {title} {Qiskit: An open-source framework for quantum computing}}
  (\bibinfo {year} {2019})\BibitemShut {NoStop}%
\bibitem [{\citenamefont {Asfaw}\ \emph {et~al.}(2020)\citenamefont {Asfaw}
  \emph {et~al.}}]{Qiskit-Textbook}%
  \BibitemOpen
  \bibfield  {author} {\bibinfo {author} {\bibfnamefont {A.}~\bibnamefont
  {Asfaw}} \emph {et~al.},\ }\href {http://community.qiskit.org/textbook}
  {\emph {\bibinfo {title} {Learn Quantum Computation Using Qiskit}}}\
  (\bibinfo {year} {2020})\BibitemShut {NoStop}%
\bibitem [{ibm()}]{ibmq_properties}%
  \BibitemOpen
  \href@noop {} {}\bibinfo {note} {See
  \url{https://quantum-computing.ibm.com/lab/docs/iql/manage/systems/properties}.}\BibitemShut
  {Stop}%
\bibitem [{\citenamefont {Funcke}\ \emph {et~al.}(2020)\citenamefont {Funcke},
  \citenamefont {Hartung}, \citenamefont {Jansen}, \citenamefont {Kühn},
  \citenamefont {Stornati},\ and\ \citenamefont
  {Wang}}]{funcke2020measurement}%
  \BibitemOpen
  \bibfield  {author} {\bibinfo {author} {\bibfnamefont {L.}~\bibnamefont
  {Funcke}}, \bibinfo {author} {\bibfnamefont {T.}~\bibnamefont {Hartung}},
  \bibinfo {author} {\bibfnamefont {K.}~\bibnamefont {Jansen}}, \bibinfo
  {author} {\bibfnamefont {S.}~\bibnamefont {Kühn}}, \bibinfo {author}
  {\bibfnamefont {P.}~\bibnamefont {Stornati}},\ and\ \bibinfo {author}
  {\bibfnamefont {X.}~\bibnamefont {Wang}},\ }\href@noop {} {\bibinfo {title}
  {Measurement error mitigation in quantum computers through classical bit-flip
  correction}} (\bibinfo {year} {2020}),\ \Eprint
  {https://arxiv.org/abs/2007.03663} {arXiv:2007.03663 [quant-ph]} \BibitemShut
  {NoStop}%
\end{thebibliography}%

\appendix

\renewcommand\thefigure{\thesection.\arabic{figure}}
\setcounter{figure}{0}

\section{Perturbation theory for the mixed-field Ising model}
\label{sec:appendix_pert}
The mixed-field Ising Hamiltonian (Eq.~\ref{eq:ising}) is analytically solvable in several limits, as discussed in the main text. In the small-$h_x$ limit, it is diagonal in the computational basis, and the ground state (assuming $J>0$ and $h_z > 0$) approaches $|00\ldots00\rangle$. The ground state energy, to second order in $h_x$ is
\begin{equation}
    E_{\rm gs} = -n\left( h_z + J + \frac{h_x^2}{2 h_z + 4 J} + O(h_x^3) \right) .
\end{equation}

Further, in the large $h_x$ limit, the ground state approaches $|++\ldots++\rangle$, where $|+\rangle = \frac{1}{\sqrt 2} (|0\rangle + |1\rangle).$ In this limit, the ground state energy is
\begin{equation}
    E_{\rm gs} = -n \left( h_x + \frac{2h_z^2 + J^2}{4h_x} + O\left(h_x^{-2}\right)\right).
\end{equation}

Finally, in the small $h_z$ limit, the Hamiltonian approaches the transverse-field Ising model, which, in the large-$n$ limit, is exactly solvable via a Jordan-Wigner transformation \cite{PFEUTY197079}. Deriving an analytical expression for the ground state energy in the small $h_z$ limit is beyond the scope of this work. Instead, we numerically compute the energy to second-order in perturbation theory in this limit (i.e. using the exact numerical eigenvectors from $h_z = 0$).

\begin{figure}[p]
    \centering
    \includegraphics[width=0.45\textwidth]{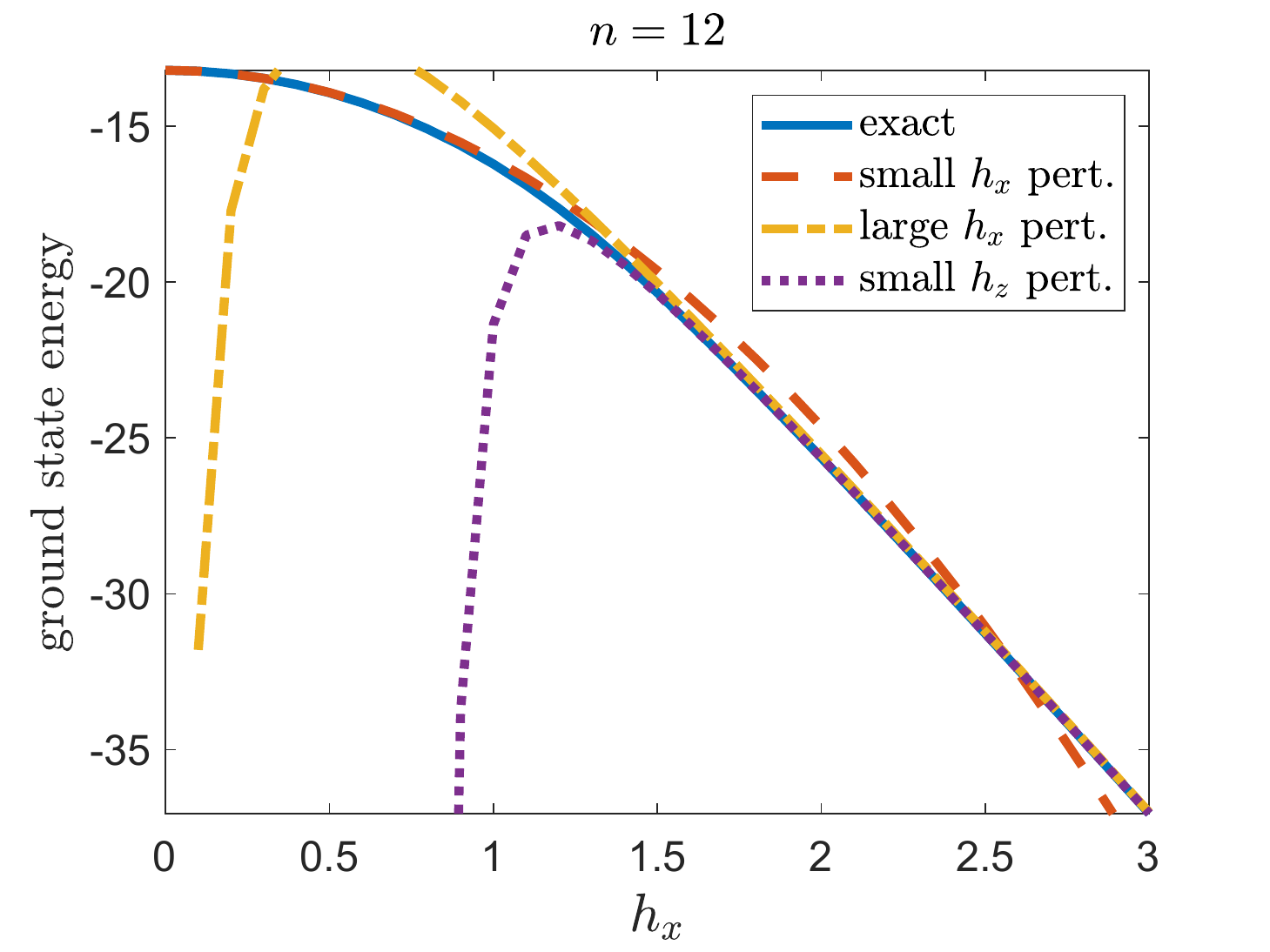}
    \includegraphics[width=0.45\textwidth]{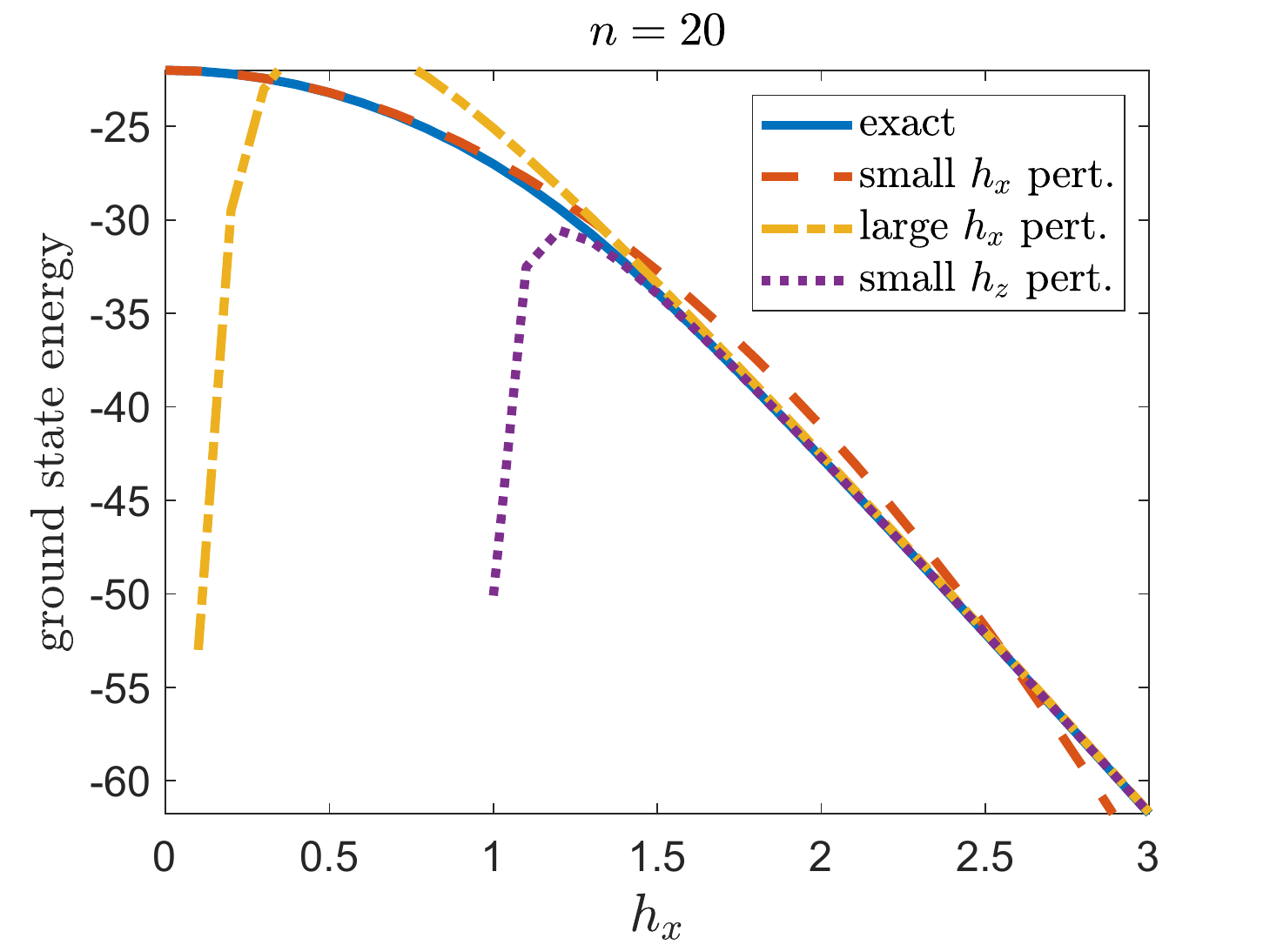}
    \caption{Exact ground state energy of the mixed-field Ising model (Eq.~\ref{eq:ising}) and predictions from second-order perturbation theory, for $h_z = 0.1$. Relevant to this work, we see that $h_x \leq 0.5$ is in the small $h_x$ perturbative regime, while $h_x = 1.5$ is outside of it (at least to second order). For the small-$h_z$ limit, we cheat slightly and use exact diagonalization to compute the eigenstates and eigenenergies when $h_z = 0$. We thus include the cyclic boundary term that is neglected in analytic treatments (cf., Eq.~2.4 in \cite{PFEUTY197079}, and the following discussion), so the small-$h_z$ line should be regarded as slightly better than one could do analytically, but approaching the analytical solution as the number of spins increases.}
    \label{fig:ising_pert}
\end{figure}

The perturbative results are compared to the exact ground state energies in Fig.~\ref{fig:ising_pert}.

\section{A note about error bars}
\label{sec:errorbars}
Throughout this paper, error bars indicate statistical errors only. There are, of course, a variety of systematic errors as well. However, the present goal is to correct these systematic errors, so it is more appropriate for error bars to only show statistical errors, which result from having a finite number of counts. We consider the variation among the qubit assignments used in the qubit assignment averaging to be a systematic error, as the same choices of qubit assignments are used in the calibration circuits as in the circuit that we wish to correct. However, we consider the variation over the different ``perturbative" points in the method introduced in Sec.~\ref{sec:pert} to be statistical error, since a larger sample of such points can always be taken.

\section{Variational ansatz circuit}
\label{sec:basis_gates}
\begin{figure}[p]
    \centering
%     \begin{equation*}
%     \Qcircuit @C=1.0em @R=0.0em @!R{
%     \qw & \gate{R_y} & \ctrl{1} & \gate{R_y} & \qw & \ctrl{5} & \gate{R_y} & \ctrl{1} & \gate{R_y} & \qw \\
%     \qw & \gate{R_y} & \targ & \gate{R_y} & \targ & \qw & \gate{R_y} & \targ & \gate{R_y} & \qw \\
%   \qw & \gate{R_y} & \ctrl{1} & \gate{R_y} & \ctrl{-1} & \qw & \gate{R_y} & \ctrl{1} & \gate{R_y} & \qw \\
%   \qw & \gate{R_y} & \targ & \gate{R_y} & \targ & \qw & \gate{R_y} & \targ & \gate{R_y} & \qw\\
%     \qw & \gate{R_y} & \ctrl{1} & \gate{R_y} & \ctrl{-1} & \qw & \gate{R_y} & \ctrl{1} & \gate{R_y} & \qw \\
%     \qw & \gate{R_y} & \targ & \gate{R_y} & \qw & \targ & \gate{R_y} & \targ & \gate{R_y} & \qw
%     }
%     \end{equation*}
    \includegraphics{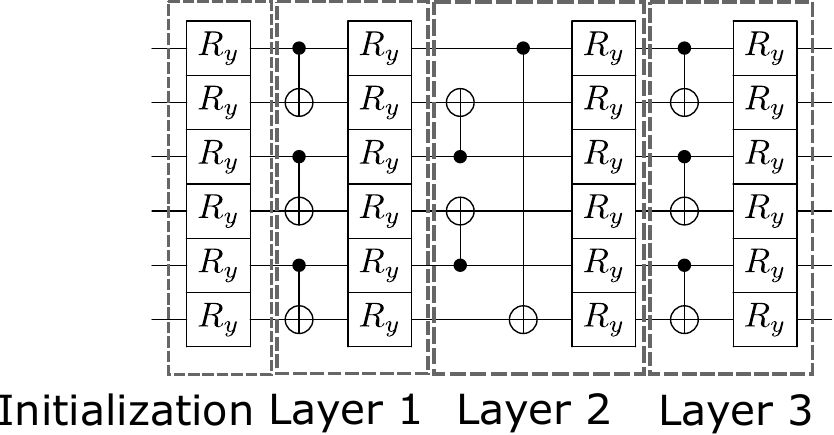}
    \caption{The Alternating Layered Ansatz that we use throughout this work, shown for 6 qubits and 3 layers. When we impose partial cyclic permutation symmetry, we demand that, within a layer of $y$ rotations, all of the rotations on even qubits have the same angle, and all of the rotations on odd qubits have the same angle.}
    \label{fig:ansatz}
\end{figure}

As described in the main text, our ansatz consists of CNOTs between adjacent qubits and Y-rotations on all qubits (Fig.\ \ref{fig:ansatz}). CNOT is a basis gate on IBM's quantum computers. However, the Y-rotations and the change of basis required to measure $X_i$ must also be written in terms of basis gates. We use the following basis-gate decompositions:
\begin{equation}
\begin{split}
    R_y(\theta) &= R_z(\pi) \sqrt{X} R_z(\pi+\theta) \sqrt{X}\\
    H R_y(\theta) &= R_z(\pi) \sqrt{X} R_z(3\pi/2 - \theta) \sqrt{X} R_z(-\pi).
\end{split}
\end{equation}
\textit{ibmq\_toronto} and \textit{ibmq\_sydney} include 12- and 20-qubit loops, the topologies studied here, so we have not needed to apply any gates between non-neighboring qubits.

\begin{figure}[p]
    \centering
    \begin{subfigure}{0.4\textwidth}
        \includegraphics[width=\textwidth]{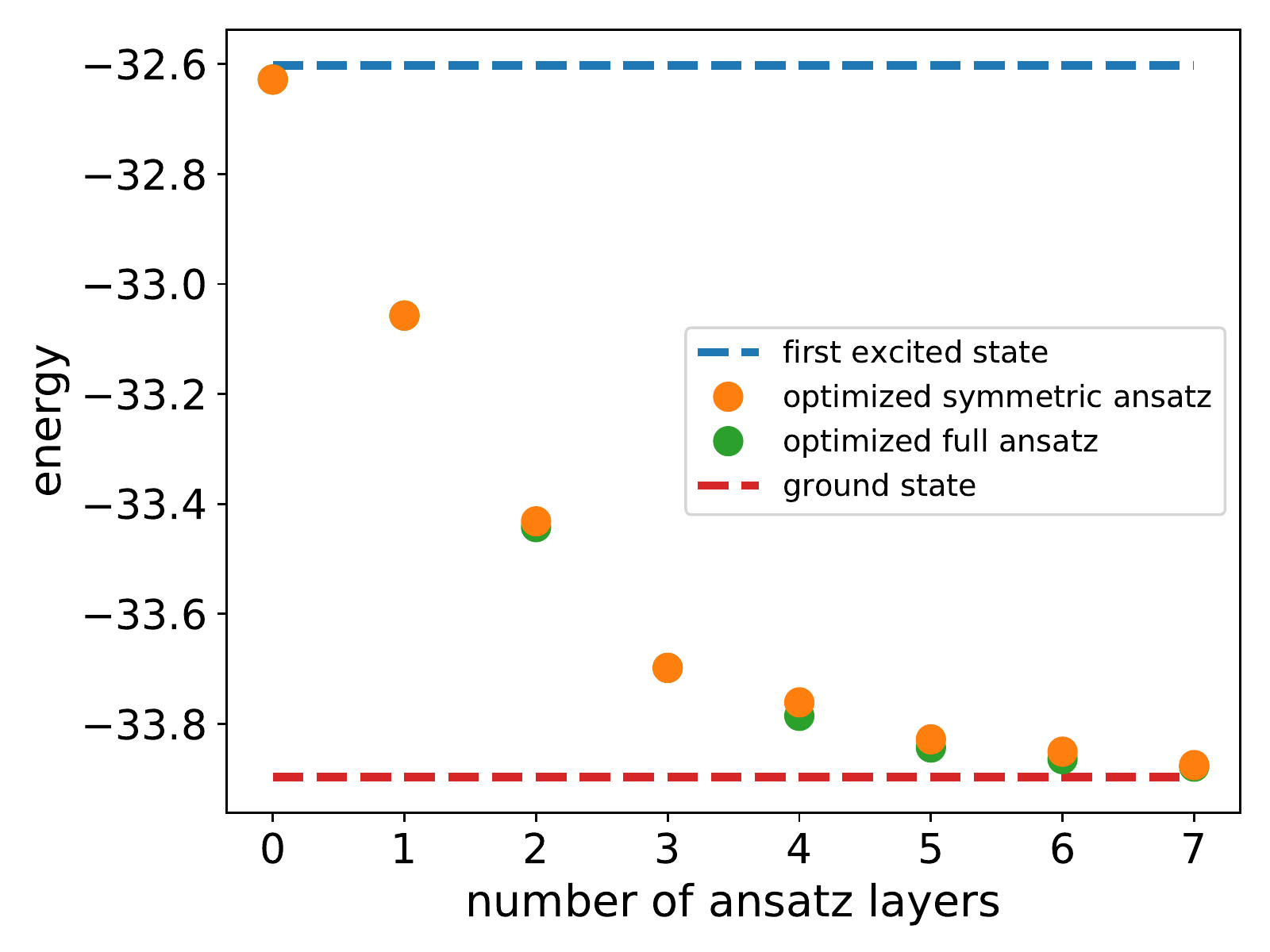}
    \end{subfigure}
    \caption{Ansatz performance in representing ground states of the mixed-field Ising Hamiltonian on 20 qubits, with $h_x = 1.5$ and $h_z = 0.1$. The symmetric ansatz is able to represent the ground states nearly as well as the full ansatz with the same number of layers. Further, relatively few layers are needed to capture the ground state. However, we experiment with many more layers in this paper so as to benchmark error mitigation techniques.}
    \label{fig:ansatz_performance}
\end{figure}

\section{Poorly performing methods}
\label{sec:poorly_performing_methods}
\begin{figure}[h]
    \centering
    \includegraphics[width=0.4\textwidth]{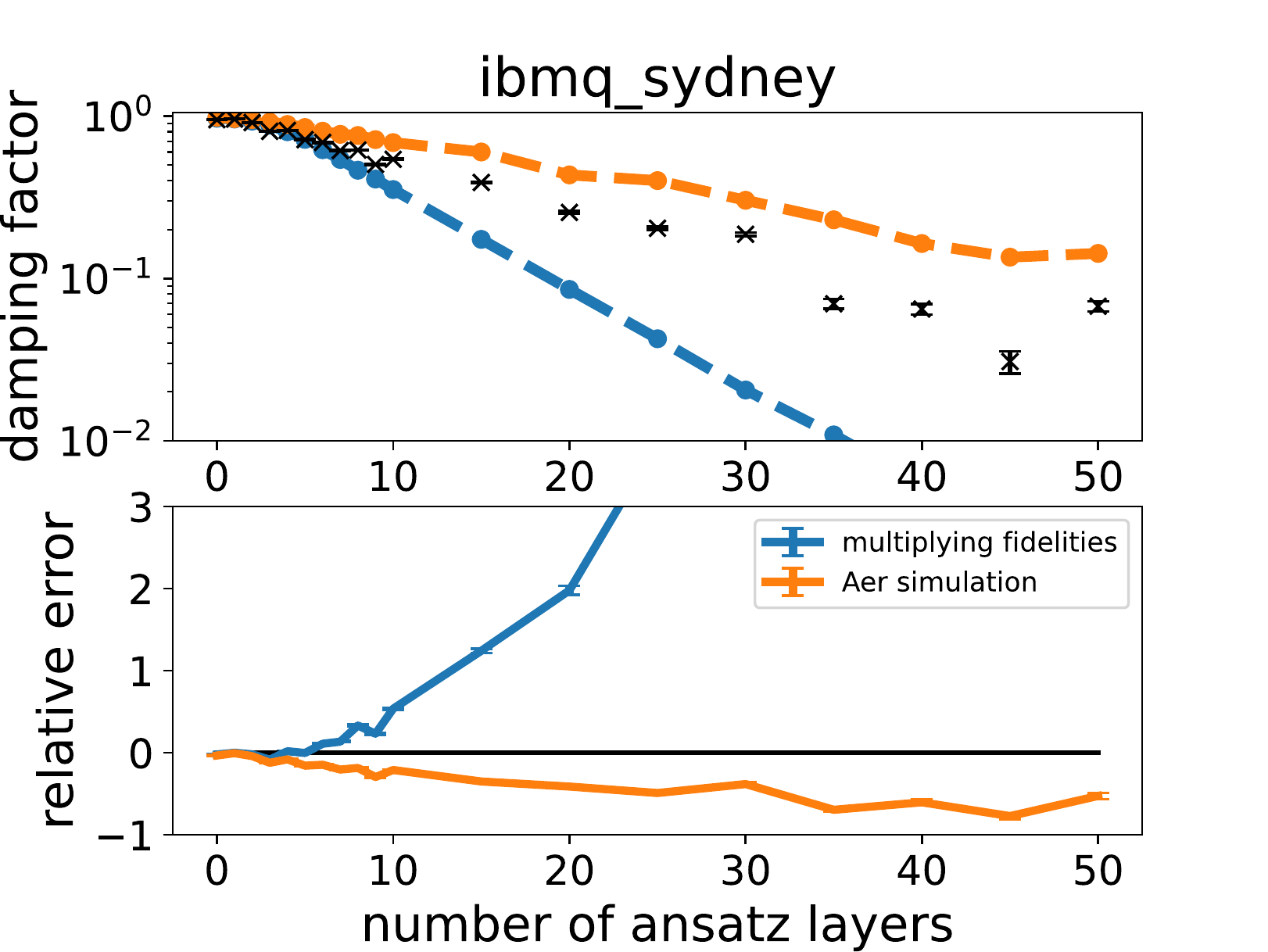}
    \includegraphics[width=0.4\textwidth]{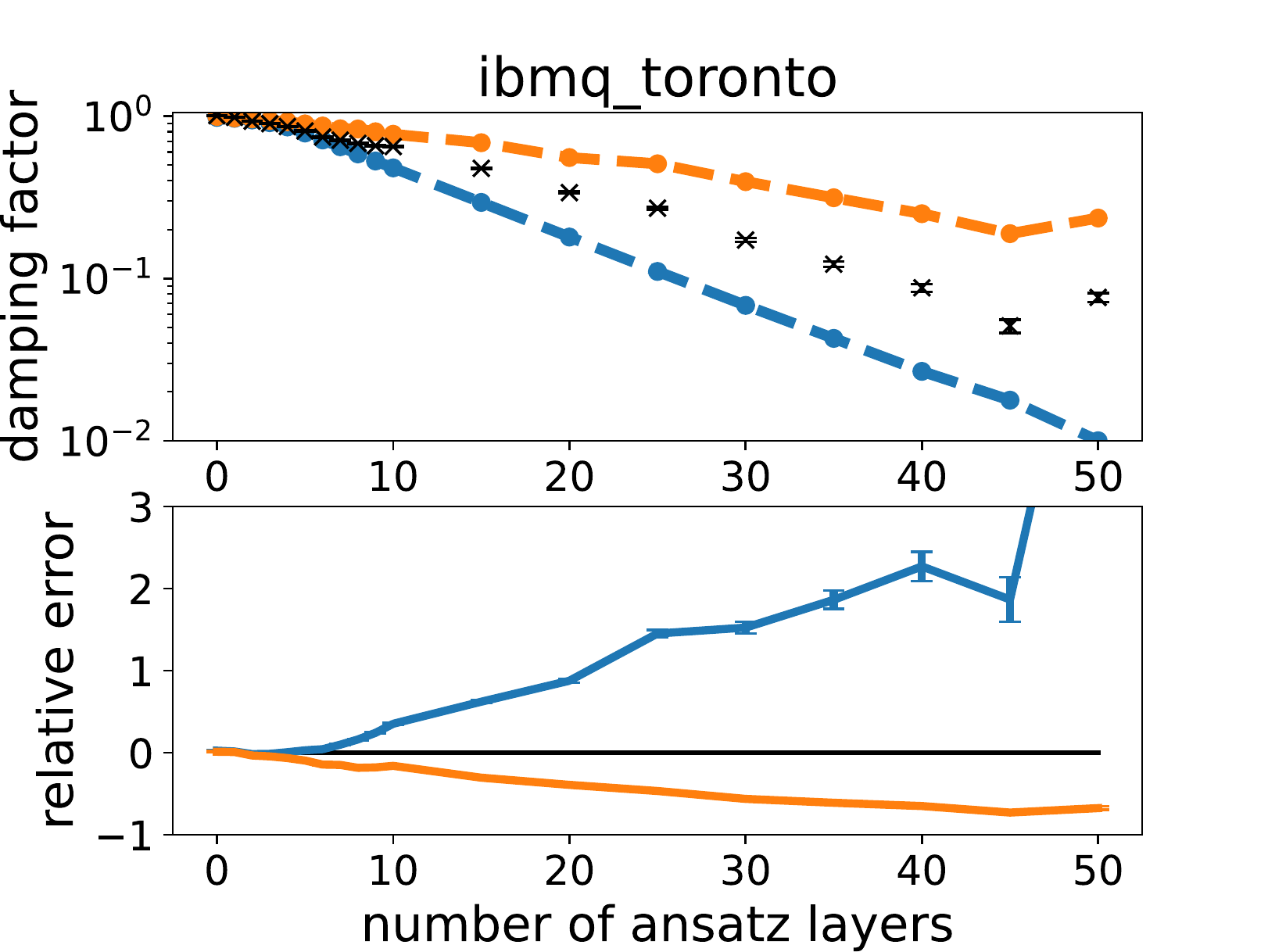}
    \caption{The measured damping factor (black ``x"s) versus the number of ansatz layers, for classically optimized circuits that are then run on \textit{ibmq\_sydney} or \textit{ibmq\_toronto}. The circuits are optimized to the ground state of the 12-qubit mixed-field Ising Hamiltonian with $h_x = 1.5$ and $h_z = 0.1$. The blue line shows the damping factor that is predicted from multiplying the fidelities of the gates in the backwards light cone of the measured qubits. The orange line shows the prediction from a Qiskit Aer noise model with local depolarizing errors, thermal relaxation errors, and readout errors. In both cases, the error rates are obtained from the device's calibration. For deep circuits, multiplying the gate fidelities consistently predicts too much suppression, whereas the Qisit Aer model with local gate errors predicts too little suppression. }
    \label{fig:from_gate_fidelities}
\end{figure}

In addition to the methods discussed in the main text, we also benchmark two methods that do not perform as well. In one of these methods, which we call ``multiplying fidelities," we multiply $(1-e_i)$, where $e_i$ is the reported error rate for gate $i$, over all of the gates appearing in the backwards light cone of the observed qubits. This is then taken as an estimate of the overall damping factor. In Fig.~\ref{fig:from_gate_fidelities}, it is clear that this method predicts more suppression than actually exists. The other poorly performing method is to use the noise model in Qiskit Aer, which includes local depolarizing channels on all of the gates, thermal relaxation errors, and readout errors. This method is not scalable since it requires simulating the circuit, but we benchmark it for comparison. It consistently predicts less suppression than exists.

\section{Readout error mitigation}
\label{sec:readout_errors_intro}
In the standard technique for readout error mitigation on $n$ measured qubits, one prepares the qubits in each of the $2^n$ computational basis states and measures the probabilities of reading out the different outcomes (e.g. \cite{Qiskit-Textbook}). This results in a $2^n\times 2^n$ matrix, which is inverted to perform the mitigation. When $n$ is large, this method becomes intractable and one must make simplifying assumptions, such as ignoring correlations among the qubits or among sets of qubits. For the Mixed-Field Ising Hamiltonian (Eq.\ \ref{eq:ising}), we only need to measure at most two qubits simultaneously, so the standard readout error mitigation method is easily tractable. Nevertheless, we assume uncorrelated readout errors and use the methods developed in Appendix \ref{sec:readout_errors} to perform the mitigation. We use the readout error rates measured by IBM during their approximately daily calibrations \cite{ibmq_properties}. We have found that this works about as well as doing the full readout error mitigation using our own calibrations.

\section{Uncorrelated readout errors}
\label{sec:readout_errors}

In this appendix, we show that the effect of uncorrelated readout errors is to reduce the expectation value of a Pauli operator by the factor $(1-2e_m)$, for each qubit measured, where $e_m = (e_0 + e_1)/2$ is the average readout error, and offset it by a small additive amount. Some similar results appear in \cite{funcke2020measurement}.

Suppose that we want to measure a Pauli string of length $N$ by applying readout gates to all of the qubits. If the readout gates were noiseless, we would measure outcome $q$ with probability $f(q)$. The noiseless expectation value of the Pauli operator is then $\langle P \rangle = \sum_q P(q) f(q)$, where $P(q) = \pm 1$ is the bit parity of $q$. Now, suppose that the measurement gates add noise such that the probability of a 0 getting measured as a 1 is $e_0$ and the probability of a 1 getting measured as a 0 is $e_1$. Let $\tilde f(q)$ be the probability distribution of the outcomes, including this readout error. Let $\tilde{\langle P \rangle} = \sum_q P(q) \tilde f(q)$.

The main result that we will establish is that
\begin{equation}
\label{eq:mainResult}
    \tilde{\langle P \rangle} = \sum_q P(q) f(q) (1-2e_0)^{n_0(q)} (1-2e_1)^{n_1(q)},
\end{equation}
where $n_0(q)$ is the number of bits in $q$ that are 0, and $n_1(q)$ is the number of bits in $q$ that are 1. If the two error rates are equal ($e_0=e_1=e_m$), Eq.~\ref{eq:mainResult} becomes
\begin{equation}
    \langle \tilde P \rangle =  \langle P \rangle (1-2e_m)^N, \hspace{10pt} e_0=e_1=e_m.
\end{equation}
However, we can also simplify Eq.\ \ref{eq:mainResult} without assuming the error rates are equal. For $N=1$, $\langle P \rangle = f(0) - f(1)$, so Eq.\ \ref{eq:mainResult} becomes
\begin{equation}
\label{eq:N1}
    \langle \tilde P \rangle = \langle P \rangle (1-e_0 -e_1) + e_1 - e_0, \hspace{10pt} N = 1.
\end{equation}

\begin{figure}
    \centering
    \includegraphics[width=0.4\textwidth]{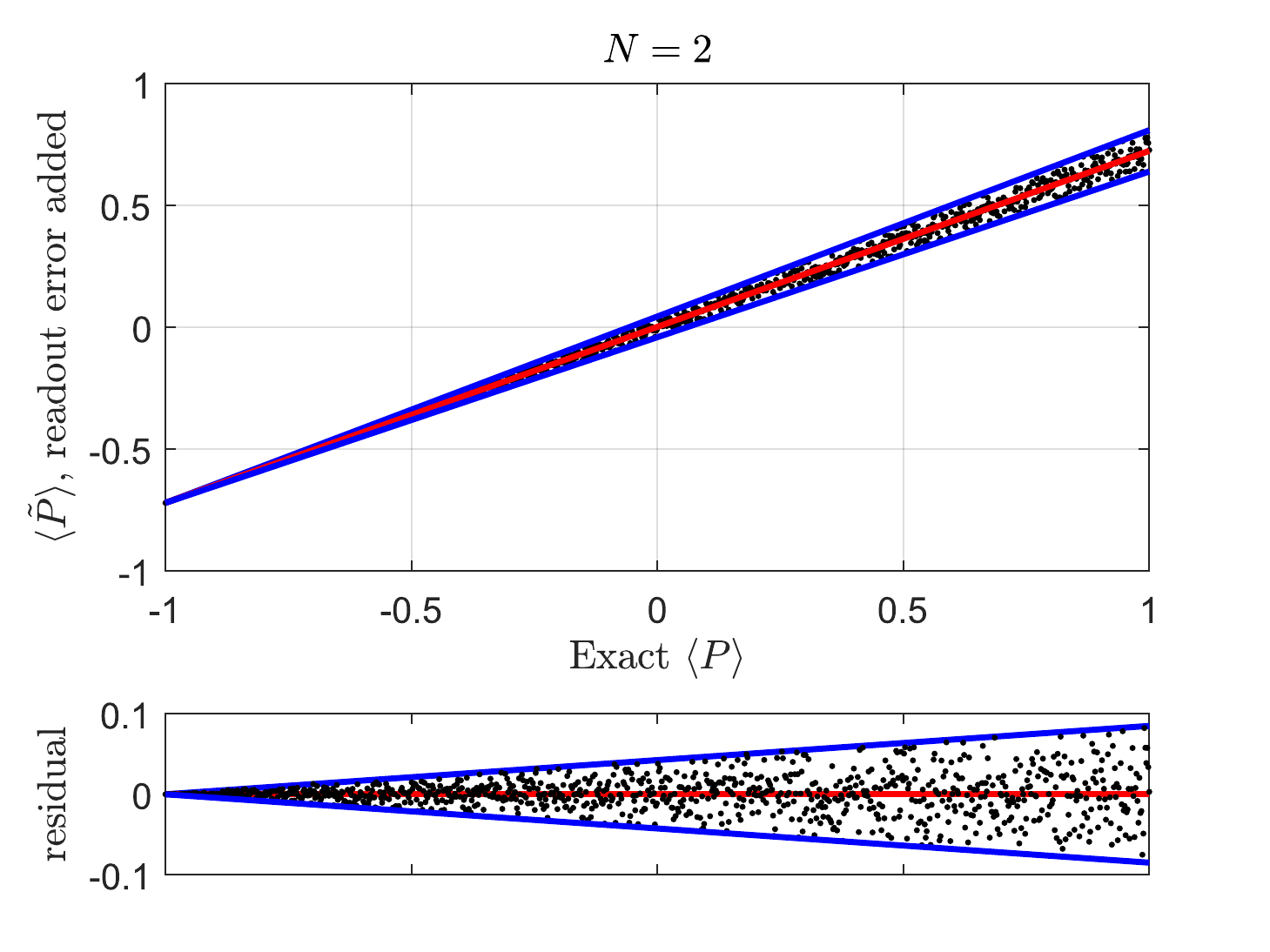}
    \caption{In this Figure and in Fig.~\ref{fig:N3}, we generate random probability distributions $f(q)$ that give the desired expected values of bit parity $\langle P \rangle$, which are equally spaced between $-1$ and $1$. $\langle \tilde P \rangle$ is then calculated using Eq.~\ref{eq:mainResult}. The resulting data are compared to Eq.~\ref{eq:N2}, and the residuals $\hat R$ are plotted. In this Figure, two qubits are measured, and the effect of the readout error is bounded by Eq.~\ref{eq:N2_ineq} (blue lines) and approximated by \ref{eq:N2} (red line). We set $e_1 = 0.1$ and $e_0 = 0.05$.}
    \label{fig:N2}
\end{figure}

\begin{figure}
    \centering
    \includegraphics[width=0.45\textwidth]{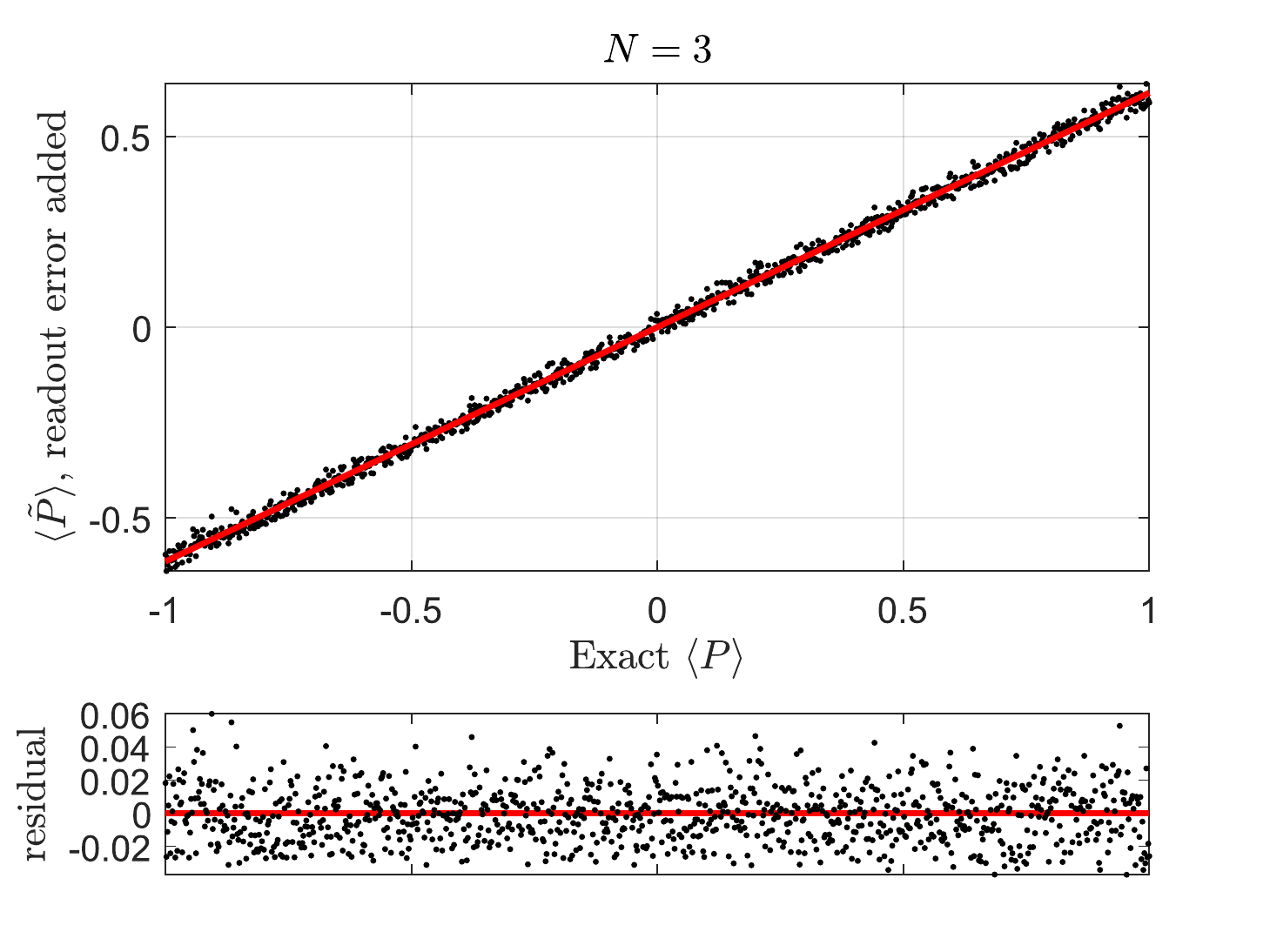}
    \includegraphics[width=0.45\textwidth]{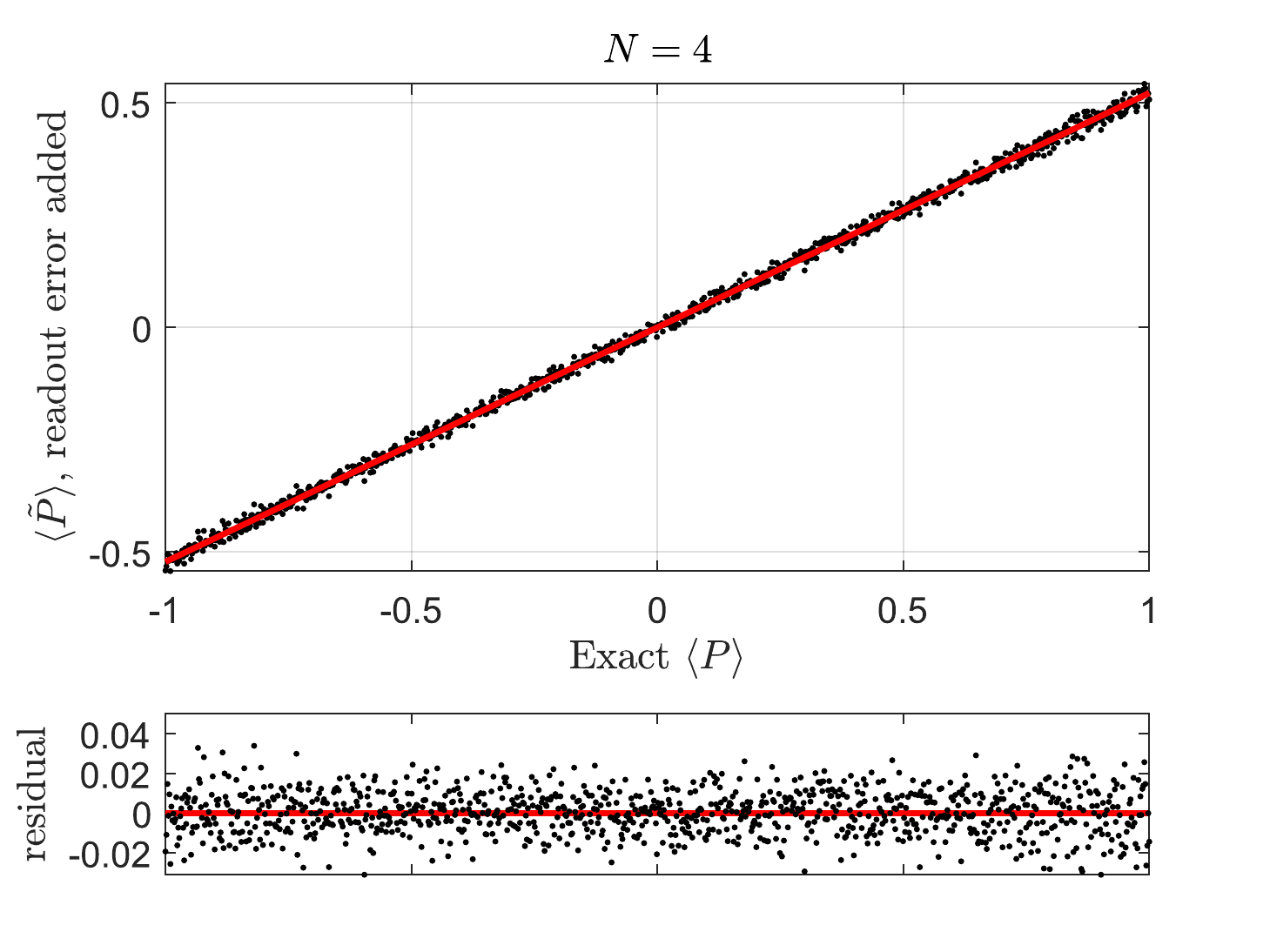}
    \includegraphics[width=0.45\textwidth]{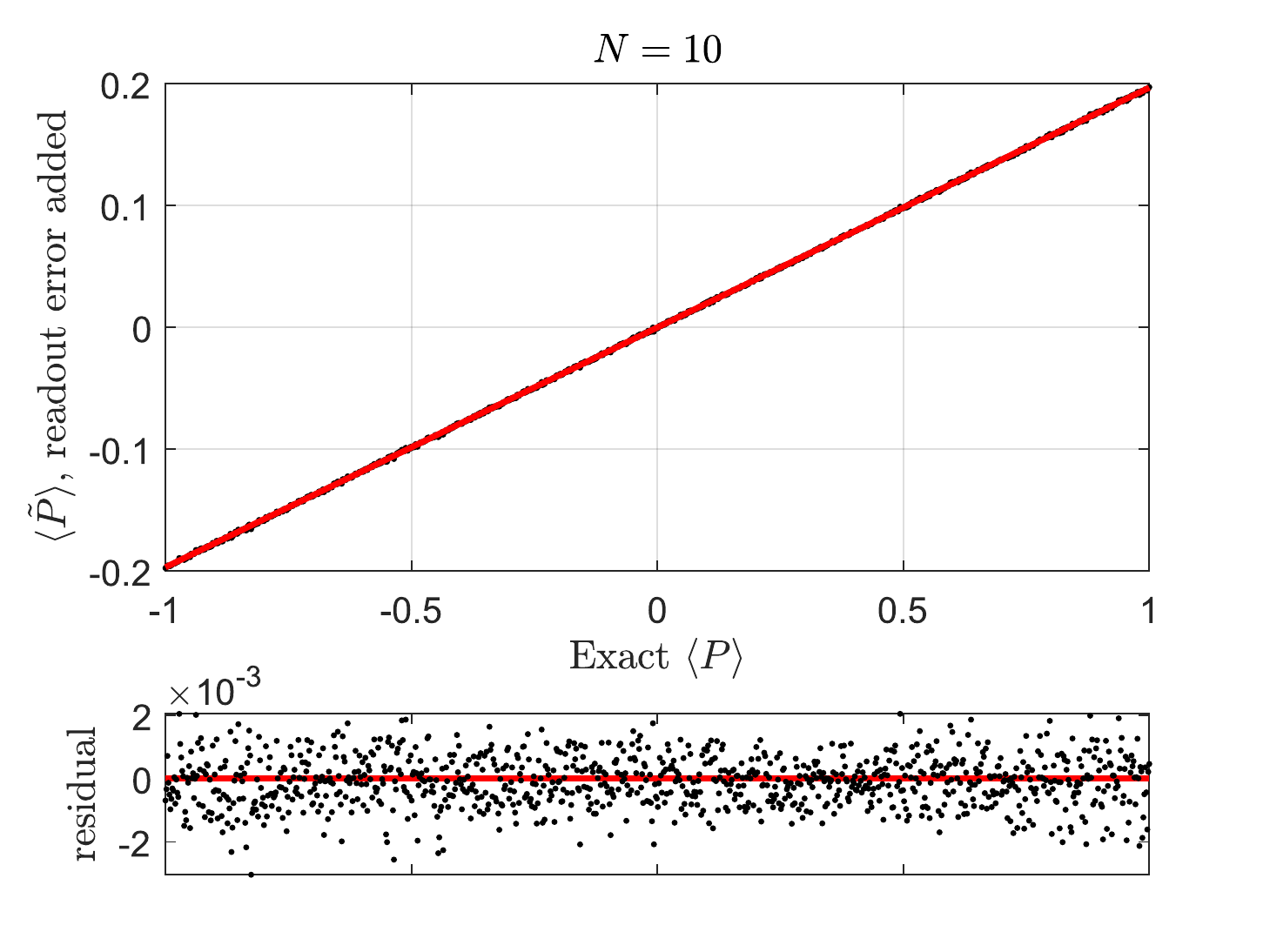}
    \includegraphics[width=0.45\textwidth]{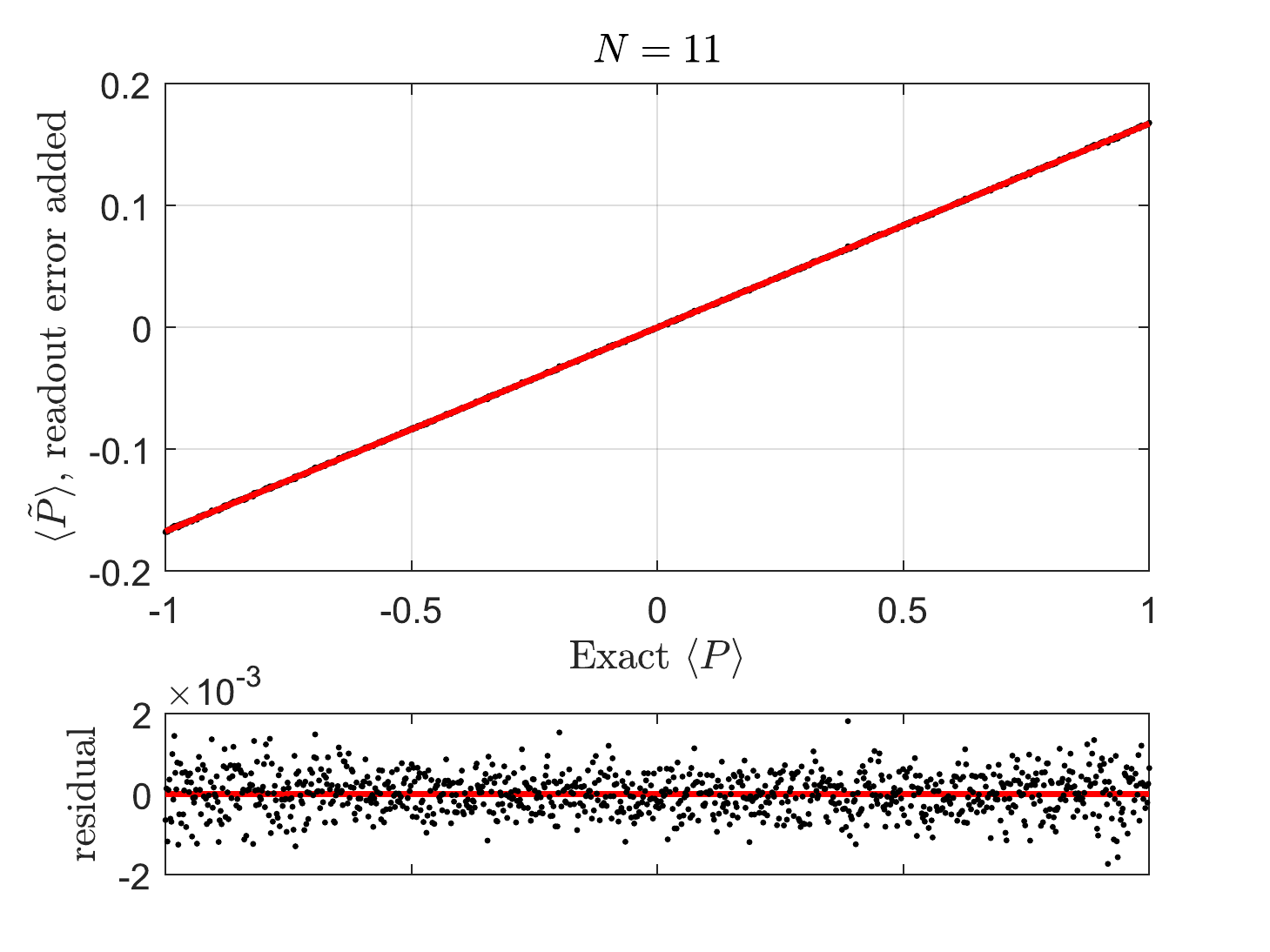}
    \caption{Same as Fig.~\ref{fig:N2} but reading out 3, 4, 10, and 11 qubits, respectively. Eq.~\ref{eq:avgResult} (red line) is a good approximation and becomes better as the number of measured qubits increases.}
    \label{fig:N3}
\end{figure}

For larger $N$, $\langle P \rangle$ does not determine $f(q)$. However, the largest and smallest $\langle \tilde P \rangle$ for a given $\langle P \rangle$ will always result from letting $f(q)$ be nonzero for at most two choices of $n_1(q)$, one with $n_1$ even and the other with $n_1$ odd. Define these values as $n_1^+$ and $n_1^-$, respectively. Define $n_0^\pm = N - n_1^\pm$ and $f_\pm$ to be the corresponding probabilities. Then $\langle P \rangle = f_+ - f_-$, and so Eq.~\ref{eq:mainResult} becomes
\begin{equation}
\begin{split}
    \langle \tilde P \rangle &= \left(\frac{1+ \langle P \rangle}{2}\right) \left( 1 - 2e_0\right)^{n_0^+} \left( 1 - 2e_1\right)^{n_1^+} - \left(\frac{1- \langle P \rangle}{2} \right) \left( 1 - 2e_0\right)^{n_0^-} \left( 1 - 2e_1\right)^{n_1^-}
\end{split}
\end{equation}
Without loss of generality, we assume that $e_1 \geq e_0$. We also assume that $e_1 < 0.5$. Then $\langle \tilde P \rangle$ is maximized by picking $n_1^+ = 0$ and $n_1^-$ to be either $N$ or $N-1$, whichever is odd. Similarly, the smallest $ \langle \tilde P \rangle$ is obtained by picking $n_1^+ = N$ or $N-1$, whichever is even, and $n_1^- = 1$. For the case of $N = 2$, this becomes
\begin{equation}
\label{eq:N2_ineq}
    (1-2e_1) \left[ \langle P \rangle (1-e_1-e_0) - (e_1-e_0)\right] \leq \langle \tilde P \rangle \leq (1-2e_0) \left[ (1-e_0-e_1) \langle P \rangle + e_1 - e_0  \right], \hspace{10pt} N = 2.
\end{equation}
The midpoint between these two bounds gives an estimate of $\langle \tilde P \rangle$ when $N = 2$:
\begin{equation}
\label{eq:N2}
    \langle \tilde P \rangle \approx \langle P \rangle (1-e_0 - e_1)^2 \langle P \rangle + (e_1 - e_0)^2, \hspace{10pt} N = 2.
\end{equation}
Note that the $y$-intercept is suppressed compared to the $N=1$ case (Eq.~\ref{eq:N1}). Indeed, it becomes more suppressed as we go to higher $N$. Numerically, we find that, for $N\gtrsim 3$, for randomly chosen $f(q)$,
\begin{equation}
\label{eq:avgResult}
    \langle \tilde P \rangle \approx (1-e_0-e_1)^N\langle P \rangle, \hspace{10pt} N \gtrsim 3
\end{equation}
where this becomes a better approximation as $N$ increases. We present numerical evidence for this result in Figs.~\ref{fig:N2}--\ref{fig:N3}.

\subsection{Proof of Eq.\ \ref{eq:mainResult}}
Consider two bit strings, both of length $N$, $q$ and $q'$. Define the following quantities:
\begin{equation}
    \begin{split}
        k_0(q,q') &= \sum_{i=0}^{N-1} \I_{q'_i=0} \I_{q_i=1}, \hspace{50pt} u_0(q,q') = \sum_{i=0}^{N-1} \I_{q'_i=0} \I_{q_i=0},\\
        k_1(q,q') &= \sum_{i=0}^{N-1} \I_{q'_i=1} \I_{q_i=0}, \hspace{50pt} u_1(q,q') = \sum_{i=0}^{N-1} \I_{q'_i=1} \I_{q_i=1}.
    \end{split}
\end{equation}

Then, upon adding readout error, the probability $\tilde f(q)$ of measuring bit string $q$ is
\begin{equation}
    \tilde f(q) = f(q)(1-e_0)^{n_0(q)}(1-e_1)^{n_1(q)} + \sum_{q'\neq q} f(q') e_0^{k_0(q,q')} e_1^{k_1(q,q')} (1-e_0)^{u_0(q,q')} (1-e_1)^{u_1(q,q')}.
\end{equation}
The expectation value of bit parity in the distribution $\tilde f$ is
\begin{equation}
    \begin{split}
        \langle \tilde P \rangle =& \sum_q P(q) f(q) (1-e_0)^{n_0(q)}(1-e_1)^{n_1(q)}
        \\ &+\sum_q \sum_{q'\neq q} f(q') P(q') (-1)^{k_0(q,q') + k_1(q,q')} e_0^{k_0(q,q')} e_1^{k_1(q,q')} (1-e_0)^{n_0(q') - k_0(q,q')}(1-e_1)^{n_1(q') - k_1(q,q')}\\
        =& \sum_q P(q) f(q) (1-e_0)^{n_0(q)}(1-e_1)^{n_1(q)}\\
        &+  \sum_{q'} \sum_{k_0=0}^{n_0(q')} \sum_{k_1=0}^{n_1(q')} \begin{pmatrix} n_0(q')\\k_0\end{pmatrix} \begin{pmatrix}n_1(q')\\k_1\end{pmatrix} f(q') P(q') e_0^{k_0} e_1^{k_1} (1-e_0)^{n_0(q') - k_0} (1-e_1)^{n_1(q') - k_1} (-1)^{k_0+k_1} \I_{k_0 + k_1 > 0}\\
        =& \sum_{q'} \sum_{k_0=0}^{n_0(q')} \sum_{k_1=0}^{n_1(q')} \begin{pmatrix} n_0(q')\\k_0\end{pmatrix} \begin{pmatrix}n_1(q')\\k_1\end{pmatrix} f(q') P(q') e_0^{k_0} e_1^{k_1} (1-e_0)^{n_0(q') - k_0} (1-e_1)^{n_1(q') - k_1} (-1)^{k_0+k_1}\\
        =& \sum_q P(q) f(q) (1-2e_0)^{n_0(q)} (1-2e_1)^{n_1(q)},
    \end{split}
\end{equation}
which is Eq.\ \ref{eq:mainResult}.

\end{document}